\newif\ifdouble
\newif\ifsingle
\newif\ifchange
\doubletrue

\ifdouble
  \documentclass[sigconf, usenames, dvipsnames]{acmart}
\else
  \documentclass[manuscript,anonymous]{acmart}
  \singletrue
\fi

\usepackage{dblfloatfix} 
\usepackage{booktabs} 
\usepackage{arydshln}
\usepackage{subcaption} 
\usepackage{hyperref}
\usepackage{lscape} 
\usepackage{xcolor}

\definecolor{blue}{RGB}{4,114,139}
\newcommand{\blue}[1]{\textcolor{blue}{#1}}
\definecolor{green}{RGB}{15,149,99}
\newcommand{\green}[1]{\textcolor{green}{#1}}
\definecolor{red}{RGB}{206,39,103}
\newcommand{\red}[1]{\textcolor{red}{#1}}

\AtBeginDocument{%
  \providecommand\BibTeX{{%
    \normalfont B\kern-0.5em{\scshape i\kern-0.25em b}\kern-0.8em\TeX}}}

\copyrightyear{2023}
\acmYear{2023}
\setcopyright{acmlicensed}\acmConference[UIST '23]{The 36th Annual ACM Symposium on User Interface Software and Technology}{October 29-November 1, 2023}{San Francisco, CA, USA}
\acmBooktitle{The 36th Annual ACM Symposium on User Interface Software and Technology (UIST '23), October 29-November 1, 2023, San Francisco, CA, USA}
\acmPrice{15.00}
\acmDOI{10.1145/3586183.3606827}
\acmISBN{979-8-4007-0132-0/23/10}

\newcommand{\system}{Augmented Math}

\begin{document}
\pagenumbering{arabic}
\pagestyle{plain}
\title{\system{}: Authoring AR-Based Explorable Explanations by Augmenting Static Math Textbooks}

\author{Neil Chulpongsatorn}
\affiliation{%
\institution{University of Calgary}
\city{Calgary}
\country{Canada}}
\email{thobthai.chulpongsat@ucalgary.ca}
\authornote{Both authors contributed equally to the paper}

\author{Mille Skovhus Lunding}
\orcid{0000-0002-5829-2208}
\affiliation{%
\institution{Aarhus University}
\city{Aarhus}
\country{Denmark}}
\affiliation{%
\institution{University of Calgary}
\city{Calgary}
\country{Canada}}
\email{milledsk@cs.au.dk}
\authornotemark[1]

\author{Nishan Soni}
\affiliation{%
\institution{University of Calgary}
\city{Calgary}
\country{Canada}}
\email{nishan.soni@ucalgary.ca}

\author{Ryo Suzuki}
\orcid{0000-0003-3294-9555} 
\affiliation{%
\institution{University of Calgary}
\city{Calgary}
\country{Canada}}
\email{ryo.suzuki@ucalgary.ca}

\renewcommand{\shortauthors}{Chulpongsatorn, et al.}
\begin{abstract}
We introduce Augmented Math, a machine learning-based approach to authoring AR explorable explanations by augmenting static math textbooks without programming. To augment a static document, our system first extracts mathematical formulas and figures from a given document using optical character recognition (OCR) and computer vision. By binding and manipulating these extracted contents, the user can see the interactive animation overlaid onto the document through mobile AR interfaces. This empowers non-technical users, such as teachers or students, to transform existing math textbooks and handouts into on-demand and personalized explorable explanations. To design our system, we first analyzed existing explorable math explanations to identify common design strategies. Based on the findings, we developed a set of augmentation techniques that can be automatically generated based on the extracted content, which are 1) dynamic values, 2) interactive figures, 3) relationship highlights, 4) concrete examples, and 5) step-by-step hints. To evaluate our system, we conduct two user studies: preliminary user testing and expert interviews. The study results confirm that our system allows more engaging experiences for learning math concepts.
\end{abstract}







\begin{teaserfigure}
\centering
\includegraphics[width=0.247\textwidth]{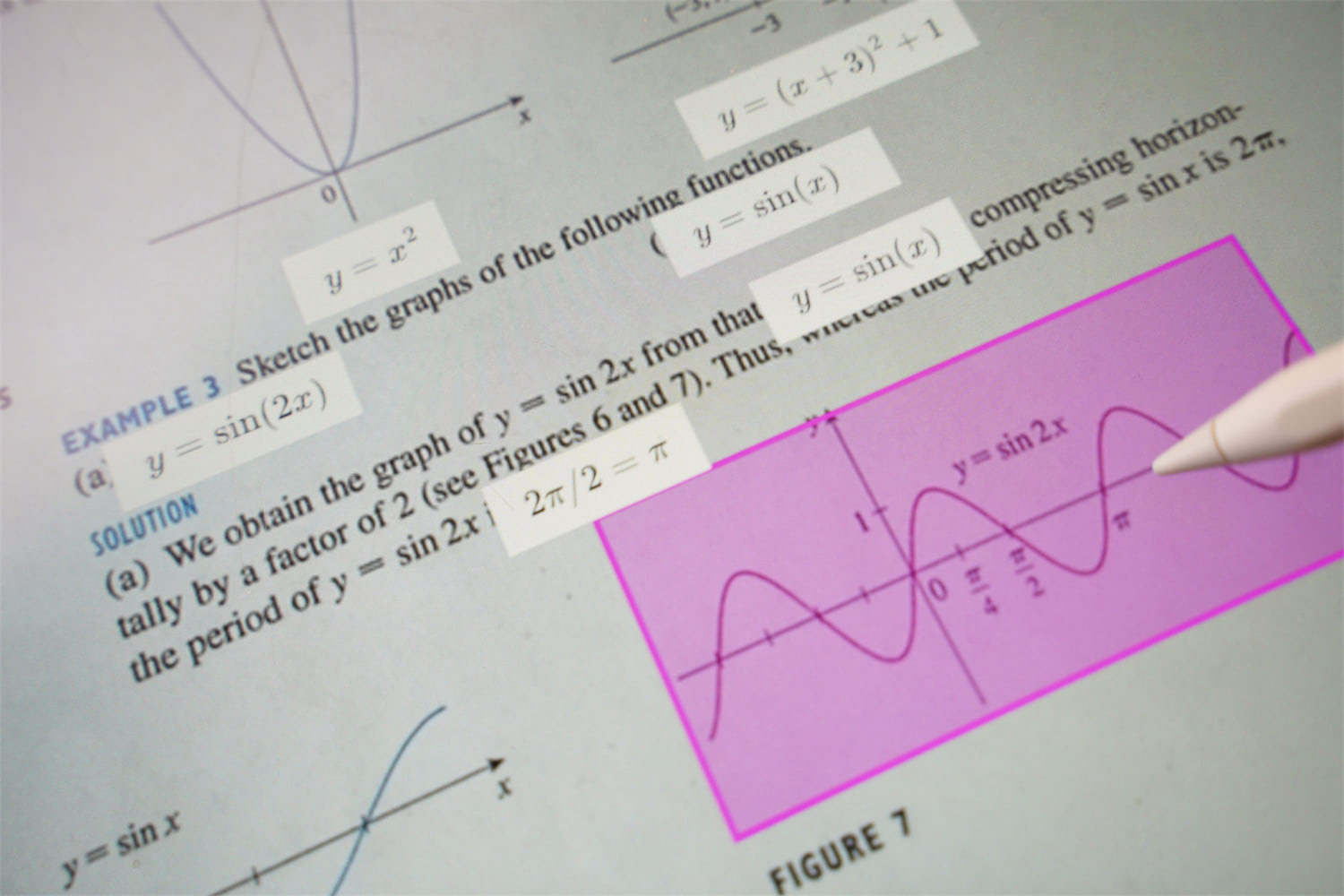}
\includegraphics[width=0.247\textwidth]{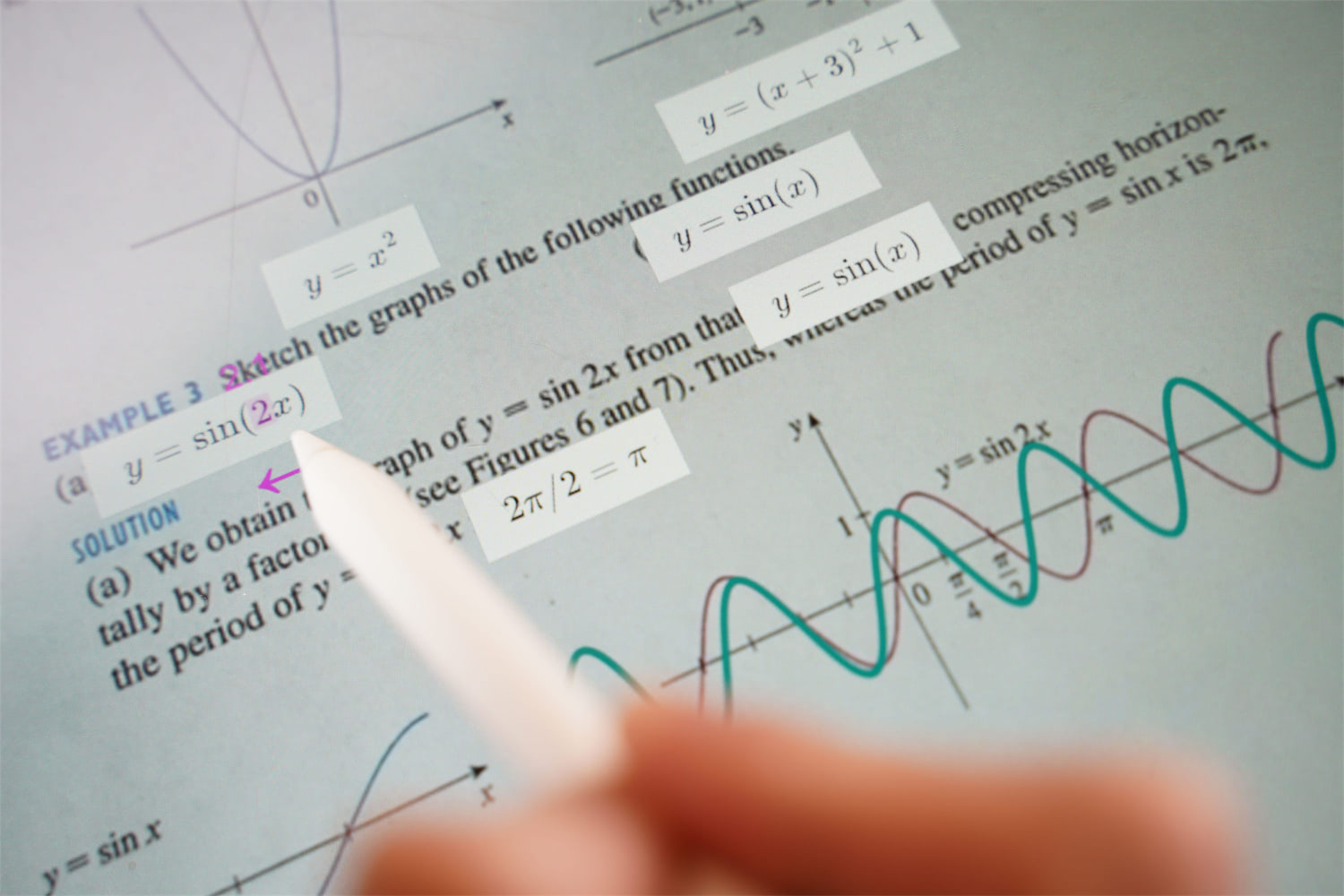}
\includegraphics[width=0.247\textwidth]{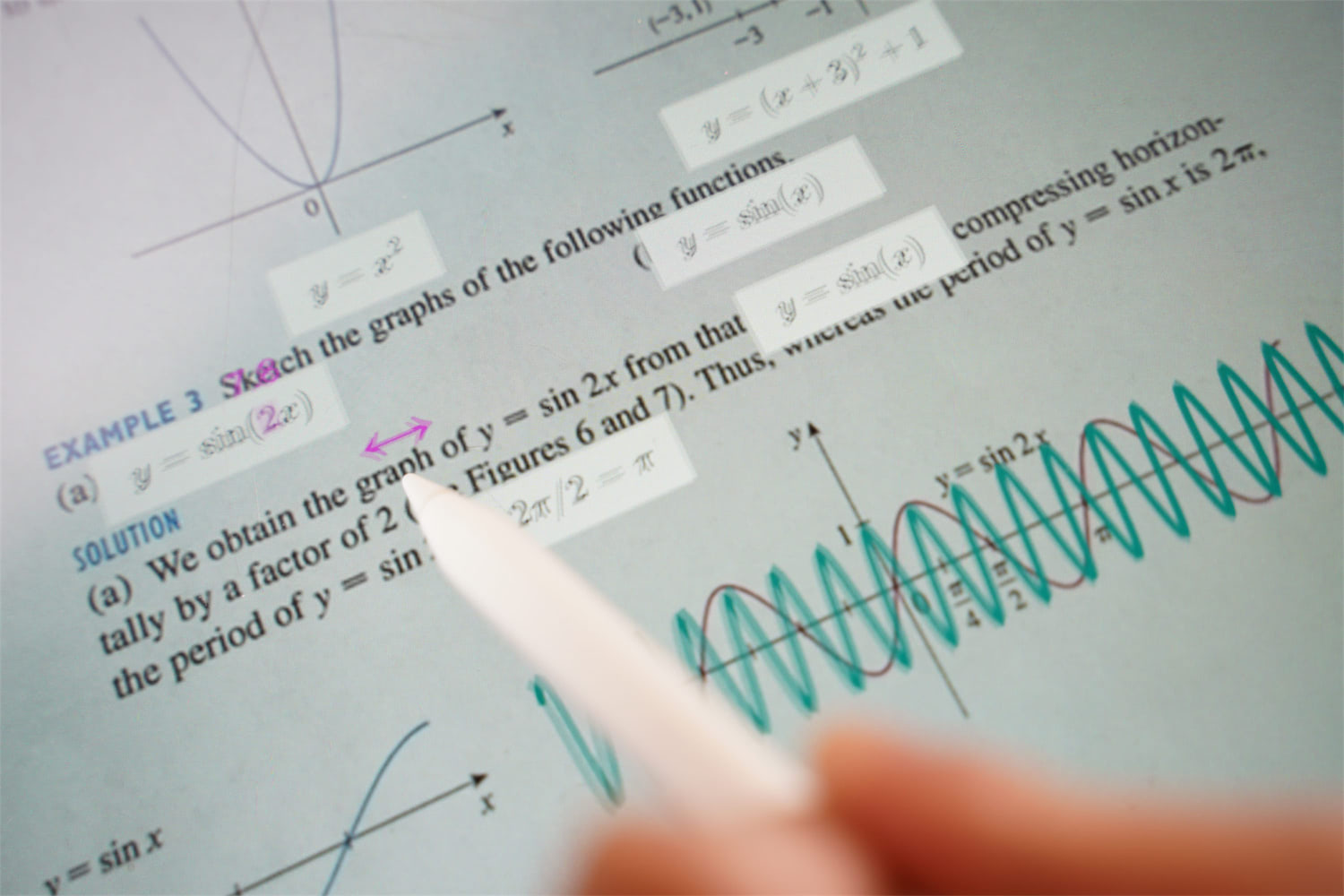}
\includegraphics[width=0.247\textwidth]{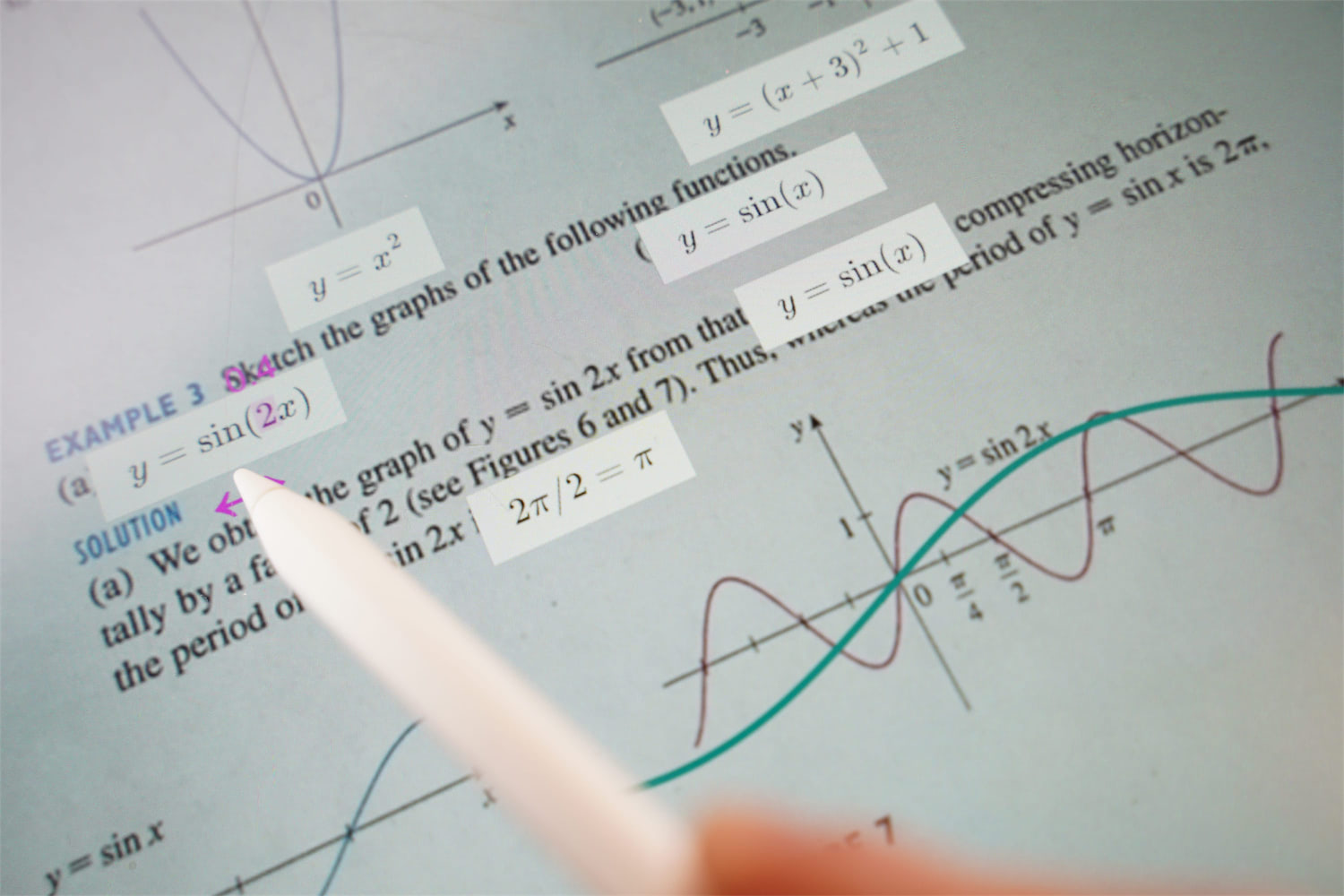}
\caption{\system{} is an authoring tool that transforms static math textbooks into on-demand AR-based explorable explanations. The figure illustrates how a user can use our system to explore changes in the sine curve graph by binding and manipulating the extracted equation in the mobile AR interface.}
\label{fig:teaser}
\end{teaserfigure}

\maketitle

\section{introduction}

\textit{``People currently think of text as information to be consumed. I want text to be used as an environment to think in.'' --- Bret Victor~\cite{victor2011explorable}}

\ \\
Today's textbooks, whether in digital or physical format, primarily consist of \textit{static explanations} that only allow users to passively consume information, without facilitating a deep understanding that comes from interactive dialogue and exploration.
The concept of \textit{explorable explanations}~\cite{victor2011explorable} has been developed to shift this paradigm, transforming text into an interactive medium to think in, rather than just a source of information to be consumed. Such dynamic and interactive explanations have enormous potential as learning aids, as they foster deeper understanding of abstract concepts through engaging and playful explorations, which is otherwise difficult to achieve through static textbooks alone. Therefore, explorable explanations have emerged as a promising approach to changing the way we learn complex concepts across various educational domains, such as math~\cite{GeoGebra95:online}, physics~\cite{perkins2006phet}, programming~\cite{guo2013online, victor2012learnable, suzuki2017tracediff}, data science~\cite{conlen2022fidyll}, and machine learning~\cite{distill, wang2020cnn, smilkov2017direct}.

However, creating these explorable explanations often requires a significant amount of time and effort, as well as substantial programming expertise. This prevents non-technical users, such as teachers or students, from creating their own interactive content tailored to their specific needs and context. As a result, they are often limited to passively utilizing existing interactive content available on the internet, while the majority of their own textbooks remain static.

In this paper, we propose \system{}, a new machine learning (ML) enabled approach to creating explorable explanations by \textit{augmenting} static math textbooks, rather than programming them from scratch. 
Our system first scans a static document to extract values, symbols, and graphs using optical character recognition (OCR) and computer vision techniques. The system then localizes the position of detected math formulas and graphs, converting them into the computer understandable mathematical expression. 
The user can then author interactive content by binding and manipulating these extracted elements.
These interactive visual outputs are embedded and overlaid on the scanned document either through mobile AR for printed paper or through a desktop interface for scanned PDFs.
This empowers non-technical end users to transform their own textbooks into on-demand, personalized, explorable explanations without requiring programming knowledge. 

To design our system, we first analyze the common design strategies widely used in the existing interactive math explanations available on various websites. We identify three high-level categories of these strategies: 1) exemplify through concrete values, 2) visualize through interactive and animated graphs, and 3) guide through contextual hints or exercise.
Based on these design strategies, we develop a set of augmentation techniques for our proof-of-concept system, including 1) dynamic values, 2) interactive figures, 3) relationship highlights, 4) concrete examples, and 5) step-by-step hints. 
These features are designed to be automatically generated based on the extracted content, while covering a wide variety of strategies identified through our analysis. 

To evaluate our system, we conduct two user studies. First, we perform an exploratory study with eleven participants, comparing our system across three conditions: 1) static paper, 2) web-based interface, and 3) mobile AR. The study results confirm that the AR-based interface is the most engaging, while the web interface provides a better system usability score (SUS). Qualitative feedback informs us that the system helps them understand mathematical concepts through interactive and playful exploration, compared to static textbooks.
In the second study, we conducted expert interviews with five math instructors to quantitatively compare our approach to existing learning resources and practices, such as handouts, videos, and existing interactive websites. In addition, we investigate how our tool could fit their potential needs and practices, seeking the opportunity for real-world use scenarios. Their feedback suggests that our approach has great potential as an on-demand and personalized learning assistant tailored for their specific context, which is not well-supported with the currently available learning practices. Based on their feedback and insights, we discuss how to expand our proposed approach beyond the current proof-of-concept prototype for future deployment.

Finally, this paper contributes to the followings: 
\begin{enumerate}
\item \system{}, a novel machine learning-based approach to creating explorable explanations by augmenting static math textbooks.
\item A set of features that consist of our augmented math textbooks, informed by a taxonomy analysis of existing practices.
\item Results and insights from preliminary user testing and expert interviews.
\end{enumerate}

\section{Related Work}

\subsection{Explorable Explanations}
Explorable explanations~\cite{victor2011explorable} have emerged as an increasingly popular practice to teach and learn abstract concepts across various educational contexts, such as math, science, and engineering (e.g. \textit{PhET}~\cite{perkins2006phet}, \textit{Distill}~\cite{distill}). Interactive explanations that leverage animations, simulations, and gaming provide more engaging experiences, naturally encouraging students to develop their understanding of complex concepts in a playful manner~\cite{adams2010student, hensberry2015effective, keller2007assessing, wang2020cnn}. Many of these math augmentation techniques were demonstrated in Bret Victor's KillMath projects~\cite{victor2011killmath}.
Various tools, such as \textit{Idyll}~\cite{conlen2018idyll}, \textit{Flapjax}~\cite{meyerovich2009flapjax}, \textit{GeoGebra}~\cite{GeoGebra95:online}, \textit{Mavo}~\cite{verou2016mavo}, and \textit{Data Theater}~\cite{lau2020data}, have been similarly developed to lower the barrier to create interactive explanations. For example, \textit{Idyll Studio}~\cite{conlen2021idyll} allows users to create data-driven explanations and \textit{GeoGebra}~\cite{GeoGebra95:online} lets users make interactive diagrams to teach various math topics such as geometry, algebra, and calculus.

However, existing tools still require the programming expertise that makes it difficult for non-technical users to adapt and create their own interactive content.
\citeauthor{Head2022}~\cite{Head2022} explicitly mention that the development of an appropriate authoring tool still requires future work, and there is a strong need for intelligent design assistants that automatically augment static math textbooks.
Towards this goal, \textit{CrossData}~\cite{chen2022crossdata} supports non-technical users in creating data-driven documents by binding text and data in the authoring process. This allows for easy updating of text, data, and graphs, but the output document is static once the authoring process is finished. 
Beyond textual explanations, some tools support non-technical users in creating interactive diagrams for math and engineering education. For example, \textit{MathPad2}~\cite{LaViola2007mathPad2} allows users to create math sketches that turn into interactive diagrams. In \textit{PhysicsBook}~\cite{cheema2012physicsbook} and \textit{PhysInk}~\cite{scott2013physink}, students can create interactive sketches that animate through given math formulas or physics simulation. Alternatively, \textit{Kitty}~\cite{kazi2014kitty} explores an authoring interface to let users make interactive animation by binding between different components, which is useful for engineering education. In these tools, however, the interactive explanations are based on sketches that are disconnected from existing material such as textbooks.

In contrast, this paper proposes to \textit{repurpose} existing static documents to turn them into explorable explanations.
This approach allows users to easily and quickly create interactive content with minimal time and effort as it does not require programming or generating content from scratch.
Additionally, to the best of our knowledge, our work is the first to investigate AR-based explorable explanations, which also have great potential for enhancing the learning experience by providing immersive and interactive content in a real-world context. This approach opens up new possibilities for engaging and personalized learning, bridging the gap between physical textbooks and digital interactivity.

\subsection{Augmented Paper}
Since Wellner's first demonstration of \textit{DigitalDesk}~\cite{wellner1991digitaldesk}, HCI researchers have actively explored the idea of \textit{augmented paper} by overlaying digital content onto physical documents~\cite{Han2021hybrid}. Researchers have investigated various ways to augment physical paper, including projecting digital information (e.g., \textit{AffinityLens}~\cite{subramonyam2019affinity}, \textit{MouseLight}~\cite{Song2010mouseLight}, \textit{Qook}~\cite{zhao2014qook}), illuminate physical paper (e.g., \textit{IllumiPaper}~\cite{klamka2017illumipaper}), and embedding content in mobile AR (e.g., \textit{Opportunistic Interfaces}~\cite{du2022opportunistic}, \textit{Teachable Reality}~\cite{monteiro2023teachable}) or see-through displays (e.g. \textit{MagicBook}~\cite{billinghurst2001magicbook}, \textit{Mixed Reality Book}~\cite{Grasset2007mixed}, \textit{HoloDoc}~\cite{Li2019holodoc}, \textit{Replicate and Reuse}~\cite{gupta2020replicate}).
However, most of these works assume prepared content, and only a few works have looked into the \textit{authoring aspect} of these augmented documents.
For example, \textit{PapARVis Designer}~\cite{chen2020augmenting} allows users to author embedded visualizations for static paper, and
\textit{Dually Noted}~\cite{qian2022dually} enables users to annotate physical documents by leveraging computer vision to recognize the layout of documents. 
Most closely related to our work, \textit{Paper Trail}~\cite{Rajaram2022papertrail} explores the authoring of augmented paper based on simple user-defined animation. 
Our work builds on top of these works to enable AR-based documents with a greater focus on interactive exploration compared to prior works.
By emphasizing explorability, we aim to create more engaging and personalized learning experiences through augmented paper.

\subsection{AR for Math Education}
Augmented reality has shown advantages in teaching math by enhancing visualizations~\cite{Ahmad2020Augmented} and creating playful learning experiences~\cite{Chen2019effect, Khan2018mathland}.
Such AR-based tools explore various math topics such as geometry~\cite{Kaufmann2002mathematics, Suselo2021using, Sarkar2019collaborative} and algebra~\cite{Chen2019effect}. 
For example, \textit{ARMath}\cite{Kang2020armath} demonstrates teaching early math skills such as counting and basic geometry by turning everyday objects into mathematical learning materials, using AR and computer vision.
Similarly, \textit{RealitySketch}\cite{suzuki2020realitysketch} also aims to visualize math and physics concepts in the real world by embedding responsive graphics through an AR sketching interface.
However, in most of these existing works, the AR contents are disconnected from existing learning materials like textbooks.
A few studies have focused on augmenting existing math textbooks. For example, Li et al.~\cite{Li2019turning} turn a math book into an interactive game to attract children's interest, however the connection between the physical book and AR content is limited to multiple-choice questions. \textit{GeoAR}~\cite{Kirner2012development} creates an augmented book to teach geometry, with AR geometric shapes appearing on the pages, but the AR content is predefined and only works with the specific book. In contrast, our work introduces a general-purpose authoring tool that enables teachers and students to create their own augmentations using existing formulas and figures in math textbooks.

\subsection{AR Authoring Tools}
Existing AR authoring tools often rely on user-sketched content (e.g. \textit{Rapido}~\cite{Leiva2021rapido}, \textit{Pronto}~\cite{Leiva2020pronto}, \textit{RealitySketch}~\cite{suzuki2020realitysketch}, \textit{Sketched Reality}~\cite{kaimoto2022sketched}, \textit{ProjectAR}~\cite{lunding2022projectar}), user-created physical models (e.g. \textit{ProtoAR}~\cite{Nebeling2018protoAR}), or pre-defined virtual models (e.g. \textit{Reality Composer}~\cite{RealityComposer}, \textit{Adobe Aero}~\cite{aero}, \textit{RealityTalk}~\cite{liao2022realitytalk}). However, there is a lack of authoring tools that enables the users to augment existing visual content such as physical paper.
Moreover, creating dynamic AR content typically requires user-defined demonstrations (e.g. \textit{Rapido}~\cite{Leiva2021rapido}, \textit{Pronto}~\cite{Leiva2020pronto}) or pre-defined animations (e.g. \textit{Reality Composer}~\cite{RealityComposer}, \textit{Adobe Aero}~\cite{aero}). While these approaches work for general-purpose AR prototyping, they are not well-suited for AR-based explorable explanations, where animation should be bound to mathematical formulas or simulations. 
To address these limitations, our proposed authoring tool leverages machine learning and data-binding techniques. This allows users to extract existing content from documents to create interactive AR explanations tailored to their needs, facilitating the creation of engaging educational materials.

\begin{figure*}[t]
\centering
\includegraphics[width=\textwidth]{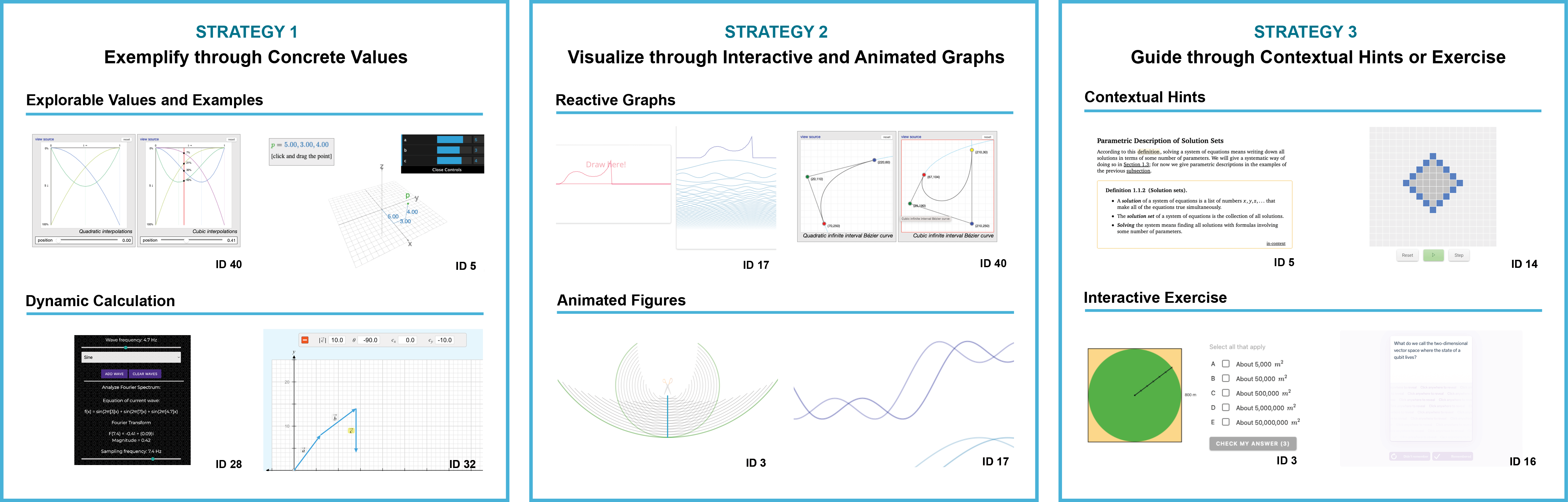}
\caption{Design strategy analysis based on existing interactive math explanations}
\label{fig:design-strategies}
\end{figure*}

\section{Design Strategies of Existing Explorable Math Explanations}

\subsection{Design Strategy Analysis}
To design our authoring interface, we first try to understand the common techniques and design strategies employed in existing explorable explanations.

\subsubsection{Motivation}
Although previous work, such as \textit{Augmented Math}~\cite{Head2022}, have presented taxonomy analyses in similar domains, their focus is broader, encompassing the readability of math formulas in books, papers, slides, and videos. 
While these analyses provide insights into low-level augmentation and annotation techniques, such as visual notations or styles, we need to understand higher-level design strategies to inform our system features. 
Therefore, our analysis specifically focuses on \textit{interactive websites} to better understand the high-level \textit{design strategies} employed in these websites. 
Our analysis complements existing taxonomy analyses presented in~\cite{Head2022} to gain a deeper understanding of explorable math explanations.

\subsubsection{Dataset}
We collected 43 existing interactive math explanations available on the internet from various sources.
These datasets were gathered from notable collections of interactive websites, such as \textit{Explorabl.es}~\cite{Explorab37:online}, \textit{Gallery of Concept Visualization}~\cite{Galleryo94:online}, \textit{GeoGebra}~\cite{GeoGebra95:online}, \textit{PhET}~\cite{PhETFree6:online}, \textit{Explained Visually}~\cite{Setosada85:online}, \textit{Visualize It}~\cite{Visualiz92:online}, \textit{Interactive Maths}~\cite{InteractiveMaths:online}, \textit{Seeing Theory}~\cite{SeeingTh51:online}, \textit{Distill}~\cite{distill}, and \textit{Awesome Interactive Math Tools}~\cite{awesome:online}, as well as through the individual searches. 
A complete list of the collected dataset with each website link is available in \autoref{appendix}.  

\subsubsection{Method}
We identified and categorized high-level design strategies through an open-coding analysis. 
After collecting the examples, two of the authors first conducted an open-coding analysis to identify initial categories and dimensions. Next, all authors reflected on the initial categorization to discuss consistency and comprehensiveness. After refining the categories, three authors performed systematic coding with individual tagging to categorize the complete dataset. Finally, the authors reflected on individual tagging to resolve discrepancies and obtain the final coding results.

\subsubsection{Results}
Our analysis led to the identification of three high-level design strategies: 1) Exemplify through concrete values, 2) Visualize through interactive and animated graphs, and 3) Guide through contextual hints and exercises. For each strategy, we also identify two common design components, as illustrated in Figure~\ref{fig:design-strategies}. 
These collected design strategies span a wide range of math topics, including \textit{Algebra} (6/43), \textit{Geometry} (9/43), \textit{Calculus} (2/43), \textit{Probability} (11/43), \textit{Arithmetic} (1/43), \textit{Applied math} (19/43), and \textit{Graph theory} (3/43). 
In addition to the high-level design strategies, we also identify other low-level design and interaction elements, which include \textit{Slider and Scrubbable} (27/43), \textit{Direct Graph Manipulation} (18/43), \textit{Options} (3/43), \textit{Text Input}  (5/43), \textit{Scroll} (2/43), \textit{Button} (24/43), and \textit{Hover} (3/43). 
The results are summarized in \autoref{appendix} (\autoref{fig:taxonomy-analysis} as well as \autoref{tab:urls1} and \autoref{tab:urls2}). 
In the following sections, we discuss each design strategy and its examples in more detail. \autoref{fig:design-strategies} also illustrates the visual summary of each strategy and design component. 

\subsection*{Strategy 1. Exemplify through Concrete Values}
\subsubsection*{Explorable Values and Examples}
The first strategy is to exemplify abstract and symbolic math representations with concrete values. 
Such examples include providing concrete examples for abstract concepts, embedding numerical values in each symbol of complex math formulas, and showing concrete probabilities with simulation experiments. 
Most of the websites allow users to dynamically update these values with interactive components such as sliders or text inputs.

\subsubsection*{Dynamic Calculation}
Another important aspect of this strategy is that when users change these values, then corresponding equations also change based on dynamic calculations and simulations.
This allows students to immediately see how each value affects the outcome, helping them gain an intuition of how symbols and equations are related to each other. 
By leveraging explorable examples and dynamic calculations, students can develop a better understanding of abstract concepts by visualizing them. 

\subsection*{Strategy 2. Visualize through Interactive and Animated Graphs}
\subsubsection*{Reactive Graphs}
While the first strategy mainly focuses on textual representation, another common strategy leverages dynamic visual representation. 
In particular, existing websites often use reactive graphs that can be dynamically changed based on corresponding interactions or value changes.
Such examples include a reactive graph of a trigonometric function (algebra), an interactive diagram for the Pythagorean theorem (geometry), and an explorable explanation of Newton's method (calculus). 
This technique is often used in conjunction with the exemplifying strategy described above. 
For example, when a user changes a value in an equation, the graphs and diagrams are also updated. 
This helps users understand the relationship between the symbolic variable and its visual representation. 

\subsubsection*{Animated Figures}
Alternatively, animated figures focus primarily on animation with less emphasis on interactivity. 
Animated figures are particularly useful for visualizing simulation results.
In this way, animation can visualize how simulation evolves with temporal representation. 
Examples include showing simulation results of probability, demonstrating various geometric transformations, and temporal visualization of sine and cosine curves.
While this representation focuses more on animation than interaction, users can still manipulate parameters such as timelines. 

\subsection*{Strategy 3. Guide through Contextual Hints or Exercises}
\subsubsection*{Contextual Hints}
Contextual hints provide additional information to students when they need it. 
Examples include referencing definitions to remind the context or breaking down complex solutions through step-by-step instructions.
This support can be customized based on a student's progress and can adapt to their individual learning needs, ensuring that they receive the most appropriate help at the right time.

\subsubsection*{Interactive Exercises}
Incorporating exercises directly into the explanations allows users to actively engage with the content. 
Exercises can come in various forms, such as multiple-choice questions for a simple quiz, fill-in-the-blank problems to solve equations, or direct manipulation of figures to answer questions.  
Providing immediate feedback on these exercises can help students identify their mistakes and reinforce their understanding of the concepts.
\begin{figure*}[h!]
\centering
\includegraphics[width=0.247\textwidth]{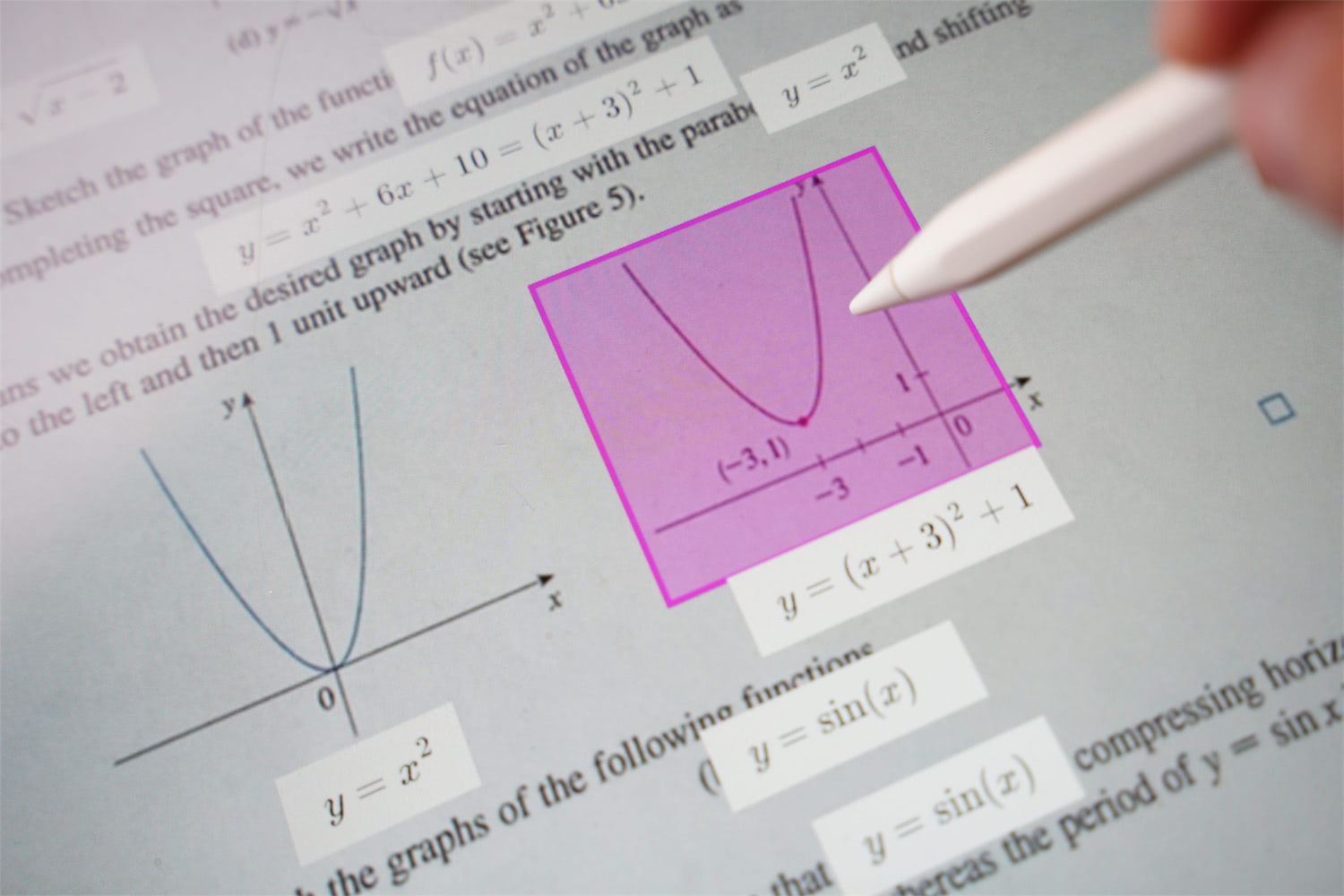}
\includegraphics[width=0.247\textwidth]{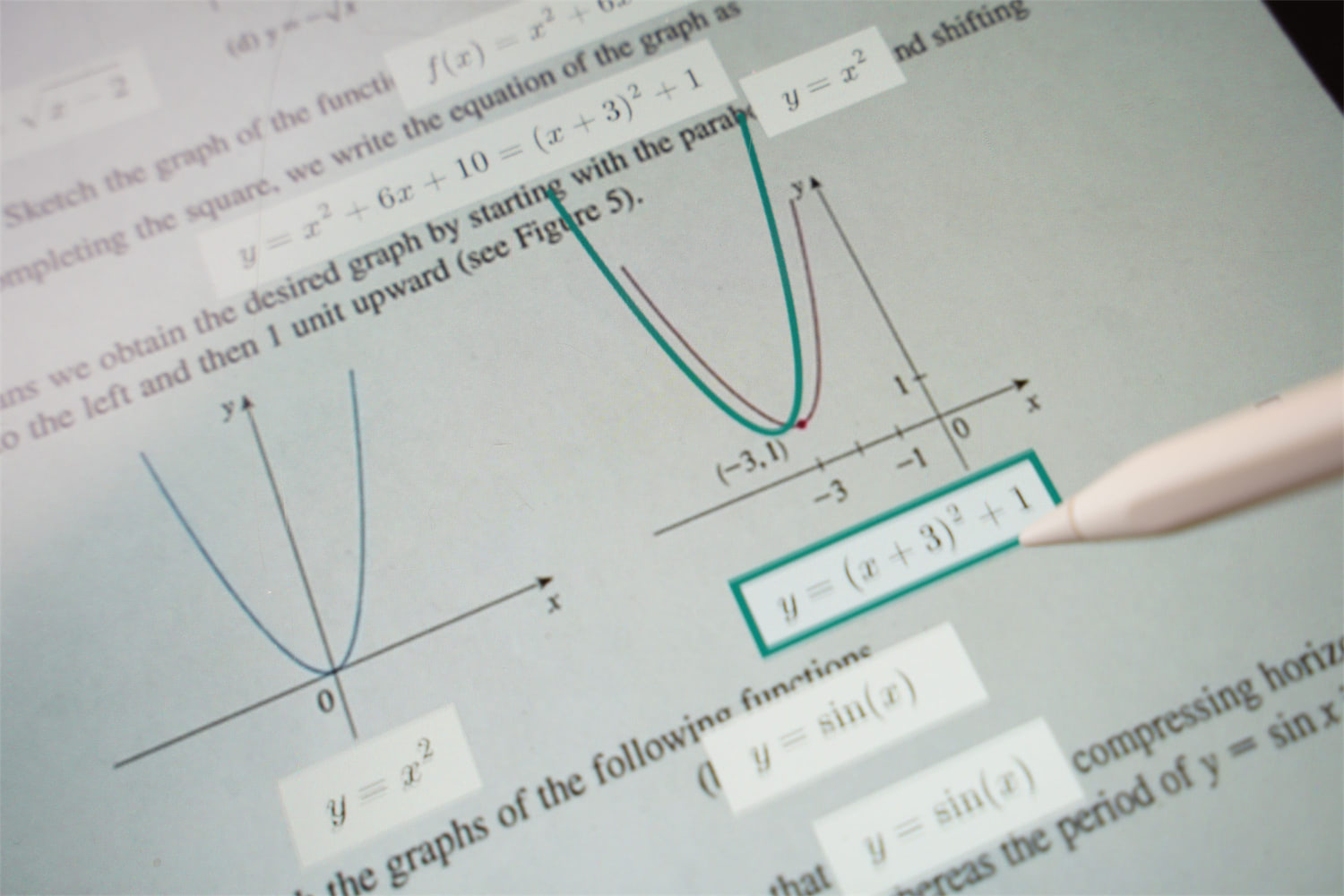}
\includegraphics[width=0.247\textwidth]{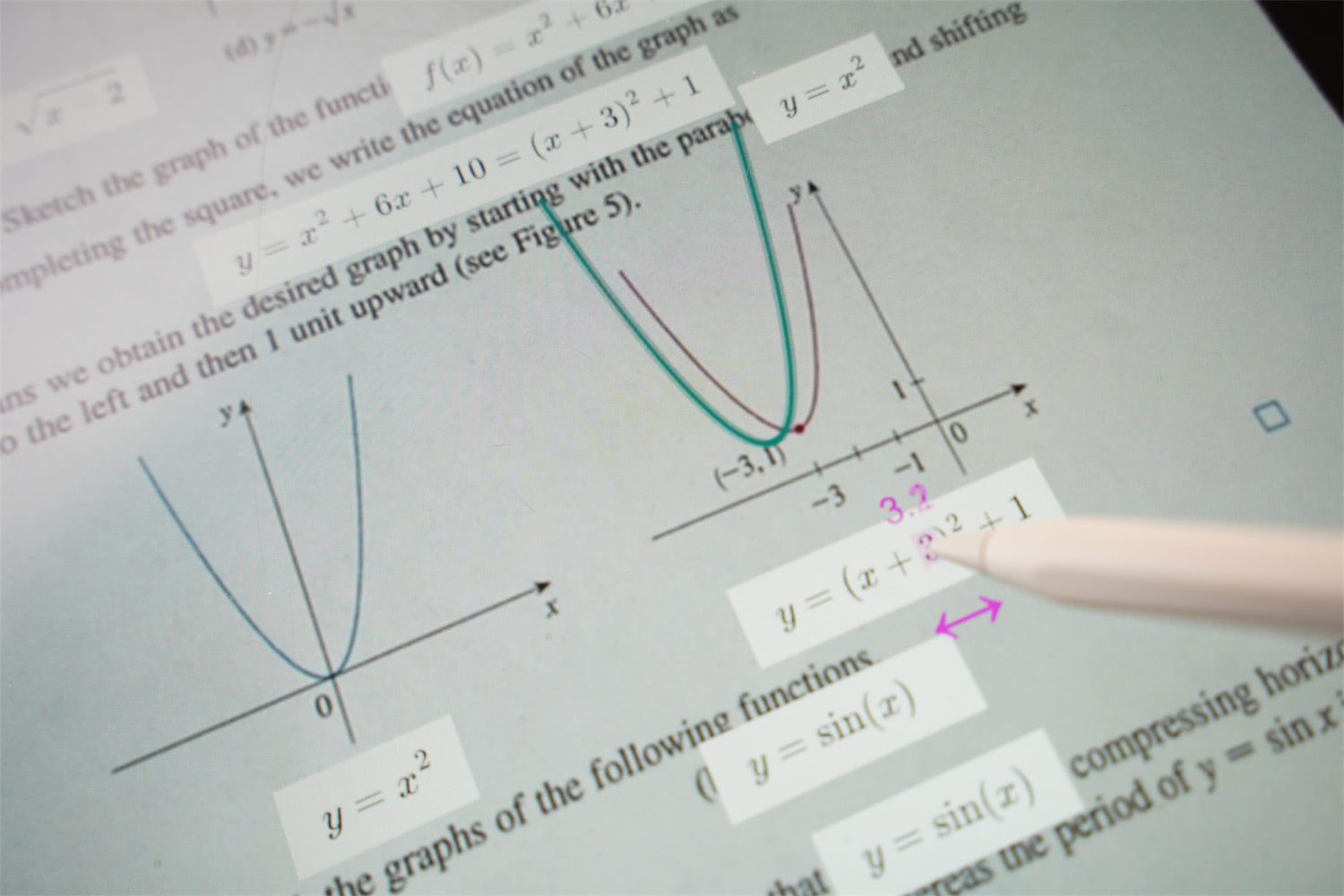}
\includegraphics[width=0.247\textwidth]{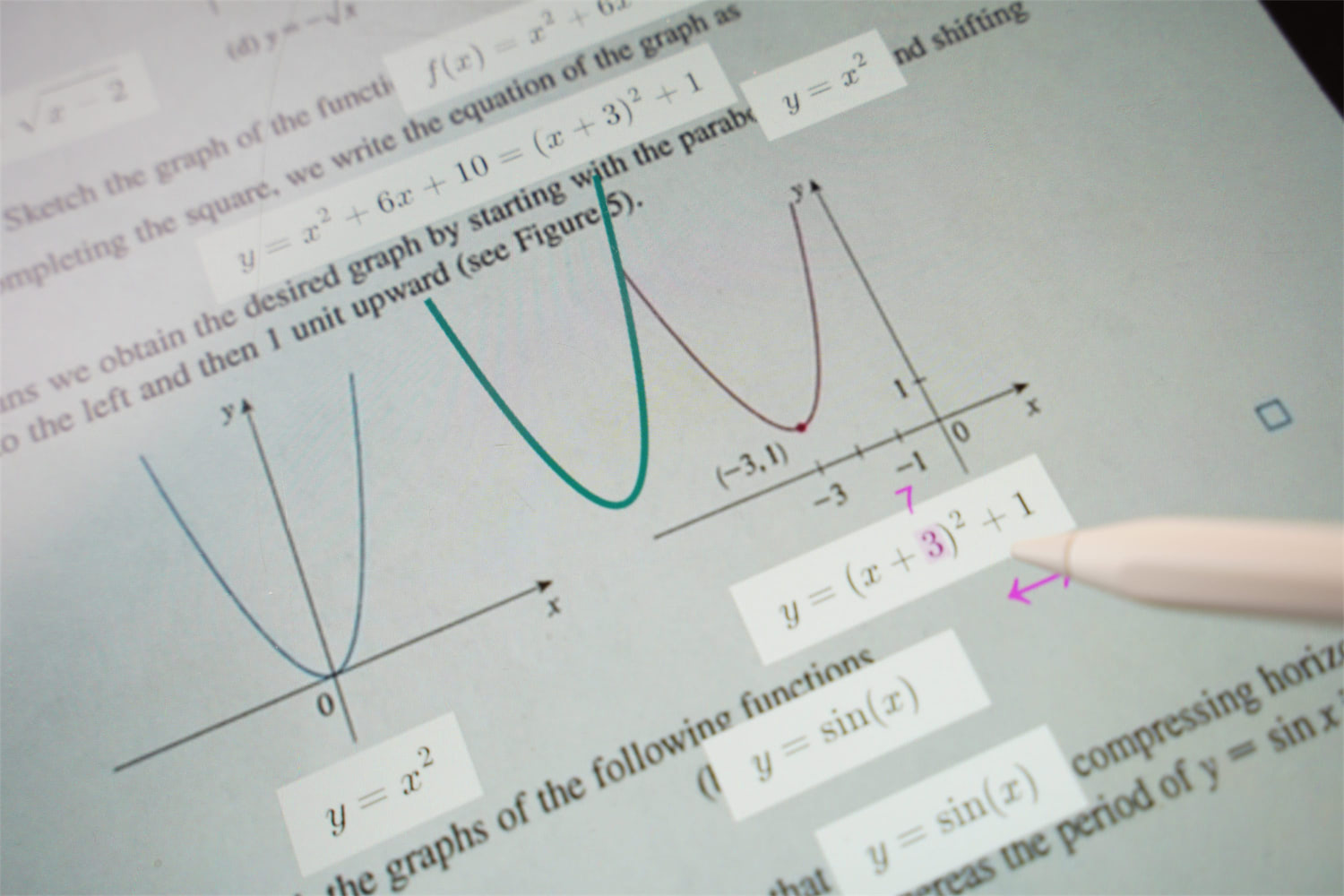}
\\
\vspace{0.05cm}
\includegraphics[width=0.247\textwidth]{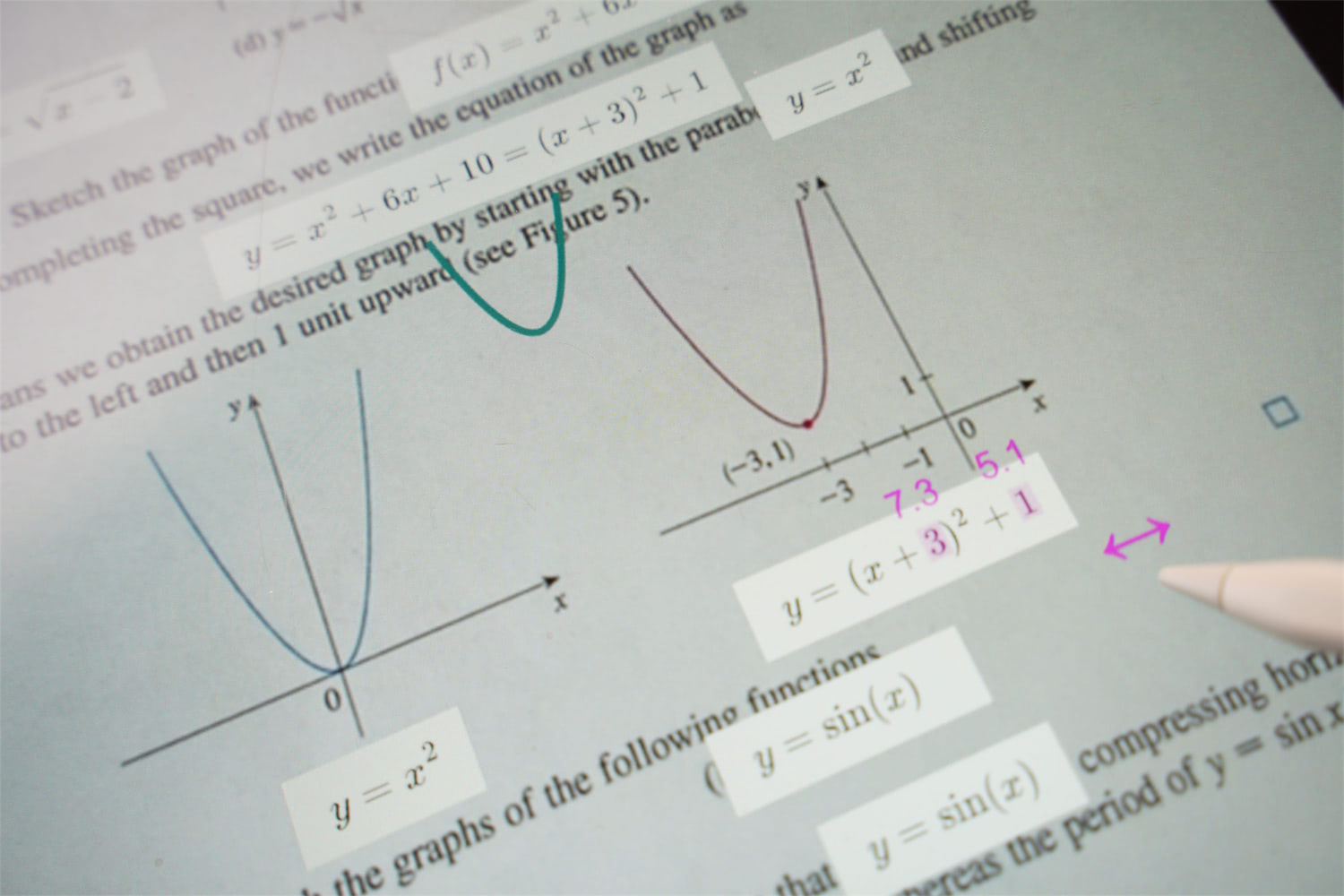}
\includegraphics[width=0.247\textwidth]{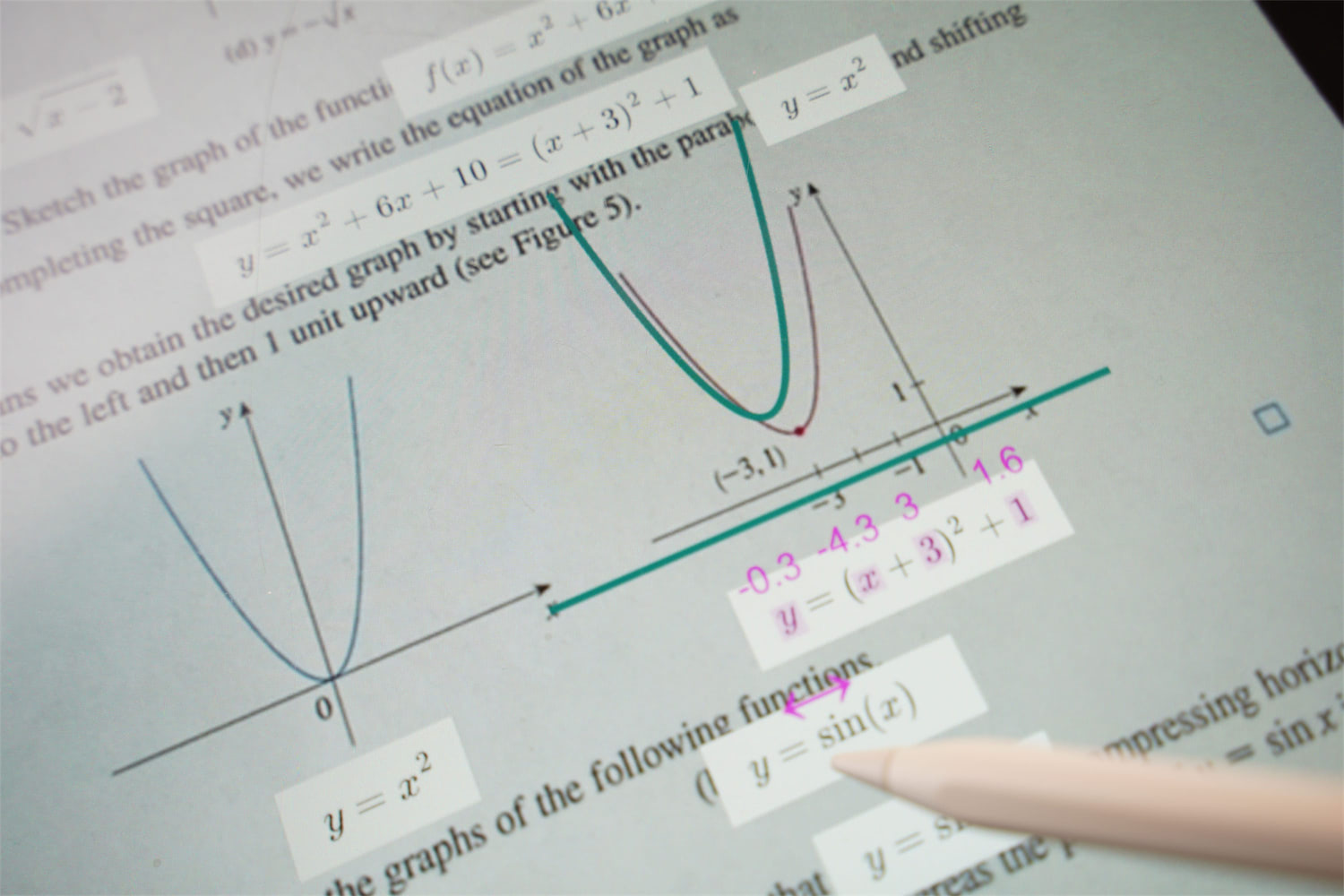}
\includegraphics[width=0.247\textwidth]{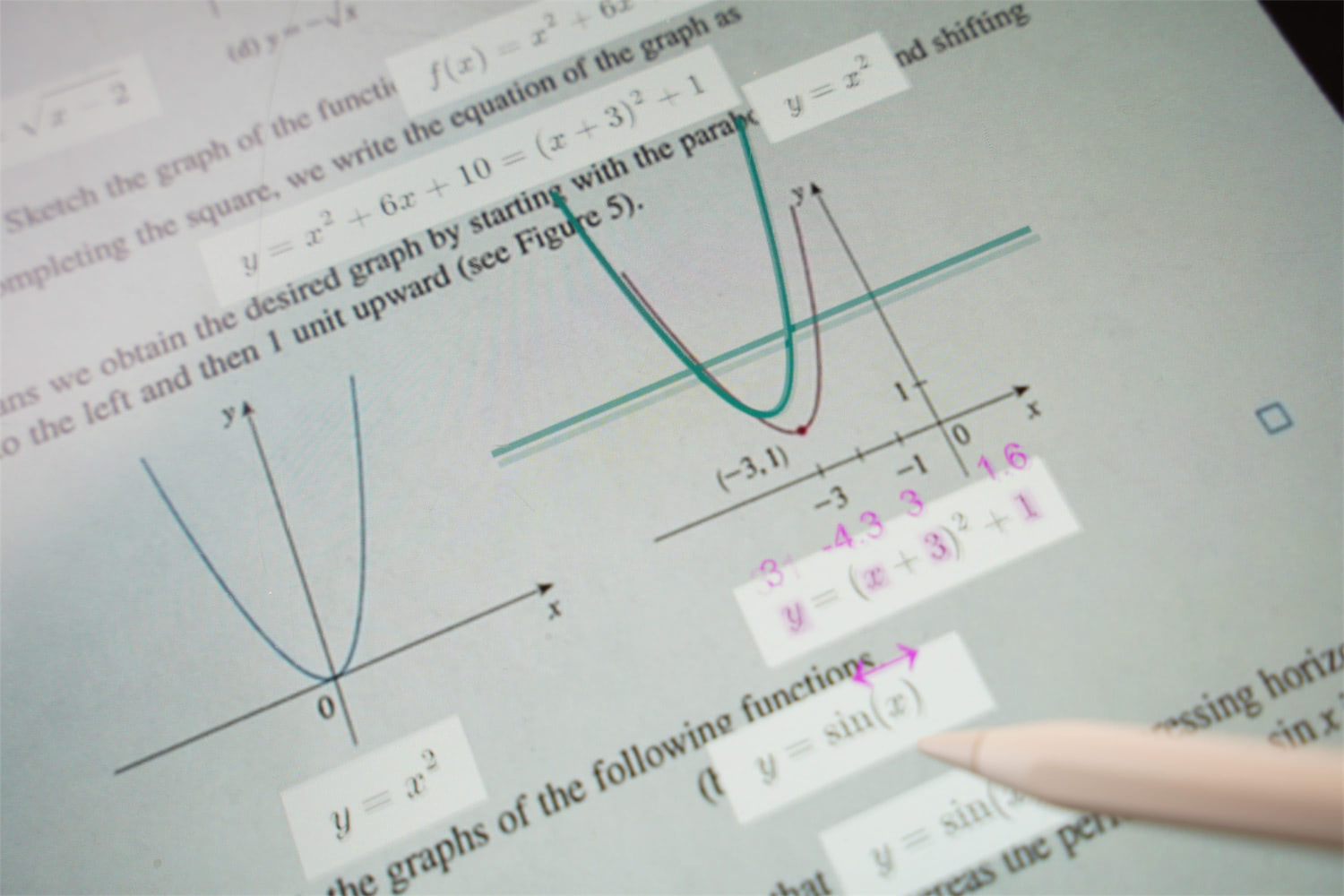}
\includegraphics[width=0.247\textwidth]{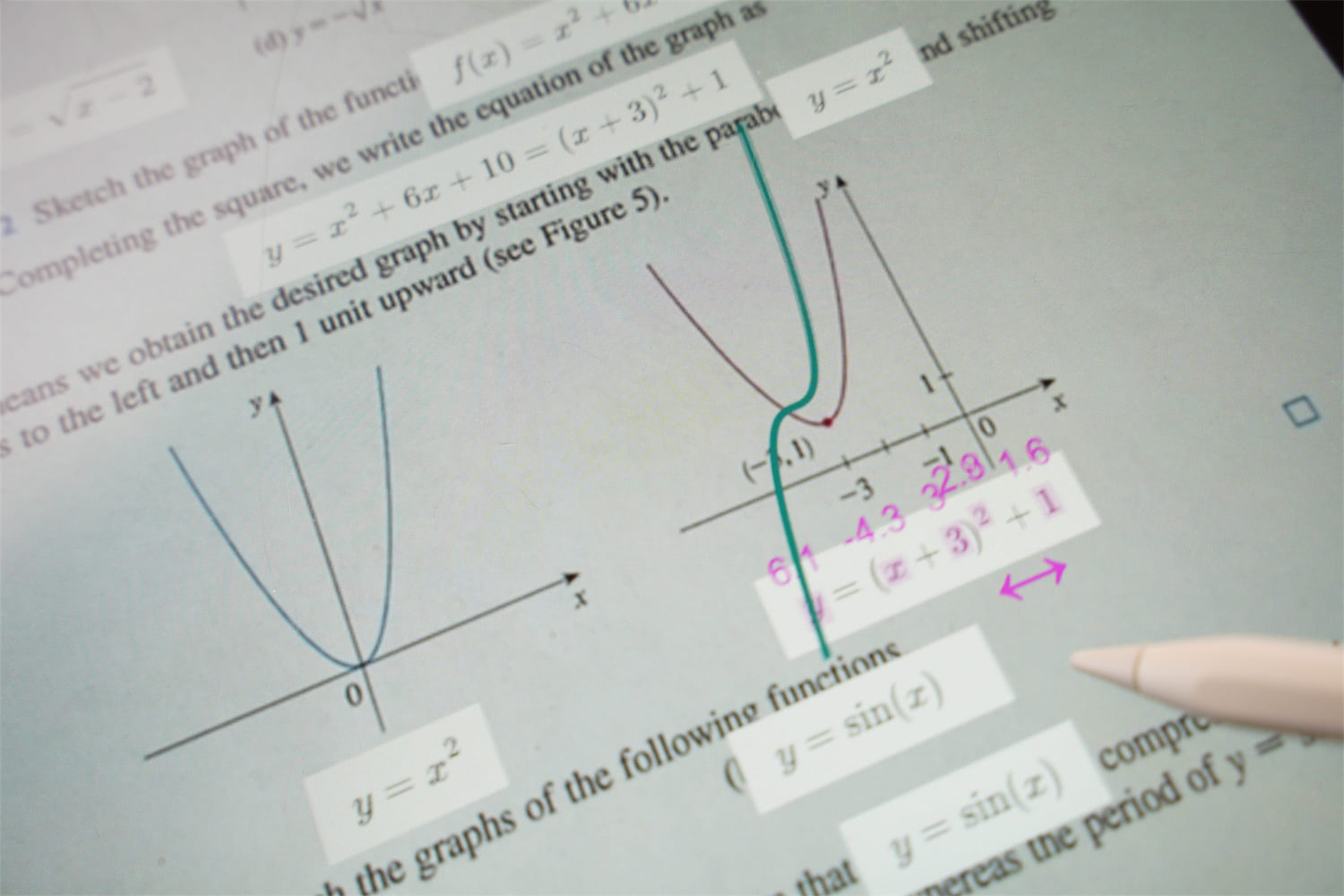}
\caption{An example authoring and interaction walkthrough to make an explorable explanation for a quadratic equation and its graph. From top left: 1) The system first extracts math equations and graphs from a scanned document and lets the user select an extracted figure, 2) When the user selects an equation to bind with the selected graph, the system automatically draws a graph given the equation, 3-4) As the user drags one of the symbols, the value, equation, and corresponding graphs update, 5-8) The user can also manipulate other symbols and elements, reflecting changes in values and graphs for interactive explorations.}
\label{fig:example-walkthrough}
\end{figure*}

\section{\system{}: System and Design}

\subsection{Overview}
This section introduces \system{}, an authoring tool for AR-based explorable explanations. 
The system aims to empower non-technical users, such as students and instructors, to create interactive explanations without programming expertise. To achieve this, our system augments static math textbooks by embedding and overlaying interactive content.
The authoring workflow comprises the following steps: 1) \textit{scan}: users first scan a math textbook or handout with a camera, 2) \textit{extract}: the system automatically extracts math formulas and graphs shown in the document using OCR and computer vision, 3) \textit{select}: users select extracted content like symbols, values, equations, and graphs to make them interactive, 4) \textit{bind}: users bind selected elements together, 5) \textit{manipulate}: users manipulate the values and data of selected elements by dragging, and 6) \textit{update}: the system propagates changes responsively updates based on data binding and manipulation.
The system supports both mobile AR and desktop interfaces for making printed paper or scanned PDFs interactive, respectively.

\subsection{Scope} 
We largely focus our investigations around high-school level mathemetics based on the United States standard curriculum. Specifically, the primary topics are: algebra, trigonometry, geometry, and calculus.  \system{} does not have features such as complex illustration nor abstract equations to support higher level topics such as statistics probability and quantum-mecahnics. 

\subsection{Example Authoring Walkthrough}
In this section, we describe an example authoring and interaction walkthrough using our system. We use an example from a high-school calculus textbook (Single Variable Calculus~\cite{stewart2015single} P.39) that presents an equation, $y = x^2 + 6x + 10 = (x + 3)^2 + 1$, and its corresponding graph. The purpose of this augmentation is to visualize how values of $a$, $b$, $c$, and $n$ affect the graph of $y = (x - a)^{n} + b$. We describe how users can transform this static content into an explorable explanation using our system. 

\subsubsection*{\textbf{Step 1: Scan and Extract Math Textbook or Handout}}
The first step is to scan a printed document or PDF from an existing math textbook. After the user takes a photo to scan it, the system automatically extracts text, math formulas, and graphs using OCR and computer vision. In \autoref{fig:example-walkthrough}-1, math formulas captured by our system are shown as extracted content. Our system converts math equations to computer-readable LaTeX expressions like \verb|y = (x + 3)^{2} + 1|, enabling semantic understanding of the given math formula. OCR extracts written text and math formulas, while custom computer vision algorithms extract figures and graphs, allowing users to select and draw a graph overlaid on the existing figure. 

\subsubsection*{\textbf{Step 2: Select Equations and Bind them to Graphs}}
After scanning and extracting the document, users can author augmented content via mobile AR or desktop interfaces. 
To make the textbook interactive, users first select extracted values, symbols, equations, and graphs. 
When a user selects a graph, the system highlights the selection as illustrated in \autoref{fig:example-walkthrough}-1. 
The system automatically extracts the x-axis, y-axis, origin, and drawn graph line using computer vision.
When the user also selects the equation $y = (x + 3)^2 + 1$, the system automatically binds and draws a graph based on the selected equation (\autoref{fig:example-walkthrough}-2).
This AR-embedded graph enables dynamic manipulation and updating based on user interaction.
Users can also bind multiple equations to a single graph to visually compare equations by overlaying them. 

\subsubsection*{\textbf{Step 3: Drag to Change the Variable Values}}
Users can dynamically change values by dragging symbols or variables.
For example, in Figure~\ref{fig:example-walkthrough}-4, when the user drags the $3$ in the equation $y = (x + 3)^2 + 1$, the system automatically replaces the value of $3$, treating it as a dynamic variable. 
In Figure~\ref{fig:example-walkthrough}-7, users can specify values for variables like $x$ and $y$, displaying horizontal and vertical lines related to the current $x$ and $y$ values on the graph.
Changing a value alters the corresponding variable and equation based on dynamic calculation.

\subsubsection*{\textbf{Step 4: Enable Bi-Directional Binding between Variable and Graph}}
Changing a value prompts the system to update the graph dynamically. 
Figure~\ref{fig:example-walkthrough}-5 and -8 show how users can modify the graph plot of $(x + a)^n + b$ by dragging the graph or the variable values of $b$ and $n$, respectively.
For example, adjusting the $a$ value of $(x + a)^2 + b$ shifts the graph horizontally, while changing the $b$ moves it vertically. Bi-directional binding allows users to drag the graph, which in turn changes the corresponding variable value. 
This reactive feature facilitates user interaction with numerical values, graphs, and charts. By adjusting variables in a formula, users can observe their effects on the final output, allowing them to develop an intuition of abstract relationships, which are otherwise hard to understand.

\subsection{Supported Augmentation Features}
To facilitate diverse interactive explorations for high-school level mathematics, we have developed the following five augmentation techniques: 1) \textit{dynamic values}, 2) \textit{interactive figures}, 3) \textit{relationship highlights}, 4) \textit{concrete examples}, and 5) \textit{step-by-step hints}.
These features are designed to be automatically generated based on the extracted content, with the goal of encompassing a wide range of strategies identified through analysis. 

\subsubsection{\textbf{Dynamic Values}}
\begin{figure}[h!]
\centering
\includegraphics[width=0.495\linewidth]{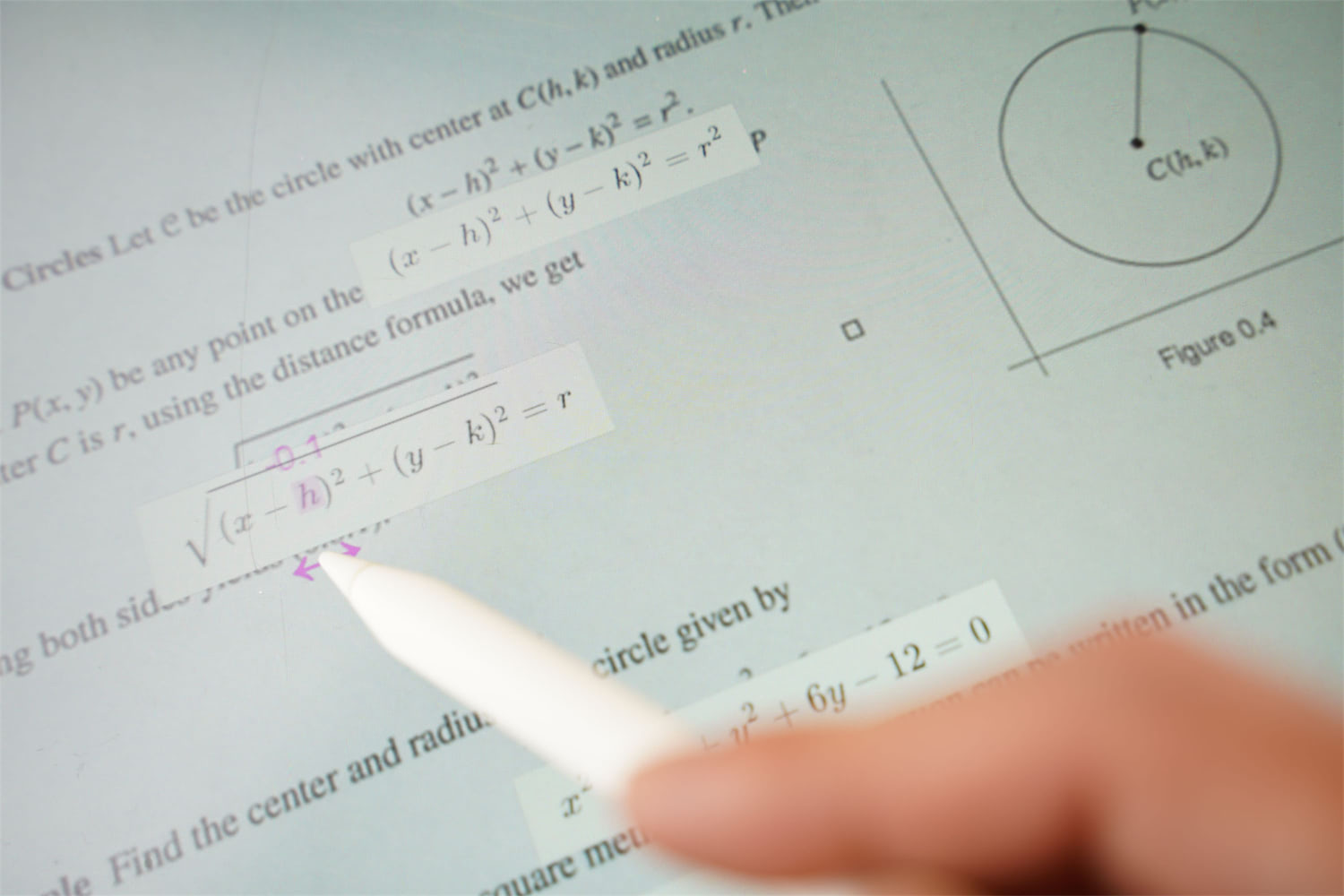}
\includegraphics[width=0.495\linewidth]{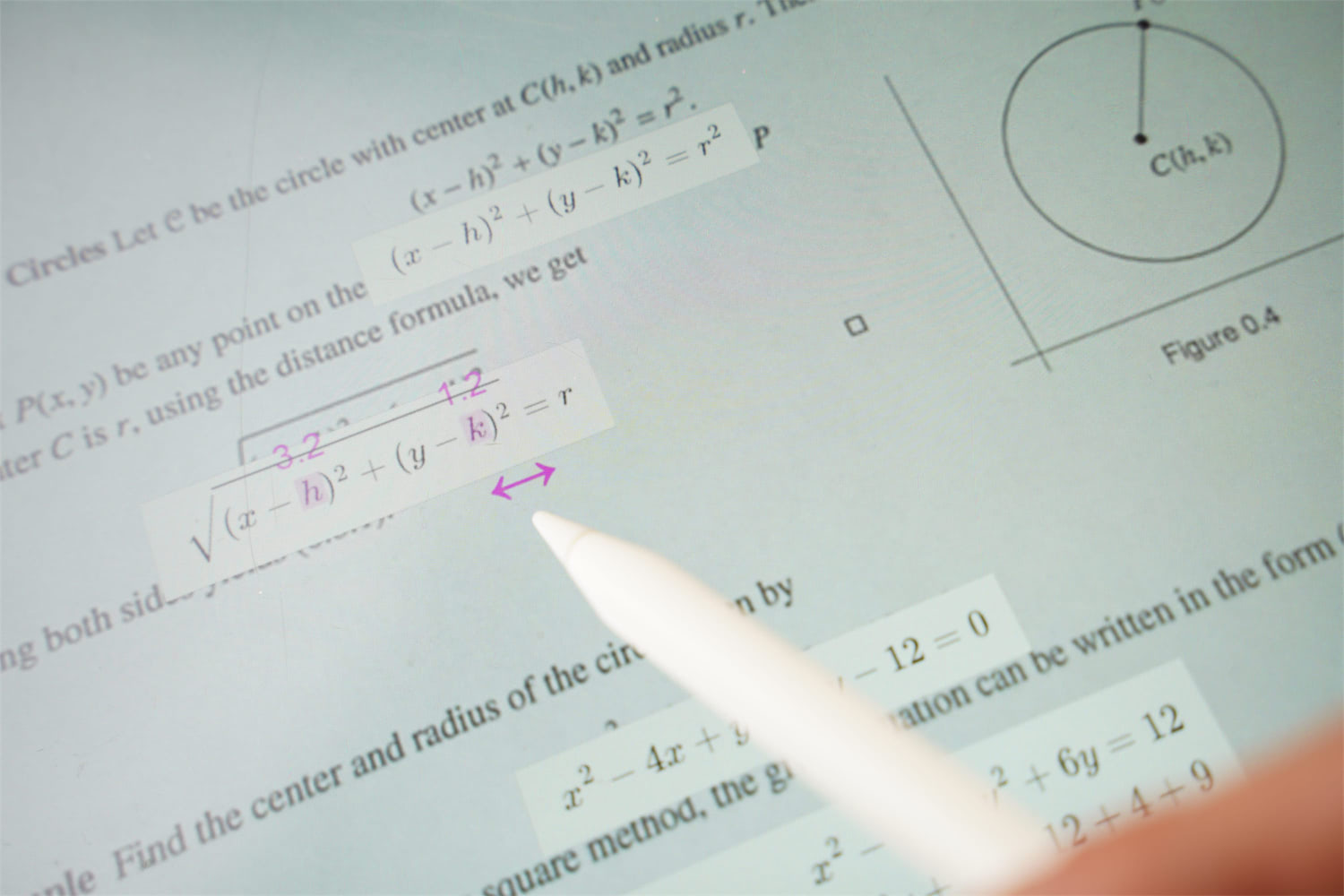}
\caption{Dynamic Values}
\label{fig:dynamic-value}
\end{figure}

Dynamic values allow users to insert and manipulate concrete values for extracted symbols and variables found in the document. 
This feature is inspired by \textit{``strategy 1: exemplify through concrete values''} from our taxonomy analysis. 
For instance, Figure~\ref{fig:dynamic-value} demonstrates users setting and adjusting values of $h$, $k$, and $r$ in the equation $\sqrt{(x-h)^2 + (y-k)^2} = r^2$. 

\noindent
Since the system processes mathematical equations as LaTeX expressions rather than plain text, it can semantically comprehend and dynamically compute equations based on the inserted values. 
Additionally, the system enables users to manipulate these values across various formulas throughout the document, generating responsive explanations that demonstrate how specific changes impact multiple equations through dynamic calculations. 
Dynamic values can be integrated with other features, such as interactive figures, which we discuss next.

\subsubsection{\textbf{Interactive Figures}}
Interactive figures offer users the opportunity to comprehend concepts through explorable visual representations. Unlike static images, these interactive figures promote intuitive understanding by allowing users to interact with them. 
This feature is inspired by \textit{``strategy 2: visualize through interactive and animated graphs''} from our taxonomy analysis. 

\begin{figure}[h!]
\centering
\includegraphics[width=0.495\linewidth]{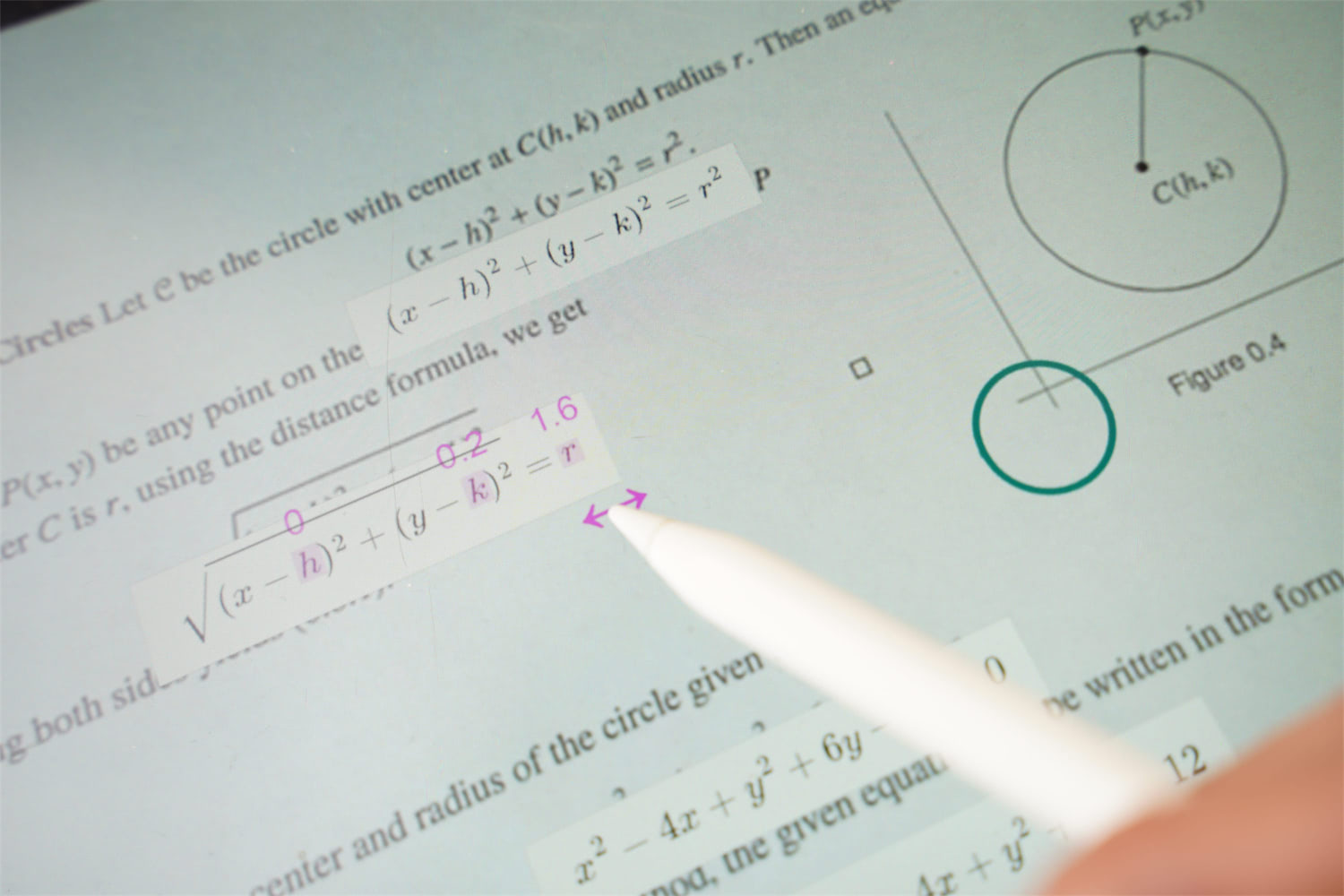}
\includegraphics[width=0.495\linewidth]{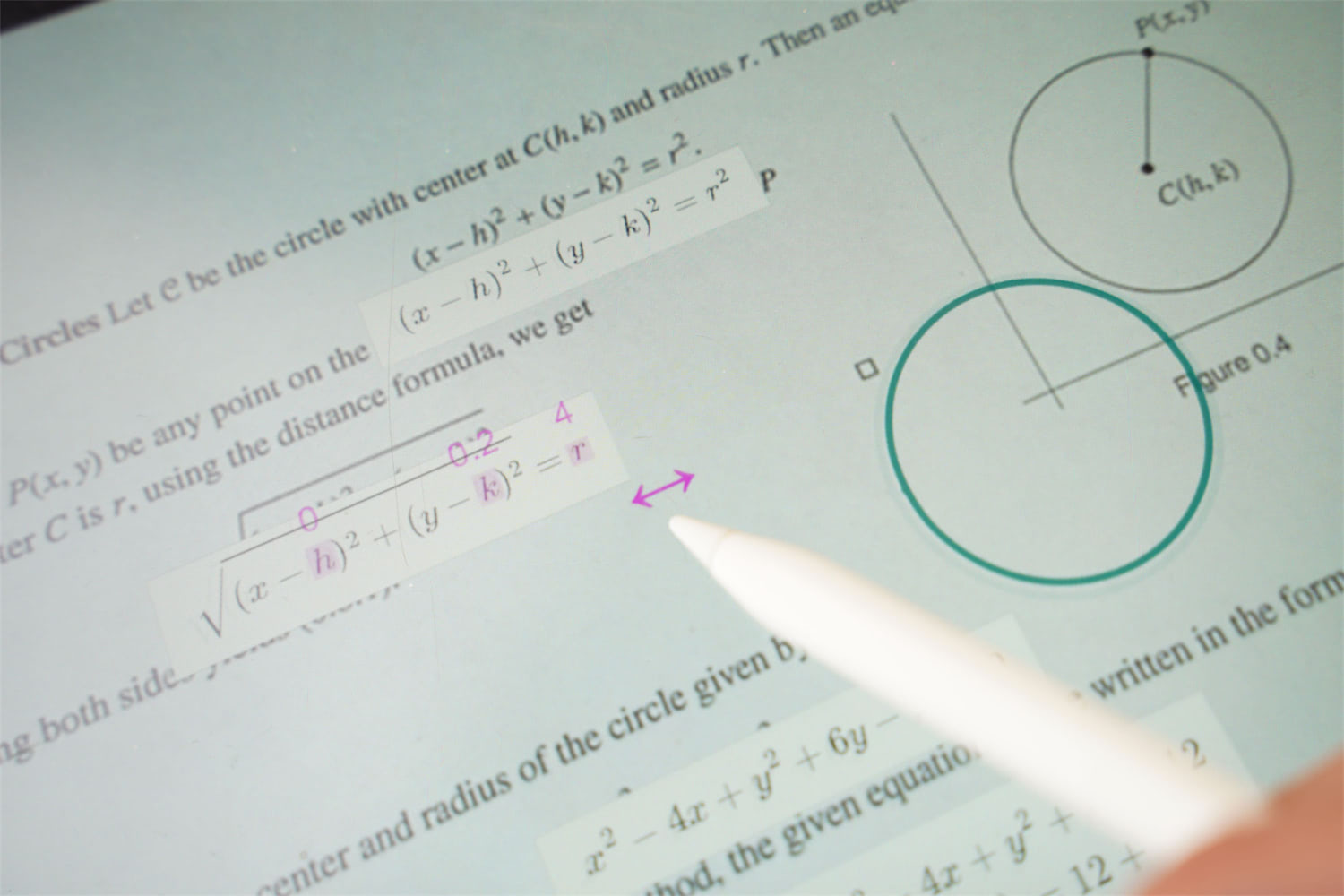}
\caption{Interactive Figures}
\label{fig:interactive-figure}
\end{figure}

\noindent
Users can make a graph interactive by selecting and binding equations to it, as described in the previous section. 
For instance, Figure~\ref{fig:interactive-figure} demonstrates how changing the value of $r$ dynamically affects the circle's radius in the graph accordingly. 

\begin{figure}[h!]
\centering
\includegraphics[width=0.495\linewidth]{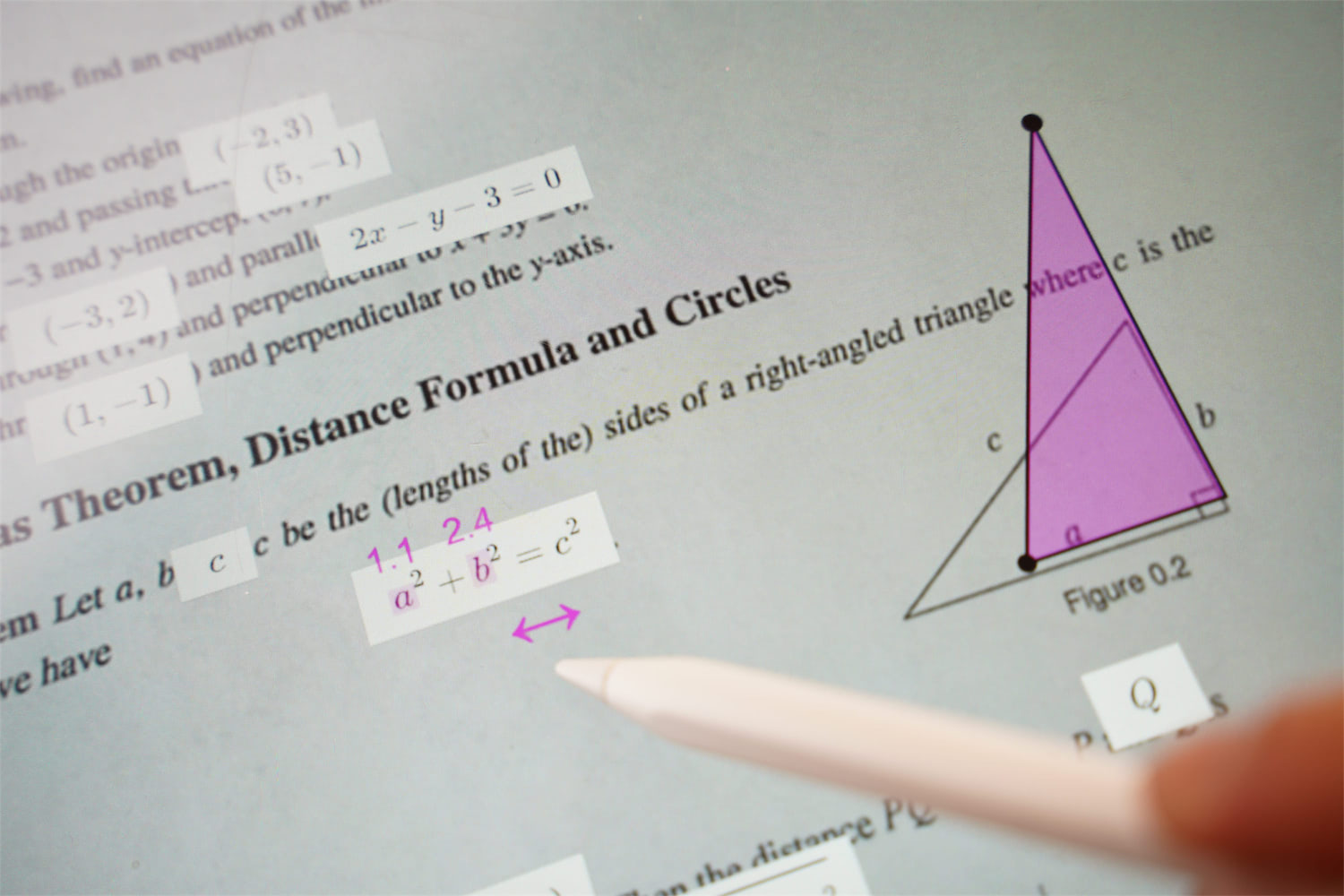}
\includegraphics[width=0.495\linewidth]{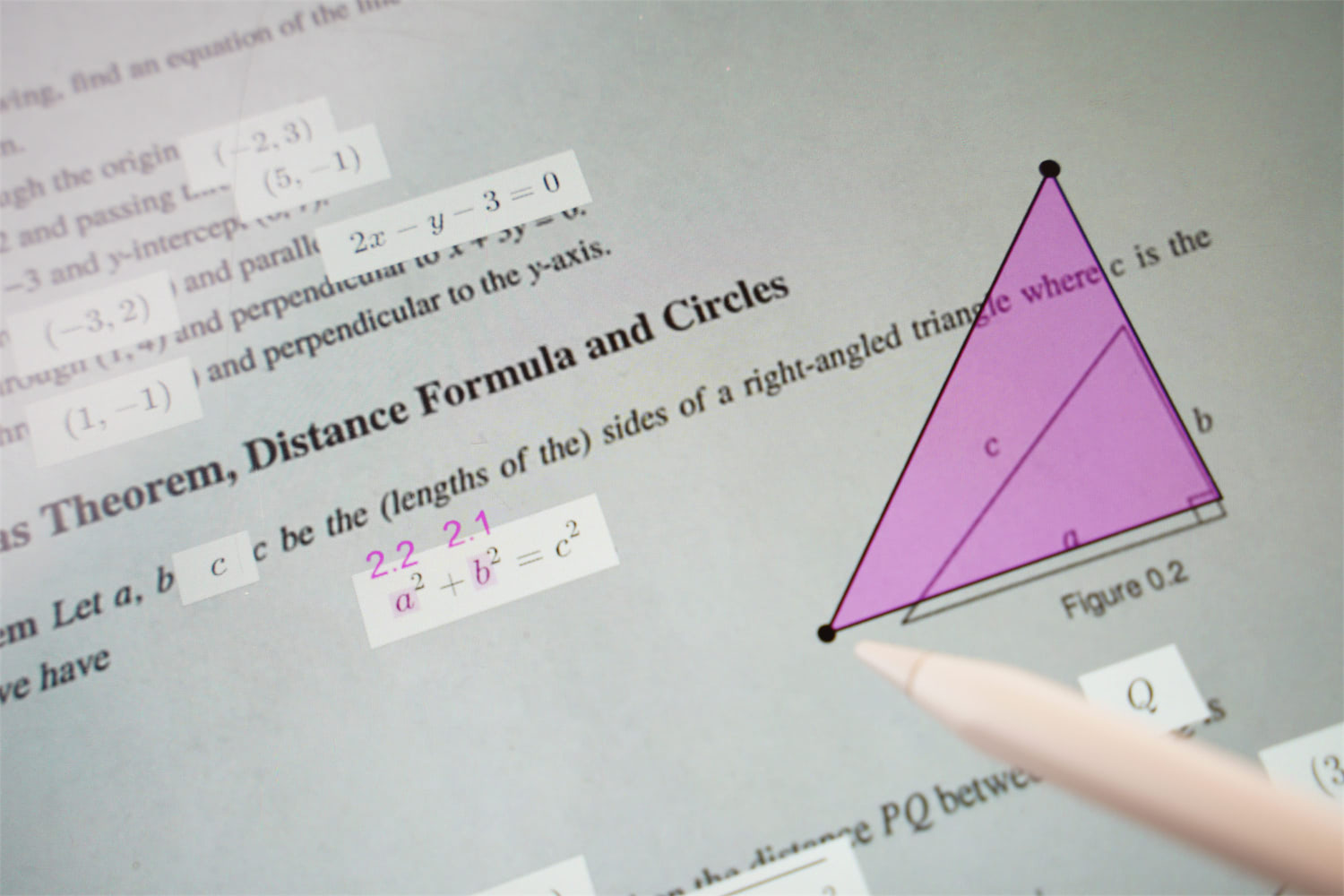}
\caption{Relationship Highlights}
\label{fig:interactive-triangle}
\end{figure} 

\noindent
Interactive figures also enhance the understanding of geometric concepts. 
For example, Figure~\ref{fig:interactive-triangle} shows how changing a variable in the equation $a^2 + b^2 = c^2$ updates the corresponding triangle shape, enabling users to grasp the geometric concept more effectively.
These interactive figures support bi-directional binding, allowing users to drag the graph or shape, which in turn updates the values accordingly.

\subsubsection{\textbf{Relationship Highlights}}
Relationship highlights enable users to visualize connections between different variables and visual references. 
This feature is inspired by \textit{``strategy 2: visualize through interactive and animated graphs''} from our taxonomy analysis. 
Relationship highlights can explicitly reveal the connection between variables and visual references,  facilitating intuitive comprehension for subjects like geometry or calculus.

\begin{figure}[h!]
\centering
\includegraphics[width=0.495\linewidth]{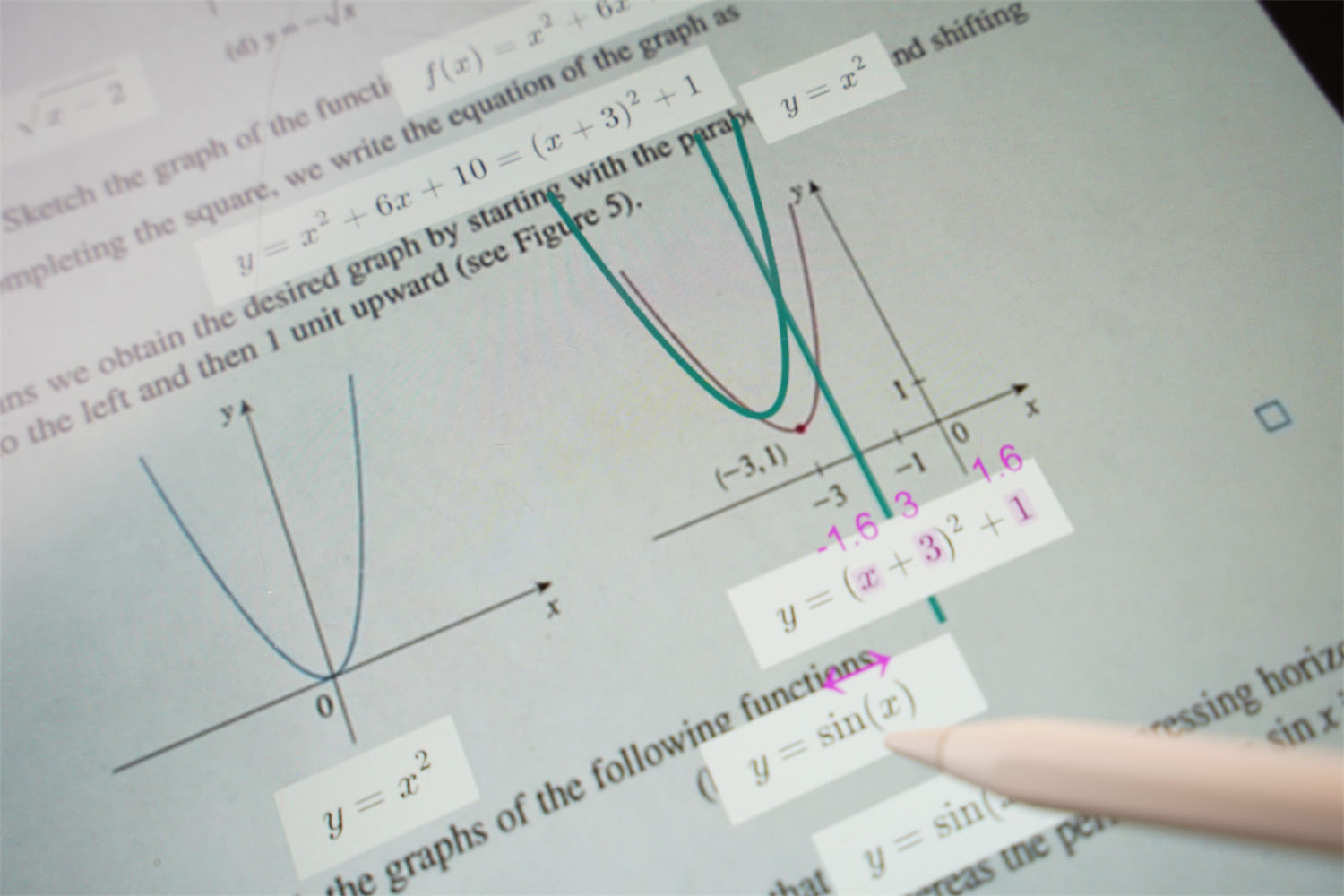}
\includegraphics[width=0.495\linewidth]{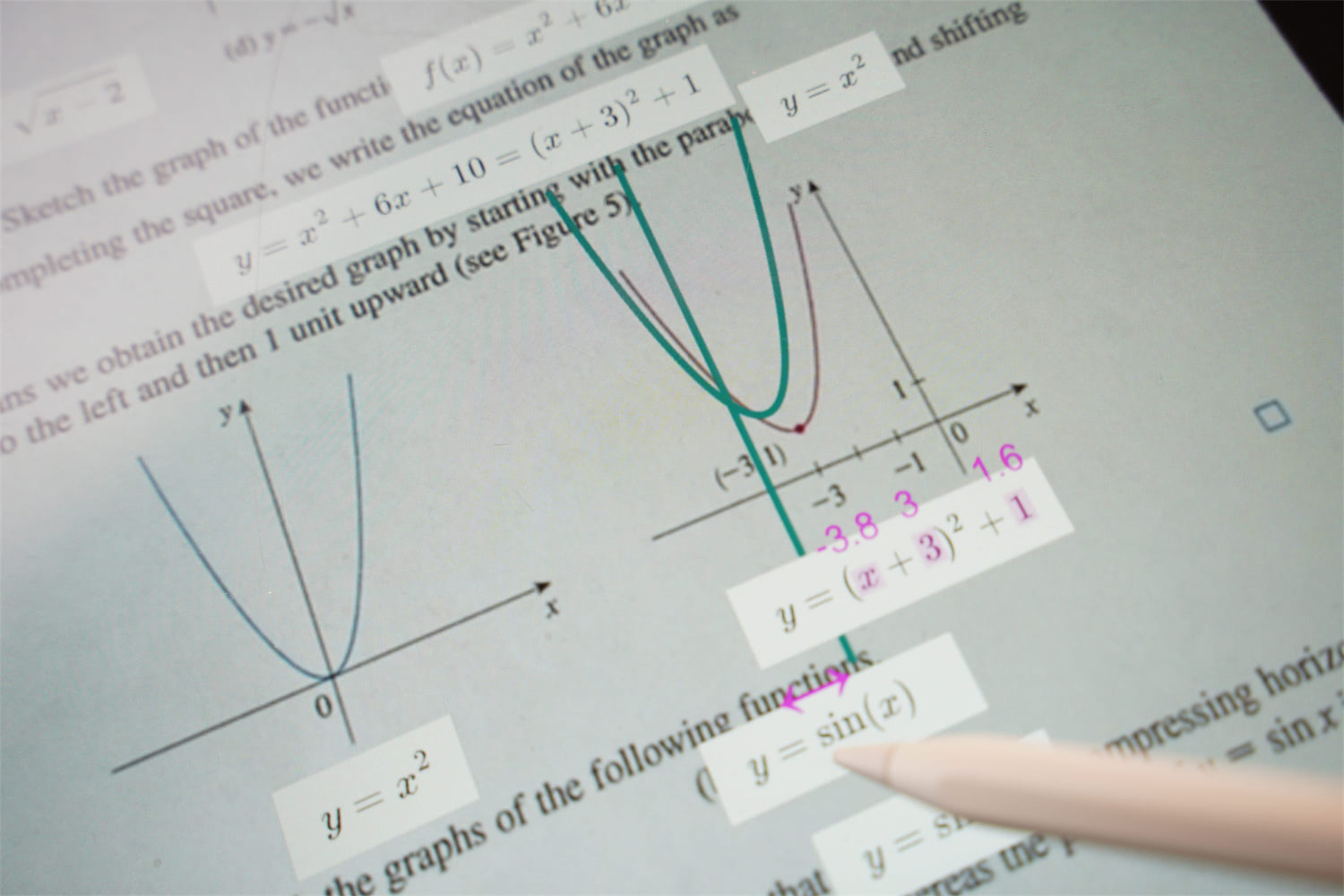}
\caption{Relationship Highlights}
\label{fig:relationship-highlights}
\end{figure}

\noindent
For example, Figure~\ref{fig:relationship-highlights} shows the relationship highlights in the graph by holding the pen over $x$ to show the current values on the graph.
Similarly, when users hold over $a$ in the triangle example (Figure~\ref{fig:interactive-triangle}), the corresponding line is highlighted, showing which variable represents the associated visual reference in the figure and graphs. 

\subsubsection{\textbf{Concrete Examples}}
Concrete examples aid users by illustrating abstract concepts through specific instances. 
This feature is inspired by both \textit{``strategy 1: exemplify through concrete values''} and \textit{``strategy 3: guide through contextual hints or exercises''} from our taxonomy analysis.
By identifying operations such as summation, the system can present concrete examples to contextualize and break down abstract formulas. 

\begin{figure}[h!]
\centering
\includegraphics[width=0.495\linewidth]{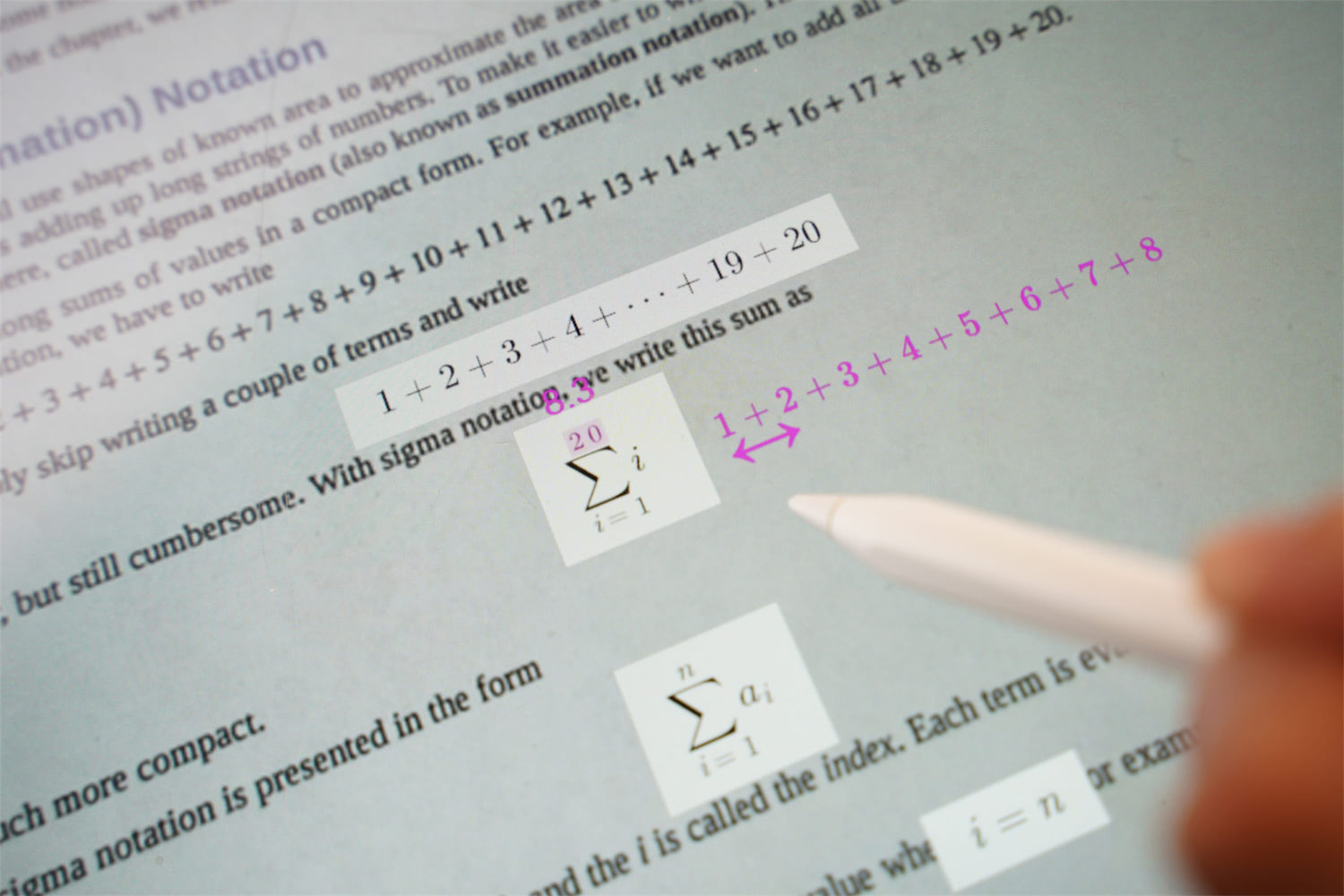}
\includegraphics[width=0.495\linewidth]{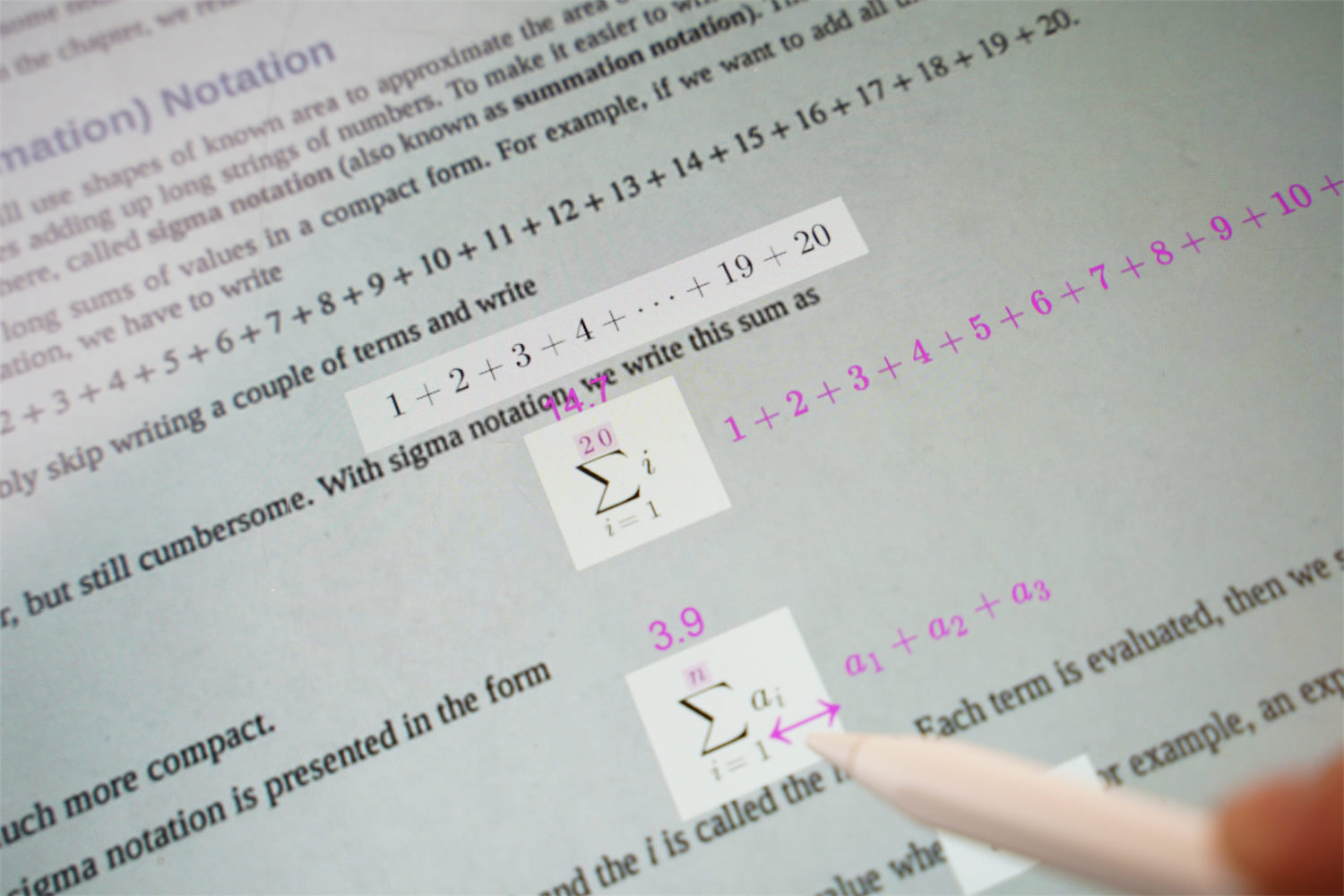}
\caption{Concrete Examples}
\label{fig:concrete-examples}
\end{figure}

\noindent
For instance, Figure~\ref{fig:concrete-examples} illustrates the user dragging the values of $20$ and $n$ in both $\sum_{i=1}^{20} i$ and $\sum_{i=1}^{n} a_i$. The system then exemplifies the formula by showing $1 + 2 + \cdots + n$ or $a_1 + a_2 + \cdots + a_n$.
Concrete examples provide users with explorable instances to enhance their understanding of abstract concepts. 

\subsubsection{\textbf{Step-by-Step Hints}}
Step-by-step hints offer contextual assistance by breaking down complex math problems or equations into a series of step-by-step instructions and solutions. 
This feature is inspired by \textit{``strategy 3: guide through contextual hints or exercises''} from our taxonomy analysis.
The system automatically computes the breakdown of a formula by performing arithmetic operations, simplifications, and decompositions.

\begin{figure}[h!]
\centering
\includegraphics[width=0.495\linewidth]{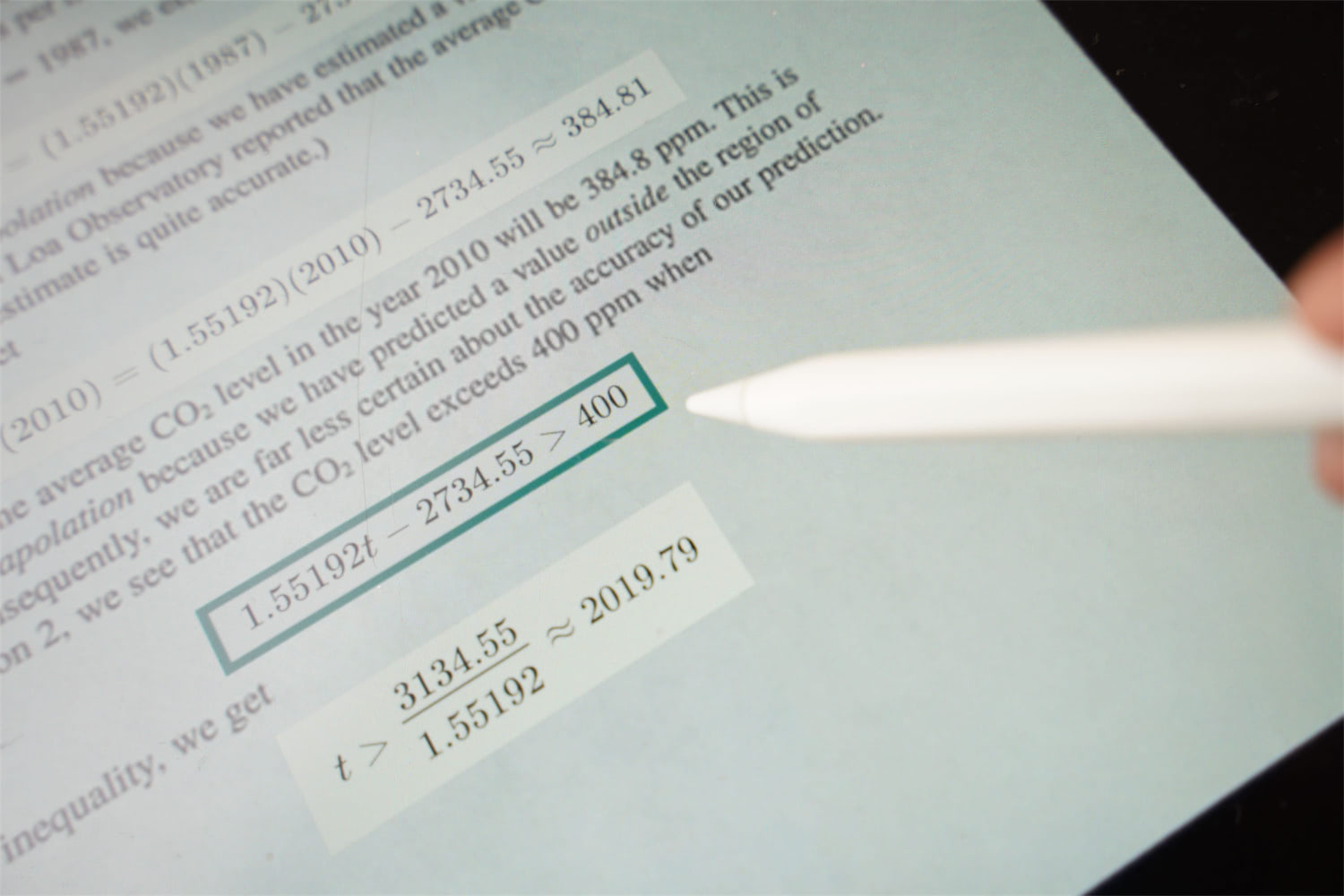}
\includegraphics[width=0.495\linewidth]{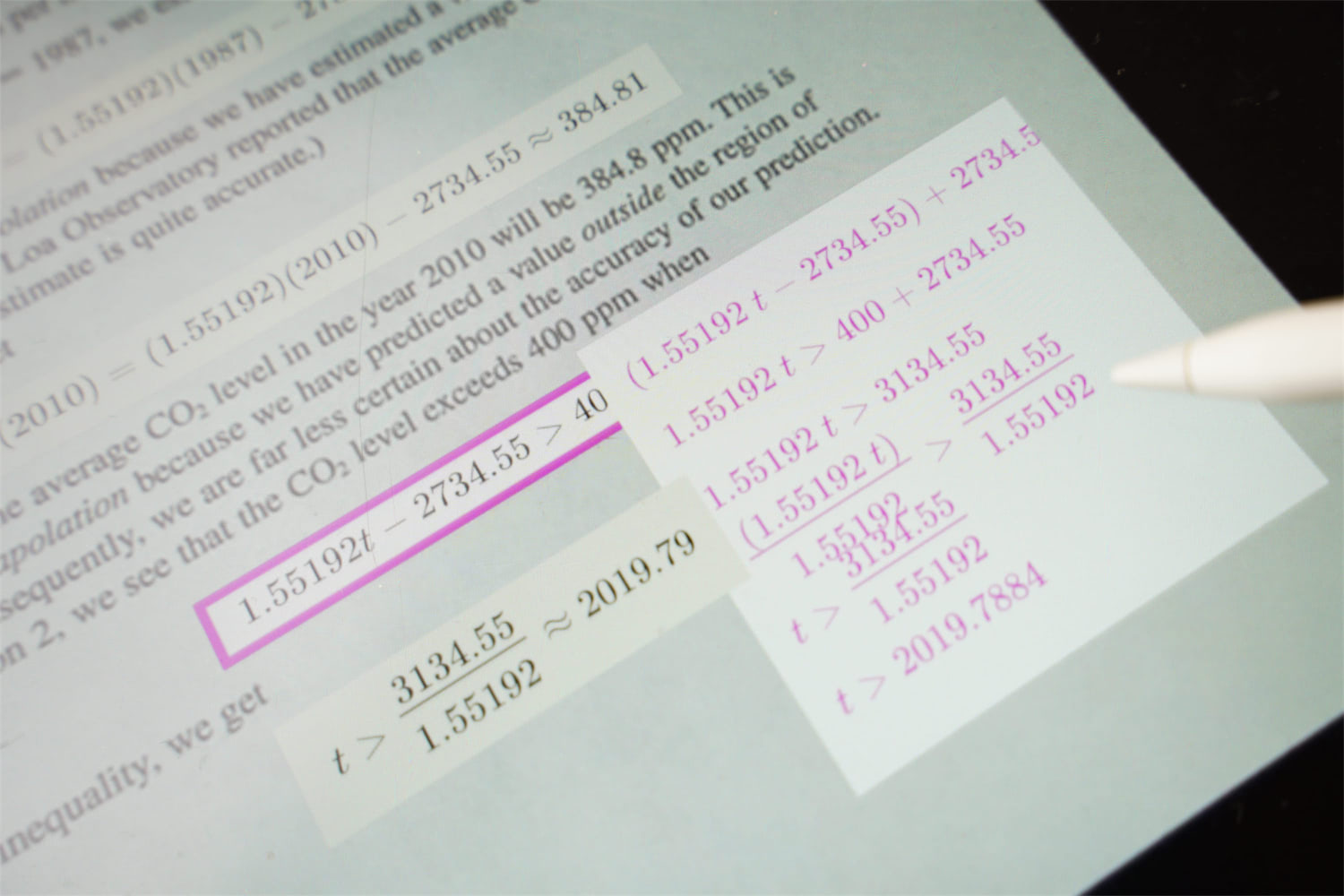}
\caption{Step-by-Step Hints}
\label{fig:step-by-step}
\end{figure}

\noindent
For example, Figure~\ref{fig:step-by-step} demonstrates the user selecting an inequality $1.55192 t - 2734.55 > 400$ and then breaking it down into a series of step-by-step operations. 
Step-by-step hints are applicable to various equations and problems, such as solving quadratic equations like $x^2 - 7x + 10 = 0 \Rightarrow (x - 5) (x - 2) = 0 \Rightarrow x = 2, 5$. 
By decomposing and breaking down the equation through automated step-by-step instructions, users receive contextual feedback that enhances their understanding.

\subsection{Implementation}
Our system is implemented using WebAR interface that leverages A-Frame, 8th Wall, and HTML Canvas. 
For graph manipulation and symbolic computation, we employ the MathJS library.
For text and math formula extraction, we utilize Google Cloud, MathPix, and CnSTD OCR.
Finally, we develop a custom computer vision algorithm based on OpenCV for figure and graph extraction.
To enhance reproducibility, we make our source code publicly available as open-source~\footnote{\url{https://github.com/ucalgary-ilab/augmented-math}} and a desktop interface is accessible through a live demo link~\footnote{\url{https://ilab.ucalgary.ca/augmented-math}}.

\subsubsection{Math Formula Extraction}
To extract text and math formulas from the input document, we employ OCR techniques. 
We first apply Google Cloud OCR to detect all text and localize the position of each letter and word. 
Google Cloud OCR outputs plain text without appropriate math expressions, making it difficult to understand the mathematical content semantically. 
Therefore, we implement other techniques to convert the scanned document to the LaTeX expression with MathPix OCR~\footnote{\url{https://mathpix.com}}.
Since MathPix does not provide location information for each detected formula, we also employ CnSTD~\footnote{\url{https://github.com/breezedeus/CnSTD}}, an OCR toolkit based on PyTorch and MXNet, that can detect the accurate position of inline and independent line math formulas within the document. 
CnSTD math formula detection algorithm is trained with CnMFD Dataset~\footnote{\url{https://github.com/breezedeus/CnMFD_Dataset}} that contains various math formulas across 17,500 pages of documents. 
Upon localizing the math in the document, we compare the similarity between detected plain text and math formulas based on the position of the localized math equation, allowing us to accurately localize the position of each math expression and overlay it on the printed document. 

\begin{figure}[h!]
\centering
\includegraphics[width=0.327\linewidth]{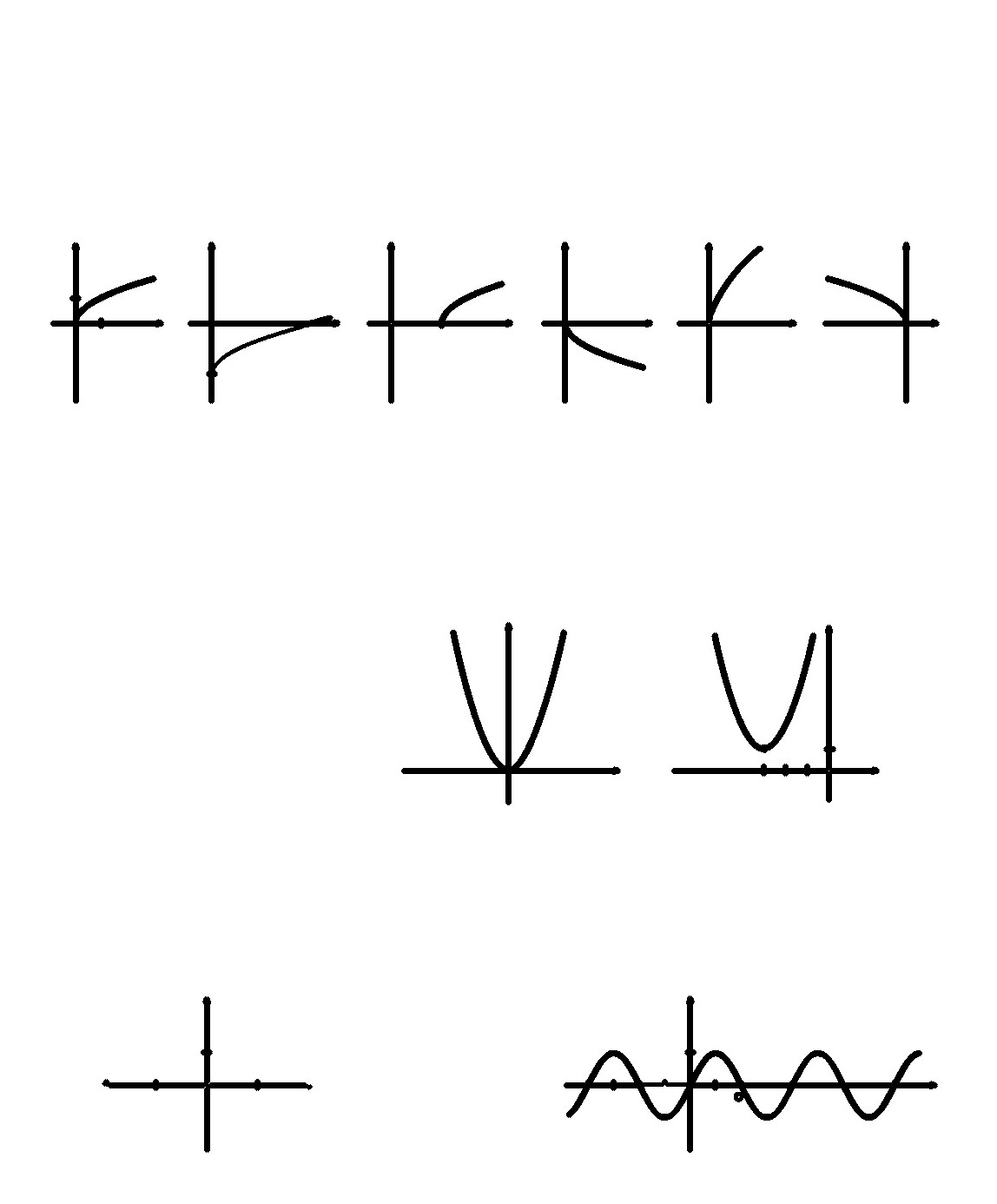}
\includegraphics[width=0.327\linewidth]{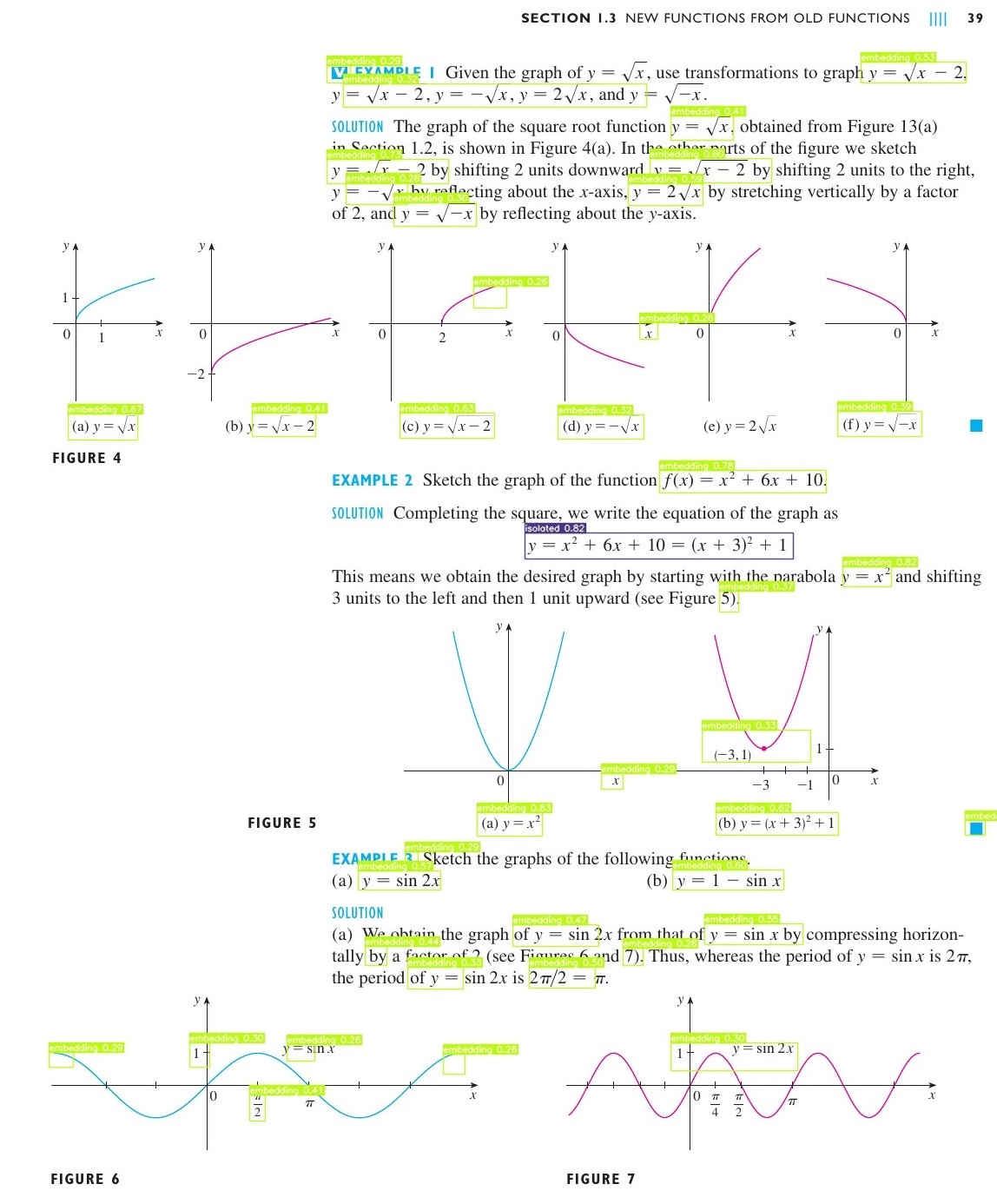}
\includegraphics[width=0.327\linewidth]{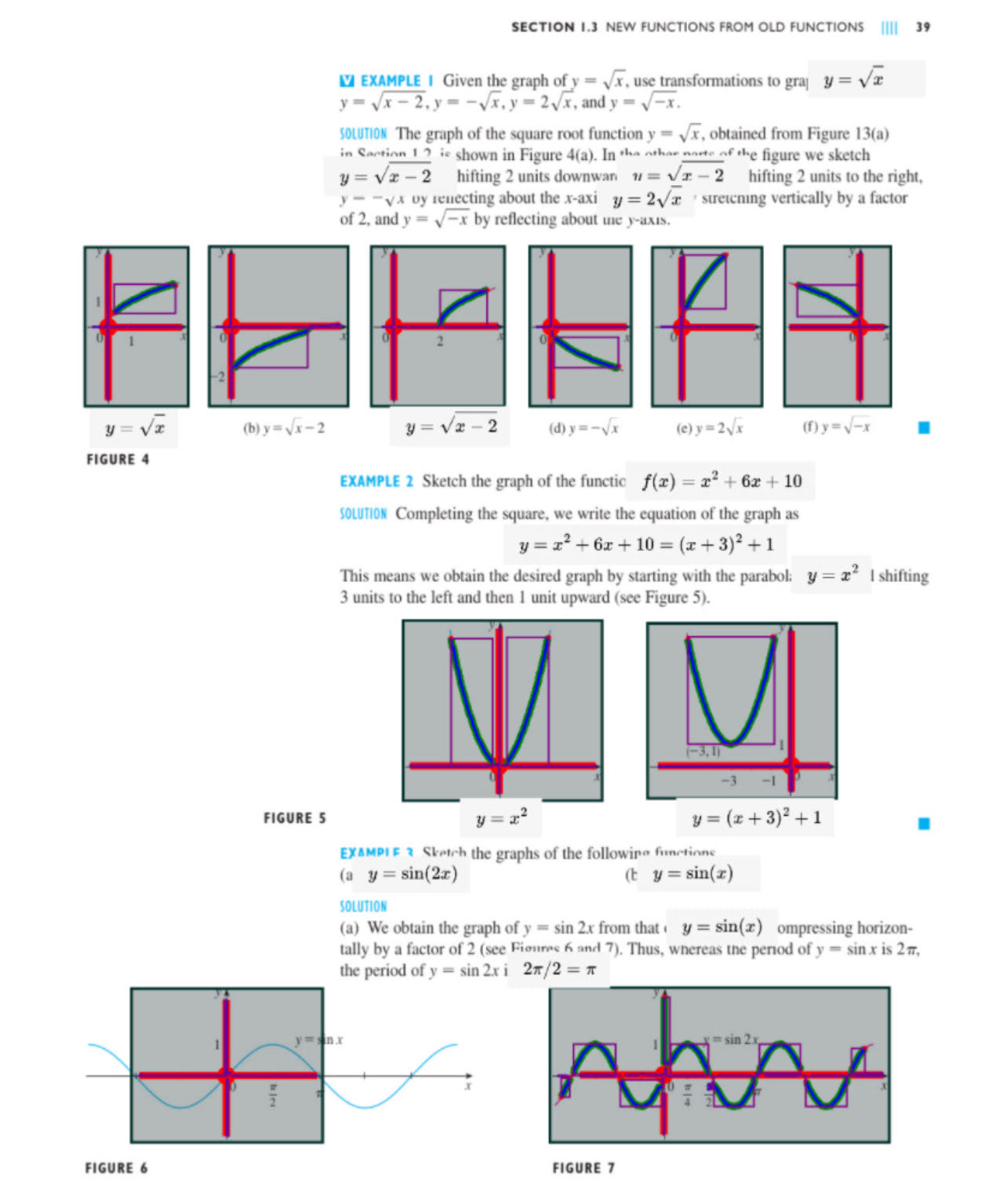}
\caption{Example results of figure extraction, math formula extraction, and graph identification.}
\label{fig:implementation}
\end{figure}

\subsubsection{Figure and Graph Extraction}
Since OCR only enables the detection and extraction of textual information, we developed a custom computer vision algorithm to accurately localize the positions  of graphs and diagrams.
To achieve this, we first remove text and extract only  diagrams from the given document by using OpenCV contour line extraction. 
By setting the minimum length of the contour line length, we filter out text and extract only graphs and diagrams.
Once the figures are extracted, the system processes the contour line data to detect bounding boxes, x-axes, y-axes, origins, and graph paths. 
To identify appropriate x-y coordinates, we extract the longest horizontal and vertical lines in the given figure and use them as the x-axis and y-axis. We then use graph paths to identify the appropriate ratio for a given pixel value. 

\subsubsection{Authoring Interface and Document Tracking}
Our authoring interface is implemented using A-Frame, a WebAR framework based on JavaScript and HTML. 
Visual elements are rendered through an HTML Canvas element embedded as a dynamic texture of a plane object in the A-Frame scene. 
We use the Konva library for Canvas manipulation. 
The detected mathematical expressions are rendered using SVG with the MathJax library. 
Graphs and diagrams are dynamically rendered through a series of lines, with points computed using the MathJS~\footnote{\url{https://mathjs.org/}} symbolic computation engine. 
We also employ the MathJS library for step-by-step hints and concrete examples features. 
The overall interface is developed using JavaScript and the React.js framework.
For document tracking, we utilize 8th Wall's image tracking features~\footnote{\url{https://www.8thwall.com/}}, which provide more reliable tracking than alternatives like MindAR and AR.js.

\subsection{Technical Evaluation}
\subsubsection{Method}
To evaluate the accuracy of our pipeline, we collected 14 different textbooks which include 5 Algebra, 4 Trigonometry, 2 Geometry, and 3 Calculus textbooks. We selected 10 random pages from each of the textbooks for our sample. We performed our detection pipeline to a total of 140 pages of the collected samples and then check the results manually (as there is no standardized data set or method).
We measured the error rate for \textbf{textual} equation recognition (MathPix, CnSTD, Google OCR), as well as \textbf{visual} graph and geometry detection (OpenCV and Custom Detection).

\subsubsection{Results}
The results suggest that the math equation recognition works fairly well with 75\% correctness. These are often very minor errors such as and extra spaces and an extra newline. Among 140 pages, there are 63 pages with graphs and 20 pages with geometries, which makes up 59\% of all the samples. Out of these 83 image samples, graph detection was accurate 48 out of 63 times (76\%), while geometry detection was only accurate 8 out of 20 times (40\%). By investigating these results, we also noticed some error patterns. For graph detection, we noticed that decoration lines or tables are detected as false positives and multiple X-Y scale lines were detected as the X-Y axis. In addition, for geometry detection, the auxiliary lines often introduce the difficulty of the appropriate detection. In general, a visually cluttered textbook with a lot of colors, shapes, or visuals often generates many errors. 

\section{User Study}
To evaluate the potential of \system{}, we conducted a user study consisting of preliminary user testing with eleven students and expert interviews with six math instructors. The purpose of the user testing was to evaluate the ease of use and user satisfaction with the system, while the expert interviews aimed to gain insights for educational use from the experts' perspectives.

\subsection{Preliminary User Evaluation}
We evaluated the usability of our system in three different conditions: the \system{} mobile AR interface, the \system{} desktop interface, and static textbook pages without AR. The same pages from a high school math textbook were used for each condition. 
We recruited 11 adult participants (8 males, 3 females, ages 18 - 39) from our local community, all of whom had at least a high school level knowledge in math and varying experiences with AR applications.
Each session lasted approximately 45 minutes, and participants were compensated with 15 CAD for their time.

\subsubsection{Method}
The study employed a within-subject evaluation design, with the three conditions randomized across participants. The AR interface was displayed on a tablet mounted on a tripod, with a pen used for interaction, while the web version ran on a laptop, with a mouse for interaction.
The study began with an introduction that included participants completing a consent form, followed by a system walkthrough. Subsequently, participants were asked to answer five math questions in each of the conditions, targeting the different features of the system: 
\begin{itemize}
\item How would the graph change if a particular constant were to be increased or decreased?
\item How would the equation change if a particular constant were to be increased or decreased?
\item How would you simplify this equation?
\item How would you expand this notation?
\item How would the variables in this equation change if we were to change the shape it represents in certain ways? 
\end{itemize}
The graphs, equations, and numbers varied slightly between the conditions. After completing all conditions, participants filled out a questionnaire that included usability questions for the AR and desktop interfaces, inquiries about each feature of the system, and an overall comparison of the three conditions. Responses were recorded on a 7-point Likert scale, and open questions allowed participants to provide additional feedback.

\subsubsection{Results} 
This section describes the results of our preliminary user study. We describe three aspects: 1) system usability between our mobile AR and desktop interfaces, 2) overall experience and engagement to compare mobile AR, desktop, and static textbooks, and 3) feedback for different features.

\subsubsection*{\textbf{Usability of Mobile AR and Desktop Interface}}
Overall, participants found both the AR interface and desktop interface easy to use, with a preference for the desktop interface. The desktop interface received an average SUS score of 81.82 (SD = 13.56), while the AR interface had an average SUS score of 75 (SD = 15.32). Many participants encountered issues with the imprecision of the AR version, including small fonts on the page that made selections difficult, especially when viewed through the camera (P1-2, P5, P9-11). This problem was exacerbated by the shakiness of participants' hands (P5) and the device itself when in use (P2, P10). In contrast, participants found the web version smoother to interact with, as the mouse provided a more precise input modality (P1, P2) and the fields were more consistently identified by the OCR (P3). However, some participants desired even greater precision, such as through text box input (P10, P11). 

\subsubsection*{\textbf{Comparison of the Three Conditions: Mobile AR, Desktop, and Static Textbook}}
\begin{figure}[h!]
\centering
\includegraphics[width=\linewidth]{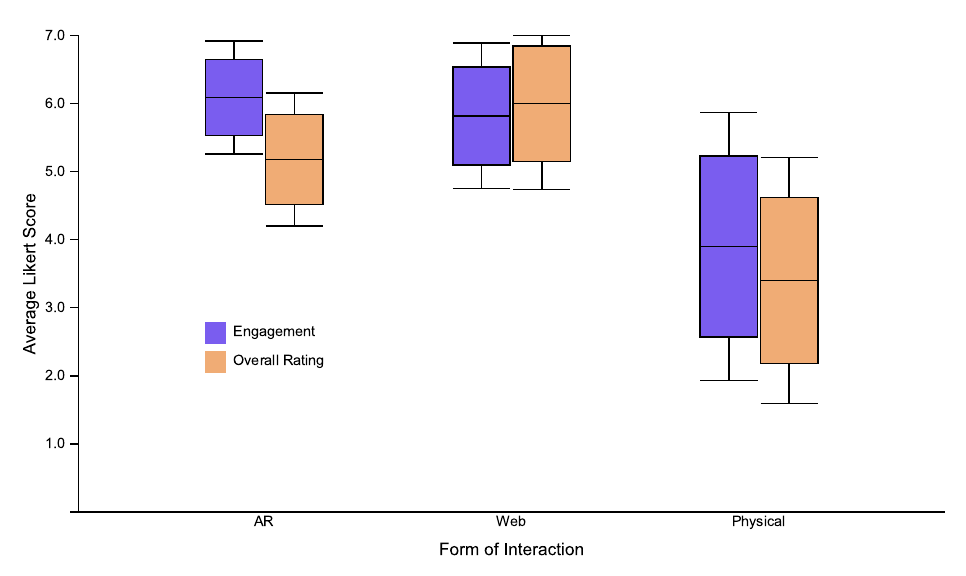}
\caption{This box plot summarizes the average scores of engagement (purple) and overall rating (orange) based on participants' preferences.}
\label{fig:boxplot}
\end{figure}

When comparing the three conditions, the desktop was the favorite among participants (M = 6.00, SD = 1.26), followed by the AR version (M = 5.18, SD = 0.98). The static textbook pages without AR were the least preferred (M = 3.40, SD = 1.81). When asked about which condition was the most engaging, participants preferred the AR interface (M = 6.09, SD = 0.83) over the desktop interface (M = 5.82, SD = 1.07) due to its novelty (P7, P10) and the intuitive physical manipulation it allowed (P11). The static textbook scored an average of 3.9 (SD = 1.97) in engagement. The results are summarized in \autoref{fig:boxplot}. 

\subsubsection*{\textbf{Comparison of the Different Features}}
\begin{figure}[h!]
\centering
\includegraphics[width=\linewidth]{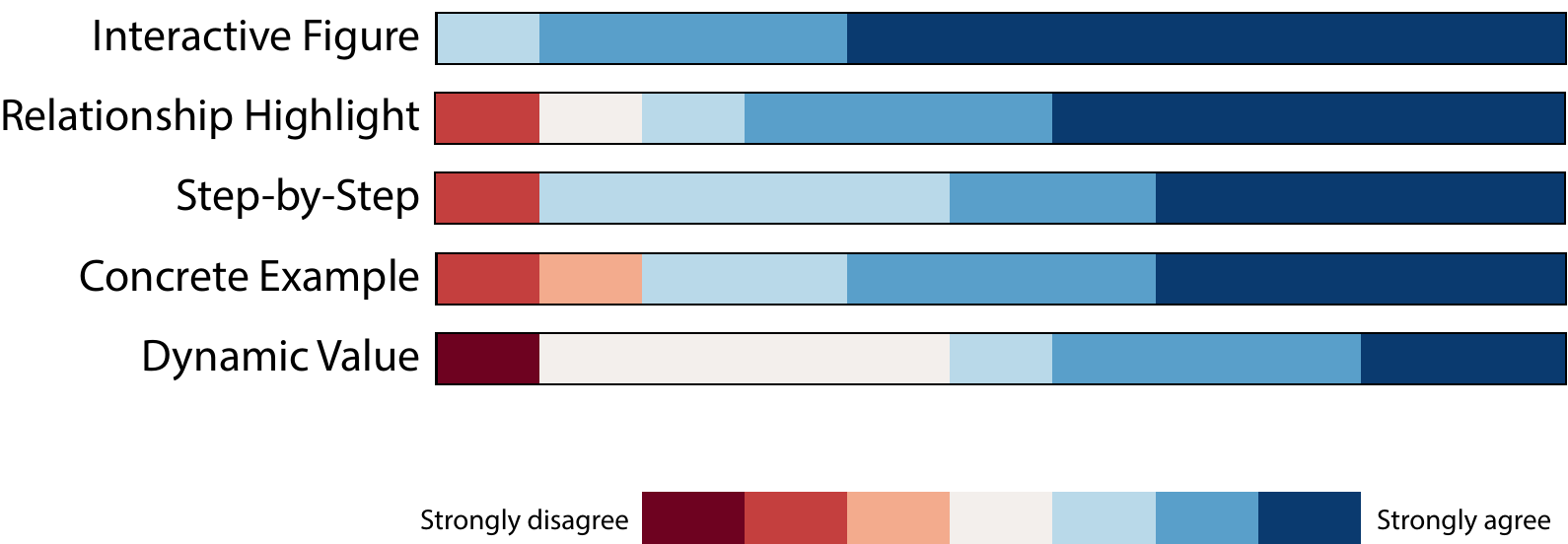}
\caption{This figure summarize our participants' feedback on the usefulness of the implemented features. From top to bottom, the features are sorted by their average score in descending order.}
\label{fig:feature}
\end{figure}

Participants generally provided positive feedback about the features. On a scale of 1 to 7, the \textit{interactive figure} received an average score of 6.55 (SD = 0.69) and was particularly popular among participants. They appreciated being able to see the cause and effect immediately on the graph as they interacted with the equation. The second favorite feature was the \textit{relationship highlight}, which scored an average of 5.81 (SD = 1.6). Participants valued how it helped them see the correlation between the shape and the equation more clearly. However, one participant (P2) did not find this feature helpful, as they believed the shape was already too abstract and disconnected from the equation. Most participants also found the \textit{step-by-step hint} helpful (M = 5.81, SD = 1.6), particularly as a convenience (P4, P7) or as a means of checking for mistakes (P9, P10). One participant (P2), however, found that it showed too many steps that they could already visualize in their head.
The \textit{exemplify} feature was well-liked (M = 5.55, SD = 1.69), as it helped participants concretize an abstract concept, and some found it easier to understand the notation through interaction (P3, P5). Some participants wanted to see it applied to more complex concepts (P7, P10), while others were more interested in the shorthand formula to solve abstracted notation rather than its detailed expansion (P2). Lastly, participants had mixed feelings about the \textit{dynamic value} feature and did not find it particularly useful compared to the other features (M = 4.91, SD = 1.76). It was more useful with long solutions (P9) or when combined with other features (P4, P7) but not as much on its own (P6). The results are summarized in \autoref{fig:feature}.

\subsection{Expert Interview}
We conducted semi-structured interviews with five math instructors to evaluate the system's potential for educational use. These instructors have teaching experience ranging from 1 to 6 years (M = 3.4 years). Three experts teach at the university level (E1, E3, E4), one teaches late elementary to middle school level (E2), and the last teaches from middle school up to early junior university level (E5). During the interviews, we asked them to compare four different educational media: textbooks, videos, interactive websites, and our AR approach. Each interview lasted approximately 60 minutes, and the experts received compensation of 15 USD.

\subsubsection{Method}
The goal of the interviews was to gather feedback on the augmented textbook and identify potential areas for improvement. We began by asking the instructors about their experience using textbooks, as well as any challenges or benefits they found in the physical medium. Next, we presented an educational video and an interactive math website covering a topic familiar to the instructors. We asked them to compare these three media as educational resources.
Finally, we demonstrated our system through an example authoring walkthrough, allowing the instructors to interact with the system. We gathered their feedback on our AR approach through a series of open-ended questions, focusing on how it compares to existing methods. 

\subsubsection{Results}
Overall, all the experts expressed enthusiasm about our system. They specifically highlighted several key benefits of our approach, including enhanced student engagement, minimal authoring efforts, adaptable content, and potential use for independent learning. Below, we outline our key findings.

\subsubsection*{\textbf{Student Engagement through Interactions}}
All the instructors we interviewed concurred that integrating interactivity into learning materials can effectively encourage students to explore and experiment with concepts at their own pace. They expressed that the interactive nature of our augmented textbook has the potential to transform mundane textbooks into engaging, game-like experiences (E4). 
Particularly, E5 mentioned that \textit{ "the ones that are fascinated by it will play around and try to learn more"}. 
The experts highlighted the value of manipulating content and observing immediate results, especially in calculus, where students grapple with unintuitive concepts like limits. As one instructor noted,
\textit{"actually seeing that change happen in real time could be helpful to understand a concept"} (E1). 
Our augmented reality approach, which combines visual and motor stimuli, may also aid students in retaining information more effectively (E5). 

While the experts generally praised the potential benefits of augmented textbooks, most (E1-E4) believed that the available interactions were not always self-explanatory and required guidance. Instructors E3 and E4 specifically emphasized the importance of incorporating built-in guidance to assist them in directing students' attention and interactions. In summary, our system can enhance engagement with traditional materials, enabling seamless integration into existing educational workflows.

\subsubsection*{\textbf{Reduced Authoring Effort}} 
During the interviews, instructors revealed that they dedicate a considerable amount of time to crafting materials for their students, often drawing from online resources or textbooks. They also expressed an interest in developing interactive teaching media but frequently lack the necessary skills (E3, E4, E5). Current authoring interfaces can be varied and complex, and instructors may not be well-versed in using them. E1 specifically stated, \textit{ "If you ask me to graph a piece wise function on GeoGebra right now, I wouldn't be able to do it"}.
Our approach holds the potential to substantially reduce the authoring effort needed to create interactive materials. E1 also noted that our approach makes it especially easy to modify existing content rather than starting from scratch. All the experts we interviewed agreed that they could utilize our tool to produce relevant materials for their students. 
Overall, our AR approach has the potential to make the creation of interactive materials more accessible and time-efficient for instructors.

\subsubsection*{\textbf{On-Demand Adaptable Content}}
Teachers frequently create customized content to align better with their curriculum and teaching objectives, building upon existing materials. \textit{"I would often just use specific images or drawings from textbooks and papers to explain a concept"} (E4). Our AR approach offers significant support in this area. During the interviews, instructors appreciated the system's ability to select from various augmentation techniques, enabling them to customize the augmented textbook according to their specific needs. This adaptability allows them to tailor the content to their course's learning objectives and their students' requirements. One expert (E1) found that integrating interactive websites into teaching can be challenging due to unfamiliarity with the site's functionalities or difficulties inputting functions into the system. Furthermore, comprehending the website's content and relating it back to the course material can also be challenging (E2). In contrast, our AR approach offers instructors more control over the material they are presenting (E1, E4). This control is particularly beneficial when teaching in front of a class, as customizing the material to suit specific teaching objectives is more desirable than using pre-animated content, such as videos (E1).
Overall, our AR approach delivers an efficient and adaptable solution for instructors to create interactive materials that align with their teaching objectives and better engage their students.

\subsubsection*{\textbf{Independent Learning}}
Instructors frequently assign materials for students to review independently, such as videos or online resources (E1, E3). Our AR approach provides an accessible solution for creating customized materials that can be easily shared with students.
During the interviews, instructors expressed enthusiasm about the ability to easily share the interactive AR textbook with their students via a web address. Moreover, having access to the materials at any time empowers students to review or advance at their own pace, allowing for more in-depth discussions and knowledge refinement during class time (E5).

\section{Limitations and Future Work}

\subsection*{Study Limitations}

\subsubsection*{\textbf{Limitations of Walkthrough Demonstration}}
While the experts were generally excited about the system, our expert interviews had certain limitations. The majority of the interview was based on a walkthrough demonstration of the AR interface, due to the limited time. Although this provided valuable insights into the system's utility, it did not evaluate the usability of the authoring flow of the system~\cite{ledo2018evaluation}. The experts briefly interacted with our system either through mobile AR (for in-person interviews) and/or desktop interfaces (for remote interviews), but a more comprehensive assessment of the authoring interface is necessary. Additionally, the experts were presented with a pre-defined math textbook and asked how they could incorporate it into their teaching. While this ensured that all experts were exposed to all system features, it would have been interesting to explore how the experts would utilize their own teaching materials.

\subsubsection*{\textbf{Real-World Deployment}}
Although the preliminary user evaluation and expert interviews suggest a potential for using \system{} in math education, deploying the system in actual teaching situations would offer further insights into its possibilities. Future work should examine which features instructors would use with their own teaching materials and whether they would discover other features that should be supported. Moreover, evaluating the system in an actual classroom setting with students would be valuable, observing how they interact with the system and exploring its potential benefits for their learning experience.

\subsection*{Improving Usability and Interaction}

\subsubsection*{\textbf{Supporting More Flexible Authoring Practices}}
In order to simplify the authoring process and enable non-technical users to create their own explorable explanations, the system relies on OCR and computer vision to automate as much of the process as possible. This includes automatically extracting math formulas and graphs. Although the system accurately detects most content, it occasionally fails.The system also automatically creates \textit{step-by-step hints} based on existing calculations. While most participants in our usability study appreciated these hints, P2 felt that there were too many steps to consume, asking for more flexible control to modify them. Moreover, E2 pointed out that the system should let users not only select from the presented equation but also directly and manually type a new equation to show on a graph.
This suggests the need for flexible authoring practice to customize the content for their own needs. Future implementations should allow users to manually input equations and graph boundaries. A detection confidence value can also be shown to inform users that they may have to adjust some things.

\subsubsection*{\textbf{Improving Interaction for Mobile AR}}
Although participants in the usability study found the AR system to be the most engaging, they rated the usability of the web system higher. Participants encountered difficulties interacting with the AR system due to various reasons, such as small text (P1, P9), shakiness of their hands (P5) and the device itself (P2, P10), and image detection issues (P3). Further work is needed to explore how AR interaction can be improved. For example, future work would explore different possibilities such as magnifying the selected content for better usability or enabling users to take snapshots of the content and interact with it flat on the table instead of directly in AR. Additionally, more work is needed to examine how AR information is presented to the user. For example, the content could be visualized outside the page instead of on top, or AR could be used to simplify the page so that only relevant information is visible.


\subsection*{Beyond Augmenting Math Textbooks}
\subsubsection*{\textbf{Adapting to Various Topics and Educational Contexts}}
Our approach has the potential to be applied in subjects beyond mathematics. In language education, for example, \textit{relationship highlight} and \textit{concrete examples} could be employed to connect abstract grammar rules with examples in the text, such as by highlighting subjects or commas in the text that pertain to the demonstrated rule. 
In social sciences, \textit{relationship highlight} could also be utilized to emphasize connections between different people and parties or dates and events. 
In music education, notes and songs on a page could be augmented with sound, enabling users to hear the tone or song. Our approach could also be used to enhance physics education by incorporating physics simulations, allowing users to interact with textbooks in more engaging ways. 

\subsubsection*{\textbf{Extensibility}}
Our system relies on computer algebra systems (CAS) like MathJS or SymPy, which are difficult to scale for more complex math. We expect that a different approach would be necessary to support broader topics. For example, an LLM-based approach like ChatGPT + Wolfram Alpha API or Google’s Minerva could be an exciting direction to integrate in the future.

\subsubsection*{\textbf{Integrating Other AR Interfaces}}
Finally, we are also interested in exploring other AR interfaces. For example, using projection mapping instead of mobile AR could enhance the teacher's live explanations by augmenting a whiteboard. This approach facilitates different affordances for collaborative learning experiences, enabling students to learn and explore with a group rather than individually. On the other hand, we are also interested in developing explorable explanations for immersive environments through mixed reality headsets. While our presented approach may not be directly applicable due to the lack of precision and resolution of the interaction, it instead opens up new opportunities for immersive explorable explanations beyond 2D interfaces. Future work should continue to investigate this exciting domain by leveraging advanced AR and machine learning techniques for the future of education.
\section{Conclusion}
We introduced \system{}, a tool for creating AR-based \textit{explorable explanations} by augmenting static math textbooks. Using OCR and computer vision, our system automatically extracts contents from documents, enabling non-technical users to transform existing math textbooks into personalized, interactive documents without programming.
Our proof-of-concept system features five augmentation techniques, based on a taxonomy analysis of existing explorable explanations. 
We conducted two user studies to evaluate our system, and the results of our studies suggest that our system fosters more engaging learning experiences.
Future work includes applying our approach to different domains beyond math textbooks and investigating the potential for classroom use through in-the-wild deployment.
\section*{Acknowledgements}
This research was funded part by the Natural Sciences and Engineering Research Council of Canada (NSERC) Discovery Grant RGPIN-2021-02857.
We also thank all of the participants for our user study.

\ifdouble
  \balance
\fi
\bibliographystyle{ACM-Reference-Format}
\bibliography{references}


\begin{thebibliography}{68}


\ifx \showCODEN    \undefined \def \showCODEN     #1{\unskip}     \fi
\ifx \showDOI      \undefined \def \showDOI       #1{#1}\fi
\ifx \showISBNx    \undefined \def \showISBNx     #1{\unskip}     \fi
\ifx \showISBNxiii \undefined \def \showISBNxiii  #1{\unskip}     \fi
\ifx \showISSN     \undefined \def \showISSN      #1{\unskip}     \fi
\ifx \showLCCN     \undefined \def \showLCCN      #1{\unskip}     \fi
\ifx \shownote     \undefined \def \shownote      #1{#1}          \fi
\ifx \showarticletitle \undefined \def \showarticletitle #1{#1}   \fi
\ifx \showURL      \undefined \def \showURL       {\relax}        \fi
\providecommand\bibfield[2]{#2}
\providecommand\bibinfo[2]{#2}
\providecommand\natexlab[1]{#1}
\providecommand\showeprint[2][]{arXiv:#2}

\bibitem[\protect\citeauthoryear{Adams}{Adams}{2010}]%
        {adams2010student}
\bibfield{author}{\bibinfo{person}{Wendy~K Adams}.}
  \bibinfo{year}{2010}\natexlab{}.
\newblock \showarticletitle{Student engagement and learning with PhET
  interactive simulations}.
\newblock \bibinfo{journal}{\emph{Il nuovo cimento C}} \bibinfo{volume}{33},
  \bibinfo{number}{3} (\bibinfo{year}{2010}), \bibinfo{pages}{21--32}.
\newblock


\bibitem[\protect\citeauthoryear{Adobe}{Adobe}{[n.d.]}]%
        {aero}
\bibfield{author}{\bibinfo{person}{Adobe}.} \bibinfo{year}{[n.d.]}\natexlab{}.
\newblock \bibinfo{title}{Create augmented reality with AR software - Adobe
  Aero}.
\newblock
  \bibinfo{howpublished}{\url{https://www.adobe.com/products/aero.html}}.
\newblock
\newblock
\shownote{(Accessed on 03/10/2023).}


\bibitem[\protect\citeauthoryear{Ahmad and Junaini}{Ahmad and Junaini}{2020}]%
        {Ahmad2020Augmented}
\bibfield{author}{\bibinfo{person}{Nur Ahmad} {and} \bibinfo{person}{Syahrul
  Junaini}.} \bibinfo{year}{2020}\natexlab{}.
\newblock \showarticletitle{Augmented Reality for Learning Mathematics: A
  Systematic Literature Review}.
\newblock \bibinfo{journal}{\emph{International Journal of Emerging
  Technologies in Learning (iJET)}} \bibinfo{volume}{15}, \bibinfo{number}{16}
  (\bibinfo{date}{August} \bibinfo{year}{2020}), \bibinfo{pages}{106--122}.
\newblock
\showISSN{1863-0383}
\urldef\tempurl%
\url{https://www.learntechlib.org/p/217969}
\showURL{%
\tempurl}


\bibitem[\protect\citeauthoryear{Apple}{Apple}{[n.d.]}]%
        {RealityComposer}
\bibfield{author}{\bibinfo{person}{Apple}.} \bibinfo{year}{[n.d.]}\natexlab{}.
\newblock \bibinfo{title}{Reality Composer on the App Store}.
\newblock
  \bibinfo{howpublished}{\url{https://apps.apple.com/us/app/reality-composer/id1462358802}}.
\newblock
\newblock
\shownote{(Accessed on 03/10/2023).}


\bibitem[\protect\citeauthoryear{Billinghurst, Kato, and Poupyrev}{Billinghurst
  et~al\mbox{.}}{2001}]%
        {billinghurst2001magicbook}
\bibfield{author}{\bibinfo{person}{Mark Billinghurst},
  \bibinfo{person}{Hirokazu Kato}, {and} \bibinfo{person}{Ivan Poupyrev}.}
  \bibinfo{year}{2001}\natexlab{}.
\newblock \showarticletitle{The MagicBook: a transitional AR interface}.
\newblock \bibinfo{journal}{\emph{Computers \& Graphics}} \bibinfo{volume}{25},
  \bibinfo{number}{5} (\bibinfo{year}{2001}), \bibinfo{pages}{745--753}.
\newblock


\bibitem[\protect\citeauthoryear{Case}{Case}{[n.d.]}]%
        {Explorab37:online}
\bibfield{author}{\bibinfo{person}{Nicky Case}.}
  \bibinfo{year}{[n.d.]}\natexlab{}.
\newblock \bibinfo{title}{Explorable Explanations}.
\newblock \bibinfo{howpublished}{\url{https://explorabl.es/}}.
\newblock
\newblock
\shownote{(Accessed on 04/03/2023).}


\bibitem[\protect\citeauthoryear{ChanRT}{ChanRT}{[n.d.]}]%
        {Visualiz92:online}
\bibfield{author}{\bibinfo{person}{ChanRT}.} \bibinfo{year}{[n.d.]}\natexlab{}.
\newblock \bibinfo{title}{Visualize It}.
\newblock \bibinfo{howpublished}{\url{https://visualize-it.github.io/}}.
\newblock
\newblock
\shownote{(Accessed on 04/03/2023).}


\bibitem[\protect\citeauthoryear{Cheema and LaViola}{Cheema and
  LaViola}{2012}]%
        {cheema2012physicsbook}
\bibfield{author}{\bibinfo{person}{Salman Cheema} {and} \bibinfo{person}{Joseph
  LaViola}.} \bibinfo{year}{2012}\natexlab{}.
\newblock \showarticletitle{PhysicsBook: a sketch-based interface for animating
  physics diagrams}. In \bibinfo{booktitle}{\emph{Proceedings of the 2012 ACM
  international conference on Intelligent User Interfaces}}.
  \bibinfo{pages}{51--60}.
\newblock


\bibitem[\protect\citeauthoryear{Chen, Tong, Wang, Bach, and Qu}{Chen
  et~al\mbox{.}}{2020}]%
        {chen2020augmenting}
\bibfield{author}{\bibinfo{person}{Zhutian Chen}, \bibinfo{person}{Wai Tong},
  \bibinfo{person}{Qianwen Wang}, \bibinfo{person}{Benjamin Bach}, {and}
  \bibinfo{person}{Huamin Qu}.} \bibinfo{year}{2020}\natexlab{}.
\newblock \showarticletitle{Augmenting static visualizations with paparvis
  designer}. In \bibinfo{booktitle}{\emph{Proceedings of the 2020 CHI
  Conference on Human Factors in Computing Systems}}. \bibinfo{pages}{1--12}.
\newblock


\bibitem[\protect\citeauthoryear{Chen and Xia}{Chen and Xia}{2022}]%
        {chen2022crossdata}
\bibfield{author}{\bibinfo{person}{Zhutian Chen} {and} \bibinfo{person}{Haijun
  Xia}.} \bibinfo{year}{2022}\natexlab{}.
\newblock \showarticletitle{Crossdata: Leveraging text-data connections for
  authoring data documents}. In \bibinfo{booktitle}{\emph{Proceedings of the
  2022 CHI Conference on Human Factors in Computing Systems}}.
  \bibinfo{pages}{1--15}.
\newblock


\bibitem[\protect\citeauthoryear{ching Chen}{ching Chen}{2019}]%
        {Chen2019effect}
\bibfield{author}{\bibinfo{person}{Yu ching Chen}.}
  \bibinfo{year}{2019}\natexlab{}.
\newblock \showarticletitle{Effect of Mobile Augmented Reality on Learning
  Performance, Motivation, and Math Anxiety in a Math Course}.
\newblock \bibinfo{journal}{\emph{Journal of Educational Computing Research}}
  \bibinfo{volume}{57}, \bibinfo{number}{7} (\bibinfo{year}{2019}),
  \bibinfo{pages}{1695--1722}.
\newblock
\urldef\tempurl%
\url{https://doi.org/10.1177/0735633119854036}
\showDOI{\tempurl}
\showeprint{https://doi.org/10.1177/0735633119854036}


\bibitem[\protect\citeauthoryear{Conlen and Heer}{Conlen and Heer}{2018}]%
        {conlen2018idyll}
\bibfield{author}{\bibinfo{person}{Matthew Conlen} {and}
  \bibinfo{person}{Jeffrey Heer}.} \bibinfo{year}{2018}\natexlab{}.
\newblock \showarticletitle{Idyll: A markup language for authoring and
  publishing interactive articles on the web}. In
  \bibinfo{booktitle}{\emph{Proceedings of the 31st Annual ACM Symposium on
  User Interface Software and Technology}}. \bibinfo{pages}{977--989}.
\newblock


\bibitem[\protect\citeauthoryear{Conlen and Heer}{Conlen and Heer}{2022}]%
        {conlen2022fidyll}
\bibfield{author}{\bibinfo{person}{Matthew Conlen} {and}
  \bibinfo{person}{Jeffrey Heer}.} \bibinfo{year}{2022}\natexlab{}.
\newblock \showarticletitle{Fidyll: A Compiler for Cross-Format Data Stories \&
  Explorable Explanations}.
\newblock \bibinfo{journal}{\emph{arXiv preprint arXiv:2205.09858}}
  (\bibinfo{year}{2022}).
\newblock


\bibitem[\protect\citeauthoryear{Conlen, Vo, Tan, and Heer}{Conlen
  et~al\mbox{.}}{2021}]%
        {conlen2021idyll}
\bibfield{author}{\bibinfo{person}{Matthew Conlen}, \bibinfo{person}{Megan Vo},
  \bibinfo{person}{Alan Tan}, {and} \bibinfo{person}{Jeffrey Heer}.}
  \bibinfo{year}{2021}\natexlab{}.
\newblock \showarticletitle{Idyll studio: A structured editor for authoring
  interactive \& data-driven articles}. In \bibinfo{booktitle}{\emph{The 34th
  Annual ACM Symposium on User Interface Software and Technology}}.
  \bibinfo{pages}{1--12}.
\newblock


\bibitem[\protect\citeauthoryear{Distill}{Distill}{2023}]%
        {distill}
\bibfield{author}{\bibinfo{person}{Distill}.} \bibinfo{year}{2023}\natexlab{}.
\newblock \bibinfo{title}{Distill}.
\newblock
\newblock
\urldef\tempurl%
\url{https://distill.pub/}
\showURL{%
\tempurl}


\bibitem[\protect\citeauthoryear{Du, Olwal, Le~Goc, Wu, Tang, Zhang, Zhang,
  Tan, Tombari, and Kim}{Du et~al\mbox{.}}{2022}]%
        {du2022opportunistic}
\bibfield{author}{\bibinfo{person}{Ruofei Du}, \bibinfo{person}{Alex Olwal},
  \bibinfo{person}{Mathieu Le~Goc}, \bibinfo{person}{Shengzhi Wu},
  \bibinfo{person}{Danhang Tang}, \bibinfo{person}{Yinda Zhang},
  \bibinfo{person}{Jun Zhang}, \bibinfo{person}{David~Joseph Tan},
  \bibinfo{person}{Federico Tombari}, {and} \bibinfo{person}{David Kim}.}
  \bibinfo{year}{2022}\natexlab{}.
\newblock \showarticletitle{Opportunistic Interfaces for Augmented Reality:
  Transforming Everyday Objects Into Tangible 6DoF Interfaces Using Ad Hoc UI}.
  In \bibinfo{booktitle}{\emph{CHI Conference on Human Factors in Computing
  Systems Extended Abstracts}}. \bibinfo{pages}{1--4}.
\newblock


\bibitem[\protect\citeauthoryear{GeoGebra}{GeoGebra}{[n.d.]}]%
        {GeoGebra95:online}
\bibfield{author}{\bibinfo{person}{GeoGebra}.}
  \bibinfo{year}{[n.d.]}\natexlab{}.
\newblock \bibinfo{title}{GeoGebra - the world’s favorite, free math tools
  used by over 100 million students and teachers}.
\newblock \bibinfo{howpublished}{\url{https://www.geogebra.org/?lang=en}}.
\newblock
\newblock
\shownote{(Accessed on 03/21/2023).}


\bibitem[\protect\citeauthoryear{Grasset, Duenser, Seichter, and
  Billinghurst}{Grasset et~al\mbox{.}}{2007}]%
        {Grasset2007mixed}
\bibfield{author}{\bibinfo{person}{Raphael Grasset}, \bibinfo{person}{Andreas
  Duenser}, \bibinfo{person}{Hartmut Seichter}, {and} \bibinfo{person}{Mark
  Billinghurst}.} \bibinfo{year}{2007}\natexlab{}.
\newblock \showarticletitle{The Mixed Reality Book: A New Multimedia Reading
  Experience}. In \bibinfo{booktitle}{\emph{CHI '07 Extended Abstracts on Human
  Factors in Computing Systems}} (San Jose, CA, USA)
  \emph{(\bibinfo{series}{CHI EA '07})}. \bibinfo{publisher}{Association for
  Computing Machinery}, \bibinfo{address}{New York, NY, USA},
  \bibinfo{pages}{1953–1958}.
\newblock
\showISBNx{9781595936424}
\urldef\tempurl%
\url{https://doi.org/10.1145/1240866.1240931}
\showDOI{\tempurl}


\bibitem[\protect\citeauthoryear{Guo}{Guo}{2013}]%
        {guo2013online}
\bibfield{author}{\bibinfo{person}{Philip~J Guo}.}
  \bibinfo{year}{2013}\natexlab{}.
\newblock \showarticletitle{Online python tutor: embeddable web-based program
  visualization for cs education}. In \bibinfo{booktitle}{\emph{Proceeding of
  the 44th ACM technical symposium on Computer science education}}.
  \bibinfo{pages}{579--584}.
\newblock


\bibitem[\protect\citeauthoryear{Gupta, Lin, Ji, Patel, and Vogel}{Gupta
  et~al\mbox{.}}{2020}]%
        {gupta2020replicate}
\bibfield{author}{\bibinfo{person}{Aakar Gupta}, \bibinfo{person}{Bo~Rui Lin},
  \bibinfo{person}{Siyi Ji}, \bibinfo{person}{Arjav Patel}, {and}
  \bibinfo{person}{Daniel Vogel}.} \bibinfo{year}{2020}\natexlab{}.
\newblock \showarticletitle{Replicate and reuse: Tangible interaction design
  for digitally-augmented physical media objects}. In
  \bibinfo{booktitle}{\emph{Proceedings of the 2020 CHI Conference on Human
  Factors in Computing Systems}}. \bibinfo{pages}{1--12}.
\newblock


\bibitem[\protect\citeauthoryear{Han, Cheng, Strachan, and Ma}{Han
  et~al\mbox{.}}{2021}]%
        {Han2021hybrid}
\bibfield{author}{\bibinfo{person}{Feng Han}, \bibinfo{person}{Yifei Cheng},
  \bibinfo{person}{Megan Strachan}, {and} \bibinfo{person}{Xiaojuan Ma}.}
  \bibinfo{year}{2021}\natexlab{}.
\newblock \showarticletitle{Hybrid Paper-Digital Interfaces: A Systematic
  Literature Review}. In \bibinfo{booktitle}{\emph{Designing Interactive
  Systems Conference 2021}} (Virtual Event, USA) \emph{(\bibinfo{series}{DIS
  '21})}. \bibinfo{publisher}{Association for Computing Machinery},
  \bibinfo{address}{New York, NY, USA}, \bibinfo{pages}{1087–1100}.
\newblock
\showISBNx{9781450384766}
\urldef\tempurl%
\url{https://doi.org/10.1145/3461778.3462059}
\showDOI{\tempurl}


\bibitem[\protect\citeauthoryear{Head, Xie, and Hearst}{Head
  et~al\mbox{.}}{2022}]%
        {Head2022}
\bibfield{author}{\bibinfo{person}{Andrew Head}, \bibinfo{person}{Amber Xie},
  {and} \bibinfo{person}{Marti~A. Hearst}.} \bibinfo{year}{2022}\natexlab{}.
\newblock \showarticletitle{Math Augmentation: How Authors Enhance the
  Readability of Formulas Using Novel Visual Design Practices}. In
  \bibinfo{booktitle}{\emph{Proceedings of the 2022 CHI Conference on Human
  Factors in Computing Systems}} (New Orleans, LA, USA)
  \emph{(\bibinfo{series}{CHI '22})}. \bibinfo{publisher}{Association for
  Computing Machinery}, \bibinfo{address}{New York, NY, USA}, Article
  \bibinfo{articleno}{491}, \bibinfo{numpages}{18}~pages.
\newblock
\showISBNx{9781450391573}
\urldef\tempurl%
\url{https://doi.org/10.1145/3491102.3501932}
\showDOI{\tempurl}


\bibitem[\protect\citeauthoryear{Hensberry, Moore, and Perkins}{Hensberry
  et~al\mbox{.}}{2015}]%
        {hensberry2015effective}
\bibfield{author}{\bibinfo{person}{Karina Hensberry}, \bibinfo{person}{Emily
  Moore}, {and} \bibinfo{person}{Katherine Perkins}.}
  \bibinfo{year}{2015}\natexlab{}.
\newblock \showarticletitle{Effective student learning of fractions with an
  interactive simulation}.
\newblock \bibinfo{journal}{\emph{Journal of Computers in Mathematics and
  Science Teaching}} \bibinfo{volume}{34}, \bibinfo{number}{3}
  (\bibinfo{year}{2015}), \bibinfo{pages}{273--298}.
\newblock


\bibitem[\protect\citeauthoryear{Kaimoto, Monteiro, Faridan, Li, Farajian,
  Kakehi, Nakagaki, and Suzuki}{Kaimoto et~al\mbox{.}}{2022}]%
        {kaimoto2022sketched}
\bibfield{author}{\bibinfo{person}{Hiroki Kaimoto}, \bibinfo{person}{Kyzyl
  Monteiro}, \bibinfo{person}{Mehrad Faridan}, \bibinfo{person}{Jiatong Li},
  \bibinfo{person}{Samin Farajian}, \bibinfo{person}{Yasuaki Kakehi},
  \bibinfo{person}{Ken Nakagaki}, {and} \bibinfo{person}{Ryo Suzuki}.}
  \bibinfo{year}{2022}\natexlab{}.
\newblock \showarticletitle{Sketched Reality: Sketching Bi-Directional
  Interactions Between Virtual and Physical Worlds with AR and Actuated
  Tangible UI}. In \bibinfo{booktitle}{\emph{Proceedings of the 35th Annual ACM
  Symposium on User Interface Software and Technology}}.
  \bibinfo{pages}{1--12}.
\newblock


\bibitem[\protect\citeauthoryear{Kang, Shokeen, Byrne, Norooz, Bonsignore,
  Williams-Pierce, and Froehlich}{Kang et~al\mbox{.}}{2020}]%
        {Kang2020armath}
\bibfield{author}{\bibinfo{person}{Seokbin Kang}, \bibinfo{person}{Ekta
  Shokeen}, \bibinfo{person}{Virginia~L. Byrne}, \bibinfo{person}{Leyla
  Norooz}, \bibinfo{person}{Elizabeth Bonsignore}, \bibinfo{person}{Caro
  Williams-Pierce}, {and} \bibinfo{person}{Jon~E. Froehlich}.}
  \bibinfo{year}{2020}\natexlab{}.
\newblock \showarticletitle{ARMath: Augmenting Everyday Life with Math
  Learning}. In \bibinfo{booktitle}{\emph{Proceedings of the 2020 CHI
  Conference on Human Factors in Computing Systems}} (Honolulu, HI, USA)
  \emph{(\bibinfo{series}{CHI '20})}. \bibinfo{publisher}{Association for
  Computing Machinery}, \bibinfo{address}{New York, NY, USA},
  \bibinfo{pages}{1–15}.
\newblock
\showISBNx{9781450367080}
\urldef\tempurl%
\url{https://doi.org/10.1145/3313831.3376252}
\showDOI{\tempurl}


\bibitem[\protect\citeauthoryear{Kaufmann and Schmalstieg}{Kaufmann and
  Schmalstieg}{2002}]%
        {Kaufmann2002mathematics}
\bibfield{author}{\bibinfo{person}{Hannes Kaufmann} {and}
  \bibinfo{person}{Dieter Schmalstieg}.} \bibinfo{year}{2002}\natexlab{}.
\newblock \showarticletitle{Mathematics and Geometry Education with
  Collaborative Augmented Reality}. In \bibinfo{booktitle}{\emph{ACM SIGGRAPH
  2002 Conference Abstracts and Applications}} (San Antonio, Texas)
  \emph{(\bibinfo{series}{SIGGRAPH '02})}. \bibinfo{publisher}{Association for
  Computing Machinery}, \bibinfo{address}{New York, NY, USA},
  \bibinfo{pages}{37–41}.
\newblock
\showISBNx{1581135254}
\urldef\tempurl%
\url{https://doi.org/10.1145/1242073.1242086}
\showDOI{\tempurl}


\bibitem[\protect\citeauthoryear{Kazi, Chevalier, Grossman, and
  Fitzmaurice}{Kazi et~al\mbox{.}}{2014}]%
        {kazi2014kitty}
\bibfield{author}{\bibinfo{person}{Rubaiat~Habib Kazi}, \bibinfo{person}{Fanny
  Chevalier}, \bibinfo{person}{Tovi Grossman}, {and} \bibinfo{person}{George
  Fitzmaurice}.} \bibinfo{year}{2014}\natexlab{}.
\newblock \showarticletitle{Kitty: sketching dynamic and interactive
  illustrations}. In \bibinfo{booktitle}{\emph{Proceedings of the 27th annual
  ACM symposium on User interface software and technology}}.
  \bibinfo{pages}{395--405}.
\newblock


\bibitem[\protect\citeauthoryear{Keller, Finkelstein, Perkins, and
  Pollock}{Keller et~al\mbox{.}}{2007}]%
        {keller2007assessing}
\bibfield{author}{\bibinfo{person}{CJ Keller}, \bibinfo{person}{ND
  Finkelstein}, \bibinfo{person}{KK Perkins}, {and} \bibinfo{person}{SJ
  Pollock}.} \bibinfo{year}{2007}\natexlab{}.
\newblock \showarticletitle{Assessing the effectiveness of a computer
  simulation in introductory undergraduate environments}. In
  \bibinfo{booktitle}{\emph{AIP Conference Proceedings}},
  Vol.~\bibinfo{volume}{883}. American Institute of Physics,
  \bibinfo{pages}{121--124}.
\newblock


\bibitem[\protect\citeauthoryear{Khan, Trujano, Choudhury, and Maes}{Khan
  et~al\mbox{.}}{2018}]%
        {Khan2018mathland}
\bibfield{author}{\bibinfo{person}{Mina Khan}, \bibinfo{person}{Fernando
  Trujano}, \bibinfo{person}{Ashris Choudhury}, {and} \bibinfo{person}{Pattie
  Maes}.} \bibinfo{year}{2018}\natexlab{}.
\newblock \showarticletitle{Mathland: Playful Mathematical Learning in Mixed
  Reality}. In \bibinfo{booktitle}{\emph{Extended Abstracts of the 2018 CHI
  Conference on Human Factors in Computing Systems}} (Montreal QC, Canada)
  \emph{(\bibinfo{series}{CHI EA '18})}. \bibinfo{publisher}{Association for
  Computing Machinery}, \bibinfo{address}{New York, NY, USA},
  \bibinfo{pages}{1–4}.
\newblock
\showISBNx{9781450356213}
\urldef\tempurl%
\url{https://doi.org/10.1145/3170427.3186499}
\showDOI{\tempurl}


\bibitem[\protect\citeauthoryear{Kirner, Reis, and Kirner}{Kirner
  et~al\mbox{.}}{2012}]%
        {Kirner2012development}
\bibfield{author}{\bibinfo{person}{Tereza~Gonçalves Kirner},
  \bibinfo{person}{Fernanda Maria~Villela Reis}, {and} \bibinfo{person}{Claudio
  Kirner}.} \bibinfo{year}{2012}\natexlab{}.
\newblock \showarticletitle{Development of an interactive book with Augmented
  Reality for teaching and learning geometric shapes}. In
  \bibinfo{booktitle}{\emph{7th Iberian Conference on Information Systems and
  Technologies (CISTI 2012)}}. \bibinfo{pages}{1--6}.
\newblock


\bibitem[\protect\citeauthoryear{Klamka and Dachselt}{Klamka and
  Dachselt}{2017}]%
        {klamka2017illumipaper}
\bibfield{author}{\bibinfo{person}{Konstantin Klamka} {and}
  \bibinfo{person}{Raimund Dachselt}.} \bibinfo{year}{2017}\natexlab{}.
\newblock \showarticletitle{IllumiPaper: Illuminated interactive paper}. In
  \bibinfo{booktitle}{\emph{Proceedings of the 2017 CHI Conference on Human
  Factors in Computing Systems}}. \bibinfo{pages}{5605--5618}.
\newblock


\bibitem[\protect\citeauthoryear{Kunin, Guo, Devlin, and Xiang}{Kunin
  et~al\mbox{.}}{[n.d.]}]%
        {SeeingTh51:online}
\bibfield{author}{\bibinfo{person}{Daniel Kunin}, \bibinfo{person}{Jingru Guo},
  \bibinfo{person}{Tyler~Dae Devlin}, {and} \bibinfo{person}{Daniel Xiang}.}
  \bibinfo{year}{[n.d.]}\natexlab{}.
\newblock \bibinfo{title}{Seeing Theory}.
\newblock
  \bibinfo{howpublished}{\url{https://seeing-theory.brown.edu/\#3rdPage}}.
\newblock
\newblock
\shownote{(Accessed on 04/03/2023).}


\bibitem[\protect\citeauthoryear{Lau and Guo}{Lau and Guo}{2020}]%
        {lau2020data}
\bibfield{author}{\bibinfo{person}{Sam Lau} {and} \bibinfo{person}{Philip~J
  Guo}.} \bibinfo{year}{2020}\natexlab{}.
\newblock \showarticletitle{Data Theater: A live programming environment for
  prototyping data-driven explorable explanations}. In
  \bibinfo{booktitle}{\emph{Workshop on Live Programming (LIVE)}}.
\newblock


\bibitem[\protect\citeauthoryear{LaViola and Zeleznik}{LaViola and
  Zeleznik}{2007}]%
        {LaViola2007mathPad2}
\bibfield{author}{\bibinfo{person}{Joseph~J. LaViola} {and}
  \bibinfo{person}{Robert~C. Zeleznik}.} \bibinfo{year}{2007}\natexlab{}.
\newblock \showarticletitle{MathPad2: A System for the Creation and Exploration
  of Mathematical Sketches}. In \bibinfo{booktitle}{\emph{ACM SIGGRAPH 2007
  Courses}} (San Diego, California) \emph{(\bibinfo{series}{SIGGRAPH '07})}.
  \bibinfo{publisher}{Association for Computing Machinery},
  \bibinfo{address}{New York, NY, USA}, \bibinfo{pages}{46–es}.
\newblock
\showISBNx{9781450318235}
\urldef\tempurl%
\url{https://doi.org/10.1145/1281500.1281557}
\showDOI{\tempurl}


\bibitem[\protect\citeauthoryear{Ledo, Houben, Vermeulen, Marquardt, Oehlberg,
  and Greenberg}{Ledo et~al\mbox{.}}{2018}]%
        {ledo2018evaluation}
\bibfield{author}{\bibinfo{person}{David Ledo}, \bibinfo{person}{Steven
  Houben}, \bibinfo{person}{Jo Vermeulen}, \bibinfo{person}{Nicolai Marquardt},
  \bibinfo{person}{Lora Oehlberg}, {and} \bibinfo{person}{Saul Greenberg}.}
  \bibinfo{year}{2018}\natexlab{}.
\newblock \showarticletitle{Evaluation strategies for HCI toolkit research}. In
  \bibinfo{booktitle}{\emph{Proceedings of the 2018 CHI Conference on Human
  Factors in Computing Systems}}. \bibinfo{pages}{1--17}.
\newblock


\bibitem[\protect\citeauthoryear{Leiva, Gr\o{}nb\ae{}k, Klokmose, Nguyen, Kazi,
  and Asente}{Leiva et~al\mbox{.}}{2021}]%
        {Leiva2021rapido}
\bibfield{author}{\bibinfo{person}{Germ\'{a}n Leiva},
  \bibinfo{person}{Jens~Emil Gr\o{}nb\ae{}k},
  \bibinfo{person}{Clemens~Nylandsted Klokmose}, \bibinfo{person}{Cuong
  Nguyen}, \bibinfo{person}{Rubaiat~Habib Kazi}, {and} \bibinfo{person}{Paul
  Asente}.} \bibinfo{year}{2021}\natexlab{}.
\newblock \bibinfo{booktitle}{\emph{Rapido: Prototyping Interactive AR
  Experiences through Programming by Demonstration}}.
\newblock \bibinfo{publisher}{Association for Computing Machinery},
  \bibinfo{address}{New York, NY, USA}, \bibinfo{pages}{626–637}.
\newblock
\showISBNx{9781450386357}
\urldef\tempurl%
\url{https://doi.org/10.1145/3472749.3474774}
\showURL{%
\tempurl}


\bibitem[\protect\citeauthoryear{Leiva, Nguyen, Kazi, and Asente}{Leiva
  et~al\mbox{.}}{2020}]%
        {Leiva2020pronto}
\bibfield{author}{\bibinfo{person}{Germ\'{a}n Leiva}, \bibinfo{person}{Cuong
  Nguyen}, \bibinfo{person}{Rubaiat~Habib Kazi}, {and} \bibinfo{person}{Paul
  Asente}.} \bibinfo{year}{2020}\natexlab{}.
\newblock \showarticletitle{Pronto: Rapid Augmented Reality Video Prototyping
  Using Sketches and Enaction}. In \bibinfo{booktitle}{\emph{Proceedings of the
  2020 CHI Conference on Human Factors in Computing Systems}} (Honolulu, HI,
  USA) \emph{(\bibinfo{series}{CHI '20})}. \bibinfo{publisher}{Association for
  Computing Machinery}, \bibinfo{address}{New York, NY, USA},
  \bibinfo{pages}{1–13}.
\newblock
\showISBNx{9781450367080}
\urldef\tempurl%
\url{https://doi.org/10.1145/3313831.3376160}
\showDOI{\tempurl}


\bibitem[\protect\citeauthoryear{Li, van~der Spek, Hu, and Feijs}{Li
  et~al\mbox{.}}{2019b}]%
        {Li2019turning}
\bibfield{author}{\bibinfo{person}{Jingya Li}, \bibinfo{person}{Erik~D. van~der
  Spek}, \bibinfo{person}{Jun Hu}, {and} \bibinfo{person}{Loe Feijs}.}
  \bibinfo{year}{2019}\natexlab{b}.
\newblock \showarticletitle{Turning Your Book into a Game: Improving Motivation
  through Tangible Interaction and Diegetic Feedback in an AR Mathematics Game
  for Children}. In \bibinfo{booktitle}{\emph{Proceedings of the Annual
  Symposium on Computer-Human Interaction in Play}} (Barcelona, Spain)
  \emph{(\bibinfo{series}{CHI PLAY '19})}. \bibinfo{publisher}{Association for
  Computing Machinery}, \bibinfo{address}{New York, NY, USA},
  \bibinfo{pages}{73–85}.
\newblock
\showISBNx{9781450366885}
\urldef\tempurl%
\url{https://doi.org/10.1145/3311350.3347174}
\showDOI{\tempurl}


\bibitem[\protect\citeauthoryear{Li, Annett, Hinckley, Singh, and Wigdor}{Li
  et~al\mbox{.}}{2019a}]%
        {Li2019holodoc}
\bibfield{author}{\bibinfo{person}{Zhen Li}, \bibinfo{person}{Michelle Annett},
  \bibinfo{person}{Ken Hinckley}, \bibinfo{person}{Karan Singh}, {and}
  \bibinfo{person}{Daniel Wigdor}.} \bibinfo{year}{2019}\natexlab{a}.
\newblock \showarticletitle{HoloDoc: Enabling Mixed Reality Workspaces That
  Harness Physical and Digital Content}. In
  \bibinfo{booktitle}{\emph{Proceedings of the 2019 CHI Conference on Human
  Factors in Computing Systems}} (Glasgow, Scotland Uk)
  \emph{(\bibinfo{series}{CHI '19})}. \bibinfo{publisher}{Association for
  Computing Machinery}, \bibinfo{address}{New York, NY, USA},
  \bibinfo{pages}{1–14}.
\newblock
\showISBNx{9781450359702}
\urldef\tempurl%
\url{https://doi.org/10.1145/3290605.3300917}
\showDOI{\tempurl}


\bibitem[\protect\citeauthoryear{Liao, Karim, Jadon, Kazi, and Suzuki}{Liao
  et~al\mbox{.}}{2022}]%
        {liao2022realitytalk}
\bibfield{author}{\bibinfo{person}{Jian Liao}, \bibinfo{person}{Adnan Karim},
  \bibinfo{person}{Shivesh~Singh Jadon}, \bibinfo{person}{Rubaiat~Habib Kazi},
  {and} \bibinfo{person}{Ryo Suzuki}.} \bibinfo{year}{2022}\natexlab{}.
\newblock \showarticletitle{RealityTalk: Real-Time Speech-Driven Augmented
  Presentation for AR Live Storytelling}. In
  \bibinfo{booktitle}{\emph{Proceedings of the 35th Annual ACM Symposium on
  User Interface Software and Technology}}. \bibinfo{pages}{1--12}.
\newblock


\bibitem[\protect\citeauthoryear{Lunding, Leiva, Gr{\o}nb{\ae}k, and
  Petersen}{Lunding et~al\mbox{.}}{2022}]%
        {lunding2022projectar}
\bibfield{author}{\bibinfo{person}{Mille~Skovhus Lunding},
  \bibinfo{person}{Germ{\'a}n Leiva}, \bibinfo{person}{Jens Emil~Sloth
  Gr{\o}nb{\ae}k}, {and} \bibinfo{person}{Marianne~Graves Petersen}.}
  \bibinfo{year}{2022}\natexlab{}.
\newblock \showarticletitle{ProjectAR: Rapid Prototyping of Projection Mapping
  with Mobile Augmented Reality}. In \bibinfo{booktitle}{\emph{Adjunct
  Proceedings of the 2022 Nordic Human-Computer Interaction Conference}}.
  \bibinfo{pages}{1--5}.
\newblock


\bibitem[\protect\citeauthoryear{Meyerovich, Guha, Baskin, Cooper, Greenberg,
  Bromfield, and Krishnamurthi}{Meyerovich et~al\mbox{.}}{2009}]%
        {meyerovich2009flapjax}
\bibfield{author}{\bibinfo{person}{Leo~A Meyerovich}, \bibinfo{person}{Arjun
  Guha}, \bibinfo{person}{Jacob Baskin}, \bibinfo{person}{Gregory~H Cooper},
  \bibinfo{person}{Michael Greenberg}, \bibinfo{person}{Aleks Bromfield}, {and}
  \bibinfo{person}{Shriram Krishnamurthi}.} \bibinfo{year}{2009}\natexlab{}.
\newblock \showarticletitle{Flapjax: a programming language for Ajax
  applications}. In \bibinfo{booktitle}{\emph{Proceedings of the 24th ACM
  SIGPLAN conference on Object oriented programming systems languages and
  applications}}. \bibinfo{pages}{1--20}.
\newblock


\bibitem[\protect\citeauthoryear{Monteiro, Vatsal, Chulpongsatorn, Parnami, and
  Suzuki}{Monteiro et~al\mbox{.}}{2023}]%
        {monteiro2023teachable}
\bibfield{author}{\bibinfo{person}{Kyzyl Monteiro}, \bibinfo{person}{Ritik
  Vatsal}, \bibinfo{person}{Neil Chulpongsatorn}, \bibinfo{person}{Aman
  Parnami}, {and} \bibinfo{person}{Ryo Suzuki}.}
  \bibinfo{year}{2023}\natexlab{}.
\newblock \showarticletitle{Teachable Reality: Prototyping Tangible Augmented
  Reality with Everyday Objects by Leveraging Interactive Machine Teaching}.
\newblock \bibinfo{journal}{\emph{arXiv preprint arXiv:2302.11046}}
  (\bibinfo{year}{2023}).
\newblock


\bibitem[\protect\citeauthoryear{Nebeling, Nebeling, Yu, and Rumble}{Nebeling
  et~al\mbox{.}}{2018}]%
        {Nebeling2018protoAR}
\bibfield{author}{\bibinfo{person}{Michael Nebeling}, \bibinfo{person}{Janet
  Nebeling}, \bibinfo{person}{Ao Yu}, {and} \bibinfo{person}{Rob Rumble}.}
  \bibinfo{year}{2018}\natexlab{}.
\newblock \bibinfo{booktitle}{\emph{ProtoAR: Rapid Physical-Digital Prototyping
  of Mobile Augmented Reality Applications}}.
\newblock \bibinfo{publisher}{Association for Computing Machinery},
  \bibinfo{address}{New York, NY, USA}, \bibinfo{pages}{1–12}.
\newblock
\showISBNx{9781450356206}
\urldef\tempurl%
\url{https://doi.org/10.1145/3173574.3173927}
\showURL{%
\tempurl}


\bibitem[\protect\citeauthoryear{of~Colorado~Boulder}{of~Colorado~Boulder}{[n.d.]}]%
        {PhETFree6:online}
\bibfield{author}{\bibinfo{person}{University of Colorado~Boulder}.}
  \bibinfo{year}{[n.d.]}\natexlab{}.
\newblock \bibinfo{title}{PhET: Free online physics, chemistry, biology, earth
  science and math simulations}.
\newblock \bibinfo{howpublished}{\url{https://phet.colorado.edu/}}.
\newblock
\newblock
\shownote{(Accessed on 04/03/2023).}


\bibitem[\protect\citeauthoryear{of~Concept~Visualization}{of~Concept~Visualization}{[n.d.]}]%
        {Galleryo94:online}
\bibfield{author}{\bibinfo{person}{Gallery of Concept~Visualization}.}
  \bibinfo{year}{[n.d.]}\natexlab{}.
\newblock \bibinfo{title}{Gallery of Concept Visualization}.
\newblock \bibinfo{howpublished}{\url{https://conceptviz.github.io/}}.
\newblock
\newblock
\shownote{(Accessed on 04/03/2023).}


\bibitem[\protect\citeauthoryear{Perkins, Adams, Dubson, Finkelstein, Reid,
  Wieman, and LeMaster}{Perkins et~al\mbox{.}}{2006}]%
        {perkins2006phet}
\bibfield{author}{\bibinfo{person}{Katherine Perkins}, \bibinfo{person}{Wendy
  Adams}, \bibinfo{person}{Michael Dubson}, \bibinfo{person}{Noah Finkelstein},
  \bibinfo{person}{Sam Reid}, \bibinfo{person}{Carl Wieman}, {and}
  \bibinfo{person}{Ron LeMaster}.} \bibinfo{year}{2006}\natexlab{}.
\newblock \showarticletitle{PhET: Interactive simulations for teaching and
  learning physics}.
\newblock \bibinfo{journal}{\emph{The physics teacher}} \bibinfo{volume}{44},
  \bibinfo{number}{1} (\bibinfo{year}{2006}), \bibinfo{pages}{18--23}.
\newblock


\bibitem[\protect\citeauthoryear{Powell and Lehe}{Powell and Lehe}{2014}]%
        {Setosada85:online}
\bibfield{author}{\bibinfo{person}{Victor Powell} {and} \bibinfo{person}{Lewis
  Lehe}.} \bibinfo{year}{2014}\natexlab{}.
\newblock \bibinfo{title}{Setosa data visualization and visual explanations}.
\newblock \bibinfo{howpublished}{\url{https://setosa.io/}}.
\newblock
\newblock
\shownote{(Accessed on 04/03/2023).}


\bibitem[\protect\citeauthoryear{Qian, Sun, Wigington, Han, Sun, Healey,
  Tompkin, and Huang}{Qian et~al\mbox{.}}{2022}]%
        {qian2022dually}
\bibfield{author}{\bibinfo{person}{Jing Qian}, \bibinfo{person}{Qi Sun},
  \bibinfo{person}{Curtis Wigington}, \bibinfo{person}{Han~L Han},
  \bibinfo{person}{Tong Sun}, \bibinfo{person}{Jennifer Healey},
  \bibinfo{person}{James Tompkin}, {and} \bibinfo{person}{Jeff Huang}.}
  \bibinfo{year}{2022}\natexlab{}.
\newblock \showarticletitle{Dually noted: layout-aware annotations with
  smartphone augmented reality}. In \bibinfo{booktitle}{\emph{Proceedings of
  the 2022 CHI Conference on Human Factors in Computing Systems}}.
  \bibinfo{pages}{1--15}.
\newblock


\bibitem[\protect\citeauthoryear{Rajaram and Nebeling}{Rajaram and
  Nebeling}{2022}]%
        {Rajaram2022papertrail}
\bibfield{author}{\bibinfo{person}{Shwetha Rajaram} {and}
  \bibinfo{person}{Michael Nebeling}.} \bibinfo{year}{2022}\natexlab{}.
\newblock \showarticletitle{Paper Trail: An Immersive Authoring System for
  Augmented Reality Instructional Experiences}. In
  \bibinfo{booktitle}{\emph{Proceedings of the 2022 CHI Conference on Human
  Factors in Computing Systems}} (New Orleans, LA, USA)
  \emph{(\bibinfo{series}{CHI '22})}. \bibinfo{publisher}{Association for
  Computing Machinery}, \bibinfo{address}{New York, NY, USA}, Article
  \bibinfo{articleno}{382}, \bibinfo{numpages}{16}~pages.
\newblock
\showISBNx{9781450391573}
\urldef\tempurl%
\url{https://doi.org/10.1145/3491102.3517486}
\showDOI{\tempurl}


\bibitem[\protect\citeauthoryear{Rodriguez-Clark}{Rodriguez-Clark}{2019}]%
        {InteractiveMaths:online}
\bibfield{author}{\bibinfo{person}{Daniel Rodriguez-Clark}.}
  \bibinfo{year}{2019}\natexlab{}.
\newblock \bibinfo{title}{Interactive Maths}.
\newblock \bibinfo{howpublished}{\url{https://www.interactive-maths.com/}}.
\newblock
\newblock
\shownote{(Accessed on 04/03/2023).}


\bibitem[\protect\citeauthoryear{{Sarkar}, {Kadam}, and {Pillai}}{{Sarkar}
  et~al\mbox{.}}{2019}]%
        {Sarkar2019collaborative}
\bibfield{author}{\bibinfo{person}{P. {Sarkar}}, \bibinfo{person}{K. {Kadam}},
  {and} \bibinfo{person}{J.~S. {Pillai}}.} \bibinfo{year}{2019}\natexlab{}.
\newblock \showarticletitle{Collaborative Approaches to Problem-Solving on
  Lines and Angles Using Augmented Reality}. In \bibinfo{booktitle}{\emph{2019
  IEEE Tenth International Conference on Technology for Education (T4E)}}.
  \bibinfo{pages}{193--200}.
\newblock
\urldef\tempurl%
\url{https://doi.org/10.1109/T4E.2019.00-24}
\showDOI{\tempurl}


\bibitem[\protect\citeauthoryear{Scott and Davis}{Scott and Davis}{2013}]%
        {scott2013physink}
\bibfield{author}{\bibinfo{person}{Jeremy Scott} {and} \bibinfo{person}{Randall
  Davis}.} \bibinfo{year}{2013}\natexlab{}.
\newblock \showarticletitle{Physink: sketching physical behavior}. In
  \bibinfo{booktitle}{\emph{Adjunct Proceedings of the 26th Annual ACM
  Symposium on User Interface Software and Technology}}.
  \bibinfo{pages}{9--10}.
\newblock


\bibitem[\protect\citeauthoryear{Smilkov, Carter, Sculley, Vi{\'e}gas, and
  Wattenberg}{Smilkov et~al\mbox{.}}{2017}]%
        {smilkov2017direct}
\bibfield{author}{\bibinfo{person}{Daniel Smilkov}, \bibinfo{person}{Shan
  Carter}, \bibinfo{person}{D Sculley}, \bibinfo{person}{Fernanda~B
  Vi{\'e}gas}, {and} \bibinfo{person}{Martin Wattenberg}.}
  \bibinfo{year}{2017}\natexlab{}.
\newblock \showarticletitle{Direct-manipulation visualization of deep
  networks}.
\newblock \bibinfo{journal}{\emph{arXiv preprint arXiv:1708.03788}}
  (\bibinfo{year}{2017}).
\newblock


\bibitem[\protect\citeauthoryear{Song, Guimbretiere, Grossman, and
  Fitzmaurice}{Song et~al\mbox{.}}{2010}]%
        {Song2010mouseLight}
\bibfield{author}{\bibinfo{person}{Hyunyoung Song}, \bibinfo{person}{Francois
  Guimbretiere}, \bibinfo{person}{Tovi Grossman}, {and} \bibinfo{person}{George
  Fitzmaurice}.} \bibinfo{year}{2010}\natexlab{}.
\newblock \showarticletitle{MouseLight: Bimanual Interactions on Digital Paper
  Using a Pen and a Spatially-Aware Mobile Projector}. In
  \bibinfo{booktitle}{\emph{Proceedings of the SIGCHI Conference on Human
  Factors in Computing Systems}} (Atlanta, Georgia, USA)
  \emph{(\bibinfo{series}{CHI '10})}. \bibinfo{publisher}{Association for
  Computing Machinery}, \bibinfo{address}{New York, NY, USA},
  \bibinfo{pages}{2451–2460}.
\newblock
\showISBNx{9781605589299}
\urldef\tempurl%
\url{https://doi.org/10.1145/1753326.1753697}
\showDOI{\tempurl}


\bibitem[\protect\citeauthoryear{Stewart}{Stewart}{2015}]%
        {stewart2015single}
\bibfield{author}{\bibinfo{person}{James Stewart}.}
  \bibinfo{year}{2015}\natexlab{}.
\newblock \bibinfo{booktitle}{\emph{Single variable calculus: Early
  transcendentals}}.
\newblock \bibinfo{publisher}{Cengage Learning}.
\newblock


\bibitem[\protect\citeauthoryear{Subramonyam, Drucker, and Adar}{Subramonyam
  et~al\mbox{.}}{2019}]%
        {subramonyam2019affinity}
\bibfield{author}{\bibinfo{person}{Hariharan Subramonyam},
  \bibinfo{person}{Steven~M Drucker}, {and} \bibinfo{person}{Eytan Adar}.}
  \bibinfo{year}{2019}\natexlab{}.
\newblock \showarticletitle{Affinity lens: data-assisted affinity diagramming
  with augmented reality}. In \bibinfo{booktitle}{\emph{Proceedings of the 2019
  CHI conference on human factors in computing systems}}.
  \bibinfo{pages}{1--13}.
\newblock


\bibitem[\protect\citeauthoryear{Suselo, W\"{u}nsche, and Luxton-Reilly}{Suselo
  et~al\mbox{.}}{2021}]%
        {Suselo2021using}
\bibfield{author}{\bibinfo{person}{Thomas Suselo}, \bibinfo{person}{Burkhard~C.
  W\"{u}nsche}, {and} \bibinfo{person}{Andrew Luxton-Reilly}.}
  \bibinfo{year}{2021}\natexlab{}.
\newblock \showarticletitle{Using Mobile Augmented Reality for Teaching 3D
  Transformations}. In \bibinfo{booktitle}{\emph{Proceedings of the 52nd ACM
  Technical Symposium on Computer Science Education}} (Virtual Event, USA)
  \emph{(\bibinfo{series}{SIGCSE '21})}. \bibinfo{publisher}{Association for
  Computing Machinery}, \bibinfo{address}{New York, NY, USA},
  \bibinfo{pages}{872–878}.
\newblock
\showISBNx{9781450380621}
\urldef\tempurl%
\url{https://doi.org/10.1145/3408877.3432401}
\showDOI{\tempurl}


\bibitem[\protect\citeauthoryear{Suzuki, Kazi, Wei, DiVerdi, Li, and
  Leithinger}{Suzuki et~al\mbox{.}}{2020}]%
        {suzuki2020realitysketch}
\bibfield{author}{\bibinfo{person}{Ryo Suzuki}, \bibinfo{person}{Rubaiat~Habib
  Kazi}, \bibinfo{person}{Li-Yi Wei}, \bibinfo{person}{Stephen DiVerdi},
  \bibinfo{person}{Wilmot Li}, {and} \bibinfo{person}{Daniel Leithinger}.}
  \bibinfo{year}{2020}\natexlab{}.
\newblock \showarticletitle{Realitysketch: Embedding responsive graphics and
  visualizations in AR through dynamic sketching}. In
  \bibinfo{booktitle}{\emph{Proceedings of the 33rd Annual ACM Symposium on
  User Interface Software and Technology}}. \bibinfo{pages}{166--181}.
\newblock


\bibitem[\protect\citeauthoryear{Suzuki, Soares, Head, Glassman, Reis,
  Mongiovi, D'Antoni, and Hartmann}{Suzuki et~al\mbox{.}}{2017}]%
        {suzuki2017tracediff}
\bibfield{author}{\bibinfo{person}{Ryo Suzuki}, \bibinfo{person}{Gustavo
  Soares}, \bibinfo{person}{Andrew Head}, \bibinfo{person}{Elena Glassman},
  \bibinfo{person}{Ruan Reis}, \bibinfo{person}{Melina Mongiovi},
  \bibinfo{person}{Loris D'Antoni}, {and} \bibinfo{person}{Bjoern Hartmann}.}
  \bibinfo{year}{2017}\natexlab{}.
\newblock \showarticletitle{Tracediff: Debugging unexpected code behavior using
  trace divergences}. In \bibinfo{booktitle}{\emph{2017 IEEE Symposium on
  Visual Languages and Human-Centric Computing (VL/HCC)}}. IEEE,
  \bibinfo{pages}{107--115}.
\newblock


\bibitem[\protect\citeauthoryear{Ubavić}{Ubavić}{[n.d.]}]%
        {awesome:online}
\bibfield{author}{\bibinfo{person}{Nikola Ubavić}.}
  \bibinfo{year}{[n.d.]}\natexlab{}.
\newblock \bibinfo{title}{GitHub - ubavic/awesome-interactive-math: A curated
  list of tools that can be used for creating interactive mathematical
  explorables.}
\newblock
  \bibinfo{howpublished}{\url{https://github.com/ubavic/awesome-interactive-math}}.
\newblock
\newblock
\shownote{(Accessed on 04/03/2023).}


\bibitem[\protect\citeauthoryear{Verou, Zhang, and Karger}{Verou
  et~al\mbox{.}}{2016}]%
        {verou2016mavo}
\bibfield{author}{\bibinfo{person}{Lea Verou}, \bibinfo{person}{Amy~X Zhang},
  {and} \bibinfo{person}{David~R Karger}.} \bibinfo{year}{2016}\natexlab{}.
\newblock \showarticletitle{Mavo: creating interactive data-driven web
  applications by authoring HTML}. In \bibinfo{booktitle}{\emph{Proceedings of
  the 29th Annual Symposium on User Interface Software and Technology}}.
  \bibinfo{pages}{483--496}.
\newblock


\bibitem[\protect\citeauthoryear{Victor}{Victor}{2011a}]%
        {victor2011explorable}
\bibfield{author}{\bibinfo{person}{Bret Victor}.}
  \bibinfo{year}{2011}\natexlab{a}.
\newblock \bibinfo{title}{Explorable explanations}.
\newblock
\newblock


\bibitem[\protect\citeauthoryear{Victor}{Victor}{2011b}]%
        {victor2011killmath}
\bibfield{author}{\bibinfo{person}{Bret Victor}.}
  \bibinfo{year}{2011}\natexlab{b}.
\newblock \bibinfo{title}{KillMath}.
\newblock
\newblock


\bibitem[\protect\citeauthoryear{Victor}{Victor}{2012}]%
        {victor2012learnable}
\bibfield{author}{\bibinfo{person}{Bret Victor}.}
  \bibinfo{year}{2012}\natexlab{}.
\newblock \showarticletitle{Learnable programming}.
\newblock \bibinfo{journal}{\emph{Worrydream. com}} (\bibinfo{year}{2012}).
\newblock


\bibitem[\protect\citeauthoryear{Wang, Turko, Shaikh, Park, Das, Hohman, Kahng,
  and Chau}{Wang et~al\mbox{.}}{2020}]%
        {wang2020cnn}
\bibfield{author}{\bibinfo{person}{Zijie~J Wang}, \bibinfo{person}{Robert
  Turko}, \bibinfo{person}{Omar Shaikh}, \bibinfo{person}{Haekyu Park},
  \bibinfo{person}{Nilaksh Das}, \bibinfo{person}{Fred Hohman},
  \bibinfo{person}{Minsuk Kahng}, {and} \bibinfo{person}{Duen Horng~Polo
  Chau}.} \bibinfo{year}{2020}\natexlab{}.
\newblock \showarticletitle{CNN explainer: learning convolutional neural
  networks with interactive visualization}.
\newblock \bibinfo{journal}{\emph{IEEE Transactions on Visualization and
  Computer Graphics}} \bibinfo{volume}{27}, \bibinfo{number}{2}
  (\bibinfo{year}{2020}), \bibinfo{pages}{1396--1406}.
\newblock


\bibitem[\protect\citeauthoryear{Wellner}{Wellner}{1991}]%
        {wellner1991digitaldesk}
\bibfield{author}{\bibinfo{person}{Pierre Wellner}.}
  \bibinfo{year}{1991}\natexlab{}.
\newblock \showarticletitle{The DigitalDesk calculator: tangible manipulation
  on a desk top display}. In \bibinfo{booktitle}{\emph{Proceedings of the 4th
  annual ACM symposium on User interface software and technology}}.
  \bibinfo{pages}{27--33}.
\newblock


\bibitem[\protect\citeauthoryear{Zhao, Qin, Liu, Liu, Zhang, and Shi}{Zhao
  et~al\mbox{.}}{2014}]%
        {zhao2014qook}
\bibfield{author}{\bibinfo{person}{Yuhang Zhao}, \bibinfo{person}{Yongqiang
  Qin}, \bibinfo{person}{Yang Liu}, \bibinfo{person}{Siqi Liu},
  \bibinfo{person}{Taoshuai Zhang}, {and} \bibinfo{person}{Yuanchun Shi}.}
  \bibinfo{year}{2014}\natexlab{}.
\newblock \showarticletitle{QOOK: enhancing information revisitation for active
  reading with a paper book}. In \bibinfo{booktitle}{\emph{Proceedings of the
  8th International Conference on Tangible, Embedded and Embodied
  Interaction}}. \bibinfo{pages}{125--132}.
\newblock


\end{thebibliography}

\appendix
\section{Analyzed Websites}\label{appendix}

In the following tables (\autoref{tab:urls1} and \autoref{tab:urls2}), we list the interactive websites analyzed in Section 3. For each site, we provide the ID, name, and link, along with the identified design strategy, input method, and domain.

\begin{figure*}[h]
\centering
\includegraphics[width=\textwidth]{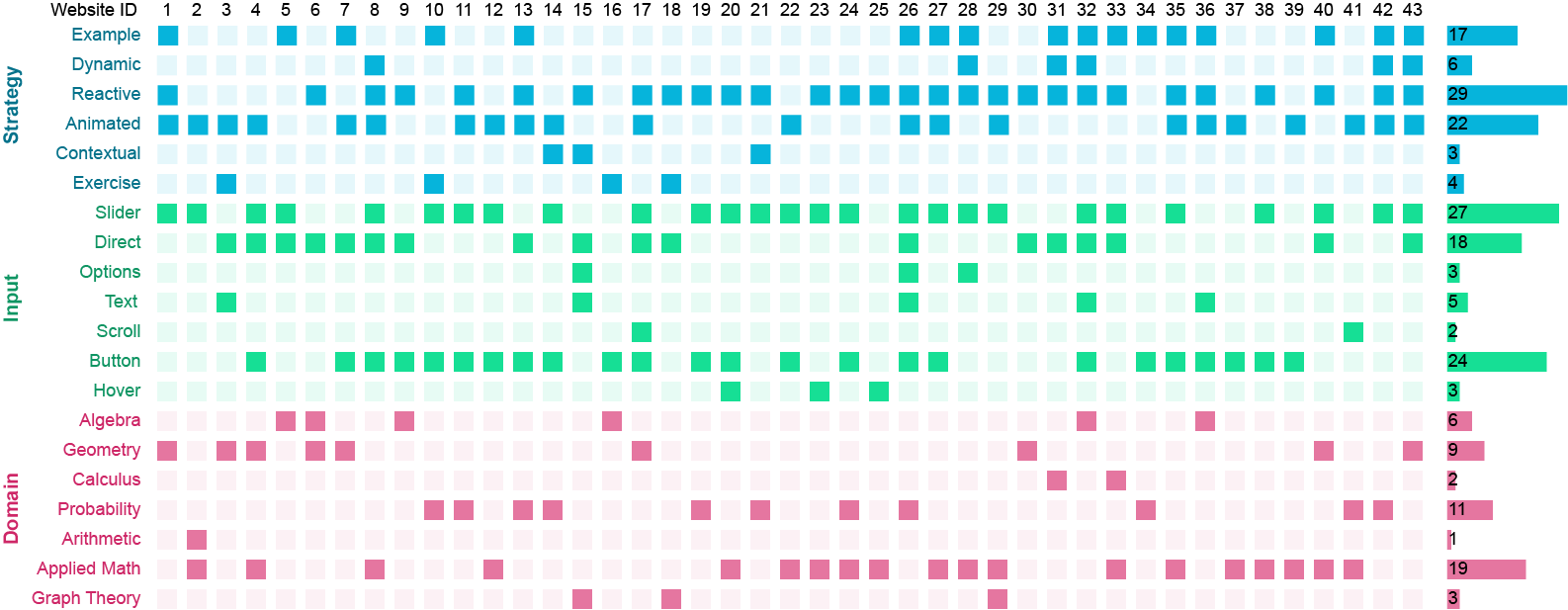}
\caption{Taxonomy analysis}
\label{fig:taxonomy-analysis}
\end{figure*}

\begin{table*}[b]
\renewcommand*{\arraystretch}{1.2}
\small
\centering
\begin{tabular}{c p{0.25\textwidth} p{0.25\textwidth} p{0.12\textwidth} p{0.12\textwidth} p{0.12\textwidth}} 
\textbf{ID} & \textbf{Name} & \textbf{Link} & \textbf{\blue{Strategy}} & \textbf{\green{Input}} & \textbf{\red{Domain}} \\
\hline
1 & Visualize It - Trigonometric Functions & \url{https://visualize-it.github.io/trig\\_functions/simulation.html} & Animated, Example, Reactive & Slider & Geometry \\
\hline
2 & Exponentiation - Explained Visually & \url{https://setosa.io/ev/exponentiation/}& Animated & Slider & Applied Math, Arithmetic \\
\hline
3 & Relating Area to Circumference & \url{https://www.geogebra.org/m/w5rczf8n\#material/bq3gwxxn}& Animated, Exercise & Direct, Text& Geometry \\
\hline
4 & Seeing circles, sines, and signals & \url{https://jackschaedler.github.io/circles-sines-signals/index.html} & Animated & Button, Direct, Slider & Applied Math, Geometry \\
\hline
5 & Systems of Linear Algebra & \url{https://textbooks.math.gatech.edu/ila/systems-of-eqns.html} & Example & Direct, Slider & Algebra \\
\hline
6 & Let's remove Quaternions from every 3D Engine & \url{http://marctenbosch.com/quaternions/} & Reactive & Direct & Algebra, Geometry\\
\hline
7 & Trigonometry for Games & \url{https://demoman.net/?a=trig-for-games} & Animated, Example & Button, Direct & Geometry \\
\hline
8 & Back to the future of handwriting recognition & \url{https://jackschaedler.github.io/handwriting-recognition/} & Animated, Dynamic Calculation, Reactive & Button, Direct, Slider & Applied Math \\
\hline
9 & Slower speed of light & \url{http://gamelab.mit.edu/games/a-slower-speed-of-light/} & Reactive & Button, Direct & Algebra \\
\hline
10 & The Taxi Cab Problem & \url{http://galgreen.com/TaxiCabProblem/\#0} & Example, Exercise & Button, Slider & Probability \\
\hline
11 & What Happens Next? COVID-19 Futures & \url{https://ncase.me/covid-19/} & Animated, Reactive & Button, Slider & Probability \\
\hline
12 & Gears – Bartosz Ciechanowski & \url{https://ciechanow.ski/gears/} & Animated & Button, Slider & Applied Math \\
\hline
13 & Seeing Theory Ch. 1: Basic Probability & \url{https://seeing-theory.brown.edu/basic-probability/index.html} & Animated, Example, Reactive & Button, Direct & Probability\\
\hline
14 & Going Critical — Melting Asphalt & \url{https://meltingasphalt.com/interactive/going-critical/}& Animated, Contextual & Button, Slider & Probability\\
\hline
15 & Graphs and Networks & \url{https://mathigon.org/course/graph-theory/introduction} & Contextual, Reactive & Direct, Options, Text & Graph Theory \\
\hline
16 & Quantum computing for the very curious& \url{https://quantum.country/qcvc} & Exercise & Button & Algebra\\
\hline
17 & An Interactive Introduction to Fourier Transforms & \url{https://www.jezzamon.com/fourier/index.html} & Animated, Reactive & Button, Direct, Scroll, Slider& Geometry \\
\hline
18 & The Wisdom and/or Madness of Crowds & \url{https://ncase.me/crowds/} & Exercise, Reactive & Direct & Graph Theory \\
\hline
19 & Complexity Explorables | Horde of the Flies & \url{https://www.complexity-explorables.org/explorables/horde-of-the-flies/} & Reactive & Button, Slider & Probability\\
\hline
20& Visualizing the Impact of Feature Attribution Baselines & \url{https://distill.pub/2020/attribution-baselines/} & Reactive & Button, Hover, Slider & Applied Math \\
\hline
21& Exploring Bayesian Optimization& \url{https://distill.pub/2020/bayesian-optimization/} & Contextual, Reactive& Slider & Probability\\
\hline
22& The Paths Perspective on Value Learning & \url{https://distill.pub/2019/paths-perspective-on-value-learning/}& Animated & Button, Slider & Applied Math \\
\hline
23& Computing Receptive Fields of CNN & \url{https://distill.pub/2019/computing-receptive-fields/} & Reactive & Hover, Slider & Applied Math \\
\hline
24& A Visual Exploration of Gaussian Processes & \url{https://distill.pub/2019/visual-exploration-gaussian-processes/} & Reactive & Button, Slider & Applied Math, Probability \\
\hline
25& Understanding RL Vision & \url{https://distill.pub/2020/understanding-rl-vision/} & Reactive & Hover& Applied Math \\
\hline
\end{tabular}
\caption{Documents analyzed in Study 1 (continued in \autoref{tab:urls2}). Each website is listed with the ID that identifies it in the paper, it's name, link, the design strategy, input method, and domain.}
\label{tab:urls1}
\end{table*}

\begin{table*}[t]
\renewcommand*{\arraystretch}{1.2}
\small
\centering
\begin{tabular}{c p{0.25\textwidth} p{0.25\textwidth} p{0.12\textwidth} p{0.12\textwidth} p{0.12\textwidth}} 
\textbf{ID} & \textbf{Name} & \textbf{Link} & \textbf{\blue{Strategy}}& \textbf{\green{Input}} & \textbf{\red{Domain}}\\
\hline
26& Seeing Theory Probability Distributions & \url{https://seeing-theory.brown.edu/probability-distributions/index.html}& Animated, Example, Reactive & Button, Direct, Options, Slider, Text & Probability\\
\hline
27& Firefly Synchronization & \url{https://visualize-it.github.io/firefly\\_synchronization/simulation.html}& Animated, Example, Reactive & Button, Slider & Applied Math \\
\hline
28& Fourier Series& \url{https://visualize-it.github.io/fourier\\_series/simulation.html} & Dynamic Calculation,Example,Reactive & Options, Slider& Applied Math \\
\hline
29& Markov Chains & \url{https://setosa.io/blog/2014/07/26/markov-chains/index.html}& Animated, Reactive & Slider & Applied Math, Graph Theory \\
\hline
30& Pythagorean Theorem & \url{https://setosa.io/pythagorean/} & Reactive & Direct & Geometry \\
\hline
31& Sine and Cosine & \url{https://setosa.io/ev/sine-and-cosine/} & Dynamic Calculation, Example, Reactive & Direct & Calculus \\
\hline
32& Vector Addition & \url{https://phet.colorado.edu/sims/html/vector-addition/latest/vector-addition\_en.html} & Dynamic Calculation, Example, Reactive & Button, Direct, Slider, Text& Algebra\\
\hline
33& Polynomial Regression& \url{https://visualize-it.github.io/polynomial\_regression/simulation.html} & Example, Reactive& Direct, Slider & Applied Math, Calculus \\
\hline
34& Random Walks & \url{https://visualize-it.github.io/random\_walk/simulation.html} & Example& Button & Probability\\
\hline
35& Porous Percolation& \url{https://visualize-it.github.io/porous\_percolation/simulation.html } & Animated, Example, Reactive & Button, Slider & Applied Math \\
\hline
36& Linear Transformations & \url{https://visualize-it.github.io/linear\_transformations/simulation.html} & Animated, Example, Reactive & Button, Text& Algebra\\
\hline
37& Mandelbrot Fractal& \url{https://visualize-it.github.io/mandelbrot\_fractal/simulation.html}& Animated & Button & Applied Math \\
\hline
38& Bernoulli Percolation& \url{https://visualize-it.github.io/bernoulli\_percolation/simulation.html} & Reactive & Button, Slider & Applied Math \\
\hline
39& Hilbert Curve & \url{https://visualize-it.github.io/hilbert\_curve/simulation.html}& Animated & Button & Applied Math \\
\hline
40& A Primer on Bézier Curves & \url{https://pomax.github.io/bezierinfo/} & Example, Reactive& Direct, Slider & Applied Math, Geometry \\
\hline
41& A visual introduction to machine learning& \url{http://www.r2d3.us/visual-intro-to-machine-learning-part-1/} & Animated & Scroll & Applied Math, Probability \\
\hline
42& Conditional probability & \url{https://setosa.io/conditional/} & Animated, Dynamic Calculation, Example, Reactive & Slider & Probability\\
\hline
43& Pi ($\pi$)& \url{https://setosa.io/ev/pi/}& Animated, Dynamic Calculation, Example, Reactive & Direct, Slider & Geometry \\
\hline 
\end{tabular}
 \caption{Documents analyzed in Study 1 (continued from \autoref{tab:urls1}). Each website is listed with the ID that identifies it in the paper, it's name, link, the design strategy, input method, and domain.}
 \label{tab:urls2}
\end{table*}

\end{document}